\documentclass[journal, letter, 12pt, onecolumn, %
]{IEEEtran}
\usepackage[utf8]{inputenc} 
\usepackage[T1]{fontenc}
\usepackage{url}
\usepackage{ifthen}
\usepackage{cite}
\usepackage{changes}
\usepackage[cmex10]{amsmath} %
\usepackage[mathcal]{euscript}
\usepackage{hyperref}  %
\usepackage{amssymb, amsfonts, amscd,dsfont, mathrsfs, mathtools}
\usepackage{xcolor}
\usepackage{tikz}
\usepackage[export]{adjustbox}
\usetikzlibrary{arrows, calc}
\usetikzlibrary{patterns}

\ifCLASSOPTIONcompsoc
 \usepackage[caption=false,font=normalsize,labelfont=sf,textfont=sf]{subfig}
\else
 \usepackage[caption=false,font=footnotesize]{subfig}
\fi

\newcommand{\msgset}[1]{\lceil #1 \rfloor}
\newcommand{\msg}{\msgset}
\newcommand{\lfa}{\leftrightarrow}

\newcommand{\ibar}{\bar{\imath}}
\newcommand{\jbar}{\bar{\jmath}}

\newcommand{\divp}{D^*}

\newcommand{\redu}{\omega} 
\newcommand{\redux}{\redu_X} 
\newcommand{\reduy}{\redu_Y} 

\newcommand{\defeq}{\triangleq}

\newcommand{\pr}{\mathbb{P}}
\newcommand{\prob}[1]{\pr\left\{#1\right\}}

\newcommand{\e}{\mathrm{e}}
\newcommand{\eps}{\epsilon}

\newcommand{\nx}{{\sf N}_X}
\newcommand{\ny}{{\sf N}_Y}

\newcommand{\Fc}{\mathcal{F}}
\newcommand{\Dec}{\Fc} 
\newcommand{\dec}{{\phi}} 
\newcommand{\deca}{{\psi}} 
\newcommand{\dect}{\tilde{\phi}} 
\newcommand{\decr}{{\varphi}} 
\newcommand{\enct}{{\theta}} 
\newcommand{\encti}{\vartheta} 

\newcommand{\enctx}{\enct_X} 
\newcommand{\encty}{\enct_Y} 

\newcommand{\enctix}{\encti_X} 
\newcommand{\enctiy}{\encti_Y} 

\newtheorem{definition}{Definition}
\newtheorem{theorem}{Theorem}
\newtheorem{proposition}{Proposition}
\newtheorem{lemma}{Lemma}
\newtheorem{corollary}{Corollary}
\newtheorem{remark}{Remark}

\newtheorem{example}{Example}
\newtheorem{fact}{Fact}

\DeclareMathAlphabet{\mathbbb}{U}{bbold}{m}{n}  

\DeclareMathOperator*{\argmin}{arg\,min}
\DeclareMathOperator*{\argmax}{arg\,max}

\DeclareMathOperator*{\minimize}{minimize}
\DeclareMathOperator*{\st}{subject~to}

\DeclareMathOperator{\interior}{int}

\newcommand{\ds}{\displaystyle}
\newcommand{\secref}[1]{Section~\ref{#1}}
\newcommand{\appref}[1]{Appendix~\ref{#1}}
\newcommand{\exref}[1]{Example~\ref{#1}}
\newcommand{\figref}[1]{Fig.~\ref{#1}}
\newcommand{\lemref}[1]{Lemma~\ref{#1}}
\newcommand{\thmref}[1]{Theorem~\ref{#1}}
\newcommand{\propref}[1]{Proposition~\ref{#1}}

\newcommand{\factref}[1]{Fact~\ref{#1}}
\newcommand{\defref}[1]{Definition~\ref{#1}}

\newcommand{\xedit}[1]{\textcolor{red}{\##1\#}}

\newcommand{\Ac}{\mathcal{A}}
\newcommand{\Ic}{\mathcal{I}}
\newcommand{\Icx}{\Ic_X}
\newcommand{\Icy}{\Ic_Y}
\newcommand{\Mc}{\mathcal{M}}
\newcommand{\kron}{\mathbbb{1}}

\newcommand{\Cc}{{\mathcal{C}}}
\newcommand{\Cct}{\tilde{\Cc}}
\newcommand{\Ec}{{\mathcal{E}}}

\newcommand{\cX}{{\mathcal{X}}}
\newcommand{\cY}{{\mathcal{Y}}}
\newcommand{\cP}{{\mathcal{P}}}
\newcommand{\Pc}{{\mathcal{P}}}
\newcommand{\Pch}{{\hat{\Pc}}}

\newcommand{\X}{{\mathcal{X}}}
\newcommand{\Y}{{\mathcal{Y}}}
\newcommand{\Z}{{\mathcal{Z}}}
\newcommand{\Nc}{\mathcal{N}}
\newcommand{\Rc}{\mathcal{R}}

\newcommand{\Ph}{\hat{P}}

\newcommand{\Qt}{\tilde{Q}}
\newcommand{\Xt}{\tilde{X}}
\newcommand{\Yt}{\tilde{Y}}

\newcommand{\Qh}{\hat{Q}}

\newcommand{\nbhdh}{\Nc_{\mathrm{H}}}

\newcommand{\Sc}{\mathcal{S}}

\newcommand{\Hc}{\mathcal{H}}
\newcommand{\Qc}{\mathcal{Q}}

\newcommand{\redset}{\tau}

\newcommand{\Ab}{\mathbf{A}}

\newcommand{\Bc}{\mathcal{B}}
\newcommand{\Tc}{\mathcal{T}}
\newcommand{\Hs}{{\sf H}}
\newcommand{\Hh}{\hat{\sf H}}
\newcommand{\Hhs}{\hat{\sf H}}
\newcommand{\bd}{\partial_*}
\newcommand{\ft}{\tilde{f}}
\newcommand{\gt}{\tilde{g}}

\newcommand{\vtx}{\mathsf{u}}
\newcommand{\vty}{\mathsf{v}}

\newcommand{\Dc}{\mathcal{D}}
\newcommand{\pdist}{\cP_{\!\star}}

\newcommand{\dmax}{d_{\max}}
\newcommand{\dmaxp}{d_{\star}}
\newcommand{\subfunc}[1]{\overset{  \raisebox{-3pt}{\resizebox{!}{3pt}{#1}}}{\raisebox{-1.2pt}{$\triangleright$}}}
\DeclareMathOperator{\sub}{\mathbin{\triangleright}}
\DeclareMathOperator{\subx}{\mathbin{\subfunc{$X$}}}
\DeclareMathOperator{\suby}{\mathbin{\subfunc{$Y$}}}

\newcommand{\projx}[1]{\Pi_X(#1)}
\newcommand{\projy}[1]{\Pi_Y(#1)}

\newcommand{\redus}{\redu^*} 
\newcommand{\Decc}{\bar{\Omega}} 
\newcommand{\Deccn}{{\Omega}} 
\newcommand{\Decs}{\Deccn^{(0)}} 
\newcommand{\Decd}{\Deccn^{(1)}} 

\newcommand{\chibar}{\bar{\chi}}

\definecolor{mygray}{gray}{0.3}

\newcommand{\drawgrid}[2][mygray]{ %
  \adjustbox{valign=c}{
    \begin{tikzpicture}
  \fill[#1]
    \foreach \row [count=\y] in {#2} {
      \foreach \cell [count=\x] in \row {
        \ifnum\cell=1 %
          (\x-1, -\y+1) rectangle ++(1, -1)
        \fi
        \pgfextra{%
          \global\let\maxx\x
          \global\let\maxy\y
        }%
      }
    }
  ;
  \draw[line width = .2em] (0, 0) grid[step=1] (\maxx, -\maxy);
  \end{tikzpicture}
}}

\newcommand{\gheight}{0.2 em}
\newcommand{\putgrid}[2][\gheighteq]{\raisebox{\gheight}{\resizebox{!}{#1}{#2}}}

\newcommand{\union}{\cup}

\usetikzlibrary{fit}
\usetikzlibrary{shapes}
\tikzstyle{color D0}=[fill opacity = .9, fill = green!20]
\tikzstyle{color D0L}=[fill opacity=0.2, fill = green!20, ]
\tikzstyle{color D1}=[fill opacity = .9, fill = blue!50]
\tikzstyle{my pattern} = [pattern=north west lines, pattern color=black]
\tikzstyle{my pattern dots} = [pattern=dots, pattern color=gray]

\begin{document}

\title{On Distributed Learning with Constant Communication Bits}

\author{
Xiangxiang Xu,~\IEEEmembership{Member,~IEEE}
and~Shao-Lun Huang,~\IEEEmembership{Member,~IEEE}%
\thanks{This paper was presented in part at the Information Theory Workshop (ITW-2021), Kanazawa, Japan, Oct. 2021 \cite{xu2021distributed}.}
\thanks{X. Xu and S.-L. Huang are with the Data Science and Information Technology Research Center, Tsinghua--Berkeley Shenzhen Institute, Shenzhen 518055, China (e-mail: xiangxiangxu@ieee.org, shaolun.huang@sz.tsinghua.edu.cn).}
}

\maketitle

\begin{abstract}
  In this paper, we study a distributed learning problem constrained by constant communication bits. Specifically, we consider the distributed hypothesis testing (DHT) problem where two distributed nodes are constrained to transmit a constant number of bits to a central decoder. In such cases, we show that in order to achieve the optimal error exponents, it suffices to consider the empirical distributions of observed data sequences and encode them to the transmission bits. With such a coding strategy, we develop a geometric approach in the distribution spaces and establish an inner bound of error exponent regions. In particular, we show the optimal achievable error exponents and coding schemes for the following cases: (i) both nodes can transmit $\log_23$ bits; (ii) one of the nodes can transmit $1$ bit, and the other node is not constrained; (iii) the joint distribution of the nodes are conditionally independent given one hypothesis. Furthermore, we provide several numerical examples for illustrating the theoretical results. Our results provide theoretical guidance for designing practical distributed learning rules, and the developed approach also reveals new potentials for establishing error exponents for DHT with more general communication constraints.  
\end{abstract}

\IEEEpeerreviewmaketitle

%
%
%
%

\section{Introduction}

The rapid development of IoT (Internet of Things) technology has led to unprecedented advances in efficient data collection, where comprehensive descriptions of physical events are provided by distributed sensory nodes \cite{verbraeken2020survey}. Despite of the large amount of available samples, effectively analyzing such sensory data can be challenging in real systems \cite{mcmahan2017communication}, due to the distributed observations at different sensory nodes, and the communication constraints between nodes and centers. In this paper, we investigate the fundamental limit of such distributed learning scenarios, where we assume the nodes can only communicate to the decision center with a constant number of bits, i.e., independent of the observed sequence length. Our goal is to characterize the performance of such distributed systems by the statistical dependency of the observations at different nodes, the communication constraints, and the central fusion rule.

In particular, we consider a distributed hypothesis testing (DHT) problem, with a pair of random variables $X, Y$ and joint distributions $P^{(0)}_{XY}$ and $P^{(1)}_{XY}$. In addition, there are $n$ samples drawn in an independently, identically distributed (i.i.d.) manner from either $P^{(0)}_{XY}$ or $P^{(1)}_{XY}$, which may correspond to the two hypothesis $\Hs = 0$ and $\Hs = 1$ in statistics, or different labels in supervised learning problems. Moreover, in the distributed setup, we assume that there are two nodes, referred to as node  $\nx$ and node $\ny$, each observes only the $n$ i.i.d. samples of $X$ and the samples of $Y$, respectively, and each node sends an encoded message to a central decoder. Then, the decoder makes a decision of the hypothesis $\Hh$ according to the received messages. Specifically, we assume that the number of communication bits cannot exceed some given constants, independent of $n$, and both nodes are required to encode (compress) the observed length-$n$ sequences to the message subject to the communication constraints. Our goal is to design the encoder of each node and the central decoder to minimize the error probability of inferring the label. We focus on the asymptotic regime such that $n$ is large, and characterize the error exponent pair for both types-I and type-II errors. The rigorous mathematical formulation is presented in \secref{sec:pf}.

The general framework of such multiterminal statistical inference problems was first introduced in \cite{berger1979decentralized}. Following this proposal, the DHT problem with full side information was formulated and investigated in \cite{ahlswede1986hypothesis}, where the sequence observed by $\ny$ can be directly transmitted to the center, while $\nx$ can only send messages at some positive rate. Following this work, there have been a series of studies on DHT under different settings of communication constraints, which are typically represented as the communications rates, or equivalently, the compression rates of the encoders. Specifically, the DHT problem with zero-rate compression was first introduced in \cite{han1987hypothesis}, where the one-bit compression (also known as \emph{complete data compression}) constraint  was also discussed.
The achievable error exponent pairs under two-sided one-bit compression were later established in \cite{han1989exponential}. The DHT problem under zero-rate compression was also investigated in \cite{amari1989statistical,shalaby1992multiterminal}.  
A comprehensive survey of representative works through this line of researches can be found in \cite{han1998statistical}. 
{Recently, the studies on DHT are still fairly active \cite{watanabe2017neyman, rahman2012optimality, hadar2019error, haim2016binary, xu2021information, zhao_lai_2021}, with new analyzing tools and settings considered, e.g., DHT with interactive extensions \cite{xiang2012interactive} and sequential extensions \cite{salehkalaibar2021distributed} , DHT over relay networks \cite{salehkalaibar2019hypothesis}, and DHT over noisy channels \cite{sreekumar2019distributed}. }Despite of such massive studies, the characterizations of DHT under general communication constraints still remain open, except for several special cases, e.g., the testing against independence problem with full side information \cite{ahlswede1986hypothesis}, or the zero-rate compression setting \cite{han1998statistical}. Specifically, for DHT with constant communication bits, previous discussions were restricted to
the one-bit compression setting \cite{han1987hypothesis, han1989exponential, han1998statistical,
  escamilla2018distributed, escamilla2020distributed}.

The primary aim of this paper is to investigate the optimal error exponent pairs of DHT with constant communication bits, and the main contributions are as follows. First,
we demonstrate that %
the optimal encoding scheme depends only on the empirical distributions of the observed sequences, rather than the sequences themselves, as long as the compression rates are zeros. %
With this coding strategy, we develop a geometric approach in the distribution spaces to characterize the achievable error exponent pairs.  Using this approach, we further provide an inner bound of the error exponent region, and compare the performance under different decoders.
In addition, we show that this inner bound is tight and establish the optimal error exponents, for the following cases: (i) \emph{two-sided one-trit compression}, where both nodes can transmit one-trit (trinary digit) message; (ii) \emph{one-sided one-bit compression}, where one node can transmit one bit, and the other node is not constraint; (iii) the nodes are conditionally independent given one hypothesis. Our characterization extends
 previous studies on two-sided one-bit compression (cf. \cite{han1989exponential, han1998statistical}) and provides a novel geometric interpretation, which suggests new potentials for error exponent region characterization of DHT under general communication constraints. %

 The rest of this paper is organized as follows. In \secref{sec:pf}, we introduce the problem formulation and related notations. Then, \secref{sec:optim-type-based} presents the optimal encoding scheme, and a geometric characterization of the achievable error exponents is provided in \secref{sec:geom-char-dht}. With such characterization, we establish the error exponent region and the optimal coding schemes under different communication settings in \secref{sec:error-exp}. Finally, we present numerical examples in \secref{sec:numerical-examples}, and conclude the paper with discussions in \secref{sec:diss}.

\section{Problem Formulation and Preliminaries}
\label{sec:pf}
In this section, we introduce the mathematical formulation of DHT problem, and also provide some useful definitions and notations.%

\subsection{Problem Formulation}
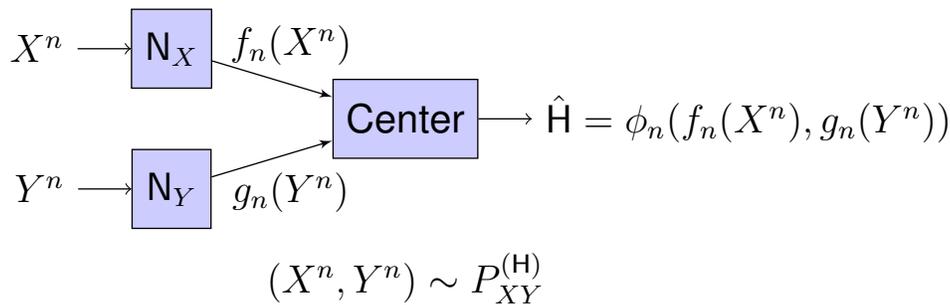
\begin{figure}[!t]
  \centering
  \resizebox{.75\textwidth}{!}{\tikzstyle{int}=[draw, fill=blue!20, minimum size=2em]
\tikzstyle{init} = [pin edge={to-,thin,black}]
\tikzstyle{initr} = [pin edge={-to,thin,black}]
\begin{tikzpicture}[node distance=2.5cm,auto,>=latex']
    \node [int, pin={[init]left:$X^n$}] (x) {$\nx$}; 
    \node [int, pin={[init]left:$Y^n$}] (y) [below of=x, node distance=1.5cm] {$\ny$};
    \coordinate (xy) at ($(x)!0.5!(y)$);
    
    \node [int] (c) [right of=xy, pin={[initr]right:$\Hhs = \dec_n(f_n(X^n), g_n(Y^n))$}] {{\sf Center }};
e    \node [] (s) [below of =c, node distance=1.7cm] {$(X^n, Y^n) \sim P^{(\Hs)}_{XY}$};
    \node [coordinate] (end) [right of=c, node distance=1.2cm]{};
    \path[->] (x) edge node [pos=.5, above, xshift = .2cm] {$f_n(X^n)$}  (c); 
    \path[->] (y) edge node [pos=.5, below, xshift = .2cm] {$g_n(Y^n)$}   (c);
\end{tikzpicture}

}
  \caption{Distributed Hypothesis Testing with Communication Constraints}  %
  \label{fig:blk}
\end{figure}

First, we assume both $X$ and $Y$ are discrete random variables, taking values from finite alphabets $\X$ and $\Y$, respectively. Then, the general setup of DHT is depicted in \figref{fig:blk}. When $\Hs = i$, $n$ i.i.d. sample pairs $\{(X_j, Y_j)\}_{j = 1}^n$ are generated from the joint distribution $P_{XY}^{(i)}$. Throughout our analyses, we assume that all entries of $P^{(0)}_{XY}$ and $P^{(1)}_{XY}$ 
are positive, i.e., for both $i = 0, 1$,
\begin{align}
   P^{(i)}_{XY}(x, y) > 0,  \quad\text{for all $x \in \X, y \in \Y$}.
   \label{eq:pos}
\end{align}

Then, node $\nx$ and node $\ny$ observe $X^n \defeq (X_1, \dots, X_n)$ and %
$Y^n \defeq (Y_1, \dots, Y_n)$, respectively, and encode their observed sequences to into messages $f_n(X^n)$ and $g_n(Y^n)$, where $f_n \colon \X^n \to \Mc^{(n)}_X$ and $g_n \colon \Y^n \to \Mc^{(n)}_Y$ are the corresponding encoders.
The encoded messages are further sent to a central machine, which makes the decision $\Hhs \defeq \dec_n(f_n(X^n), g_n(Y^n))$, with $\dec_n \colon \Mc^{(n)}_X \times \Mc^{(n)}_Y \to \{0, 1\}$ being used as the decoder.

Due to the limited communication budgets in practice, there are typically constraints on the sizes of the message sets $\Mc_X^{(n)}$ and $\Mc_Y^{(n)}$. Following the convention introduced in  \cite{han1998statistical}, we use $\|f_n\| \defeq \left|\Mc_X^{(n)}\right|$ and $\|g_n\| \defeq \left|\Mc_Y^{(n)}\right|$ to denote the cardinalities of message sets, and express the constraints on $\|f_n\|$ and $\|g_n\|$ as a pair $(R_X, R_Y)$, referred as the \emph{rate} of encoders $f_n$ and $g_n$, with $R_X, R_Y \in [0, \infty) \cup \{0_M \colon M \geq 1\}$. Specifically, each $R_X \in [0, \infty)$ indicates the constraint\footnote{Throughout, the logarithm $\log(\cdot)$ indicates the natural logarithm with base $\e$, unless otherwise specified.} %
\begin{align}
  \limsup_{n \to \infty} \frac{1}{n} \log \|f_n\| \leq R_X,
  \label{eq:rate:pos}
\end{align}
and each $R_X = 0_M$ with $M \geq 1$ indicates the constraint
\begin{align}
  \limsup_{n\to \infty}\|f_n\| \leq M,
  \label{eq:rate:0}
\end{align}
namely, the encoded message $f_n(x^n)$ is allowed to take at most $M$ distinct values\footnote{For mathematical convenience, we allow $M$ to take $1$, where no information can be transmitted from the node to center.}. %
The constraint $R_Y$ for $\|g_n\|$ is similarly defined.
Specifically, we refer to $f_n$ (or $g_n$) as a \emph{zero-rate} encoder if it satisfies the constraint $R_X = 0$ (or $R_Y = 0$), and the corresponding hypothesis testing setting is called the \emph{zero-rate compression} regime. In this paper, we consider the DHT problem with constant communication bits, 
also referred to as \emph{constant-bit compression} regime, where we have $R_X \in \{0_M \colon M \geq 1\}$ or $R_Y \in \{0_M \colon M \geq 1\}$.  In particular, we will focus on the constant-bit communication constraint $(0_{M_X}, 0_{M_Y})$ with $M_X, M_Y \geq 1$, i.e., node $\nx$ and node $\ny$ can transmit at most $\log_2 M_X$ and $\log_2 M_Y$ bits to the center, respectively.

Then, each coding scheme can be characterized as a tuple $\Cc_n = (f_n, g_n, \dec_n)$ of encoder and decoder functions. In addition, for each given $\Cc_n$, 
we define the type-I error $\pi_0(\Cc_n)$ and type-II error $\pi_1(\Cc_n)$ associated with $\Cc_n$ as
   $ \pi_i(\Cc_n) \defeq \prob {\Hhs \neq i \middle| \Hs = i}$
  for $i = 0, 1$, where $\prob{\cdot}$ denotes the probability 
with respect to the i.i.d. sampling process over $n$ sample pairs.

In particular, we consider the asymptotic regime such that $n$ is large and characterize the achievable error exponents, defined as follows.

%
%
%
%

%
%
%

  %

%

%

  \begin{definition}[Error Exponent Region]
    \label{def:exp:reg}
    Given a rate pair $(R_X, R_Y)$, an error exponent pair $(E_0, E_1)$ is achievable under $(R_X, R_Y)$, if there exists a sequence of coding schemes $\{\Cc_n = (f_n, g_n, \dec_n)\}_{n\geq 1}$ such that the encoders $f_n$ and $g_n$ satisfy the rate constraints $(R_X, R_Y)$, and 
    \begin{align}
      \lim_{n\to \infty} \frac1n\log\pi_i(\Cc_n) = -E_i,\quad i = 0, 1.
      \label{eq:exp:def}
    \end{align}
    Then, we define the error exponent region $\Ec(R_X, R_Y)$ as the closure of the set of all achievable error exponent pairs under the rate constraints. Specifically, under constant-bit compression, if the coding schemes $\Cc_n$'s in \eqref{eq:exp:def} have a common decoder $\dec$ for all $n \geq 1$, 
   we call an error exponent pair $(E_0, E_1)$ is achievable under decoder $\dec$.   %
   Then, we use $\Ec[\dec]$ to denote the closure of the set of all such pairs.
  \end{definition}

   Our goal is to characterize the error exponent region
 under constant-bit compression regime and the coding schemes to achieve the error exponents.

  \subsection{Definitions and Notations}
  Given an alphabet $\Z \in \{\X, \Y, \X \times \Y\}$, we use $\Pc^\Z$ to denote the set of distributions supported on $\Z$. Then, for a joint distribution $Q_{XY} \in \cP^{\X \times \Y}$, the corresponding marginal distributions are denoted by $[Q_{XY}]_X \in \cP^\X$ and $[Q_{XY}]_Y  \in \cP^\Y$. In particular, for each  $i = 0, 1$, we denote  $P^{(i)}_{X} \defeq [P^{(i)}_{XY}]_X, P^{(i)}_{Y} \defeq [P^{(i)}_{XY}]_Y$.

 In addition, a sequence $(z_1, \dots, z_n) \in \Z^n$ is denoted by $\{z_i\}_{i = 1}^n$ or simply $z^n$, and we use $\Ph_{z^n} \in \cP^\Z$ to denote its empirical distribution (type), defined as%
$  \Ph_{z^n}(z') \defeq \frac1n \sum_{i = 1}^n \kron_{\{z_i = z' \}}$  for all $z' \in \Z$,
 where $\kron_{\{\cdot\}}$ denotes the indicator function. Specifically, the set of all empirical distributions of sequences in $\Z^n$ is denote as
$\ds  \Pch_n^\Z \defeq \left\{\Ph_{z^n} \colon z^n \in \Z^n\right\}$.

Furthermore, we use  $\pdist \defeq \cP^\X \times \cP^\Y$ to denote the product space of marginal distributions.
For each $i = 0, 1$ and $t > 0$, we define the subsets $\Dc_i(t)$ of $\pdist$ as
\begin{align}
  \Dc_i(t) \defeq \{(Q_X, Q_Y) \in \pdist\colon \divp_i(Q_X, Q_Y) < t\},
  \label{eq:def:Dc}
\end{align}
where the function  $\divp_i \colon \pdist \to \mathbb{R}$ is defined as
\begin{align}
  \divp_i(Q_X, Q_Y) \defeq  \min_{\substack{Q_{XY}\colon [Q_{XY}]_X = Q_X \\\qquad\,[Q_{XY}]_Y = Q_Y}}
  D(Q_{XY}\|P^{(i)}_{XY}),
  \label{eq:def:divp:i}
\end{align}
where $D(\cdot\|\cdot)$ denotes the Kullback-Leibler (KL) divergence between distributions.

In addition, we define several useful operations on $\pdist$ as follows. For a given $\Ac \subset \pdist$, we define its projections $\projx{\Ac}$  on $\cP^\X$ and $\projy{\Ac}$ on $\cP^\Y$, as
\begin{subequations}
\begin{align}
  \projx{\Ac} &\defeq \{Q_X \in \cP^\X \colon
                (Q_X, Q_Y') \in \Ac%
                    \text{ for some }Q_Y' \in \cP^\Y\},\\
  \projy{\Ac} &\defeq \{Q_Y \in \cP^\Y \colon (Q_X', Q_Y) \in \Ac%
      \text{ for some }Q_X' \in \cP^\X\}.
\end{align}
\label{eq:def:proj}
\end{subequations}
Then, we have the following definition.
\begin{definition}
  \label{def:sub}
  The binary operator ``$\,\sub$'' on $\pdist$ is defined as
$ 
    \Ac \sub \Ac' \defeq \{(Q_X, Q_Y) \in \Ac \colon Q_X \in \projx{\Ac'}, Q_Y \in \projy{\Ac'}\}$, for all $\Ac, \Ac' \subset \pdist$.
  In addition, for each $k \geq 0$, we define the operator ``$\,\mathop{\sub_{k}}$'' as
   $ \Ac \sub_0 \Ac' \defeq \Ac$, $ \Ac \sub_1 \Ac' \defeq \Ac'$, and $
    \Ac \mathop{\sub_{k+2}} \Ac' \defeq (\Ac \sub_k \Ac') \sub\, (\Ac \sub_{k+1} \Ac')$ for $k \geq 0$.
\end{definition}

We also define operators ``$\subx$'', ``$\suby$'' as
\begin{align}
  \Ac \subx \Ac' &\defeq \{(Q_X, Q_Y) \in \Ac \colon Q_X \in \projx{\Ac'}\},   \label{eq:def:subx}
\\
  \Ac \suby \Ac' &\defeq \{(Q_X, Q_Y) \in \Ac \colon Q_Y \in \projy{\Ac'}\}. \label{eq:def:suby}
\end{align}

\begin{figure}[!t]
  \centering
  \subfloat[{$\Ac_0 \sub \Ac_1, \Ac_0 \subx \Ac_1$, and $\Ac_0 \suby \Ac_1$}]{
    \resizebox{!}{5cm}{\tikzstyle{D shape}=[minimum width=3.5cm, minimum height=1.7cm]
\tikzstyle{DL shape}=[minimum width=5.4cm, minimum height=2.7cm]

\begin{tikzpicture}[line width = .1em]
  \coordinate (C0) at (-.8, .8);
  \coordinate (C1) at (1, -1);
  
  \node [draw, color D0, ellipse, rotate=30, DL shape, align=center, name = D03, label={[xshift = -1.1cm, yshift = -1.3cm, text = black!50!gray, scale = 1.3]$\Ac_0$}] at (C0)   {}; 

  \node [draw, color D1, ellipse, rotate=30, D shape, align=center,  name = D1,   label={[xshift = 2cm, yshift = -2cm, scale = 1.3]$\Ac_1$}] at (C1)   {}; 
  
  \node [draw = none, rectangle, minimum width=3.17cm, minimum height=2.3cm, align=center, name = B1] at   (C1)   {}; 
  \node [draw = none, anchor=north west, rectangle, minimum width=.89cm, minimum height=0.57cm, align=center, name = B] at   (B1.north west)   {};

  \draw [opacity = .3] (B1.north east) -- ++ (-6, 0);
  \draw [opacity = .3] (B1.south west) -- ++ (0, 5);
  

    \begin{scope}
    \clip ($(B.north west) + (0, 2.5)$) rectangle ($(B.north west) + (2.3, -1.7)$); 
    \node [draw,
    pattern= vertical lines,
    , ellipse, rotate=30, DL shape, align=center] at (C0)   {}; 
  \end{scope}

  \begin{scope}
    \clip ($(B.north west) + (-3, 0)$) rectangle ($(B.north west) + (2, -2)$); 
    \node [draw, 
    pattern= horizontal lines,
    , ellipse, rotate=30, DL shape, align=center] at (C0)   {}; 
  \end{scope}

  
  \matrix [xshift = .5cm, yshift = -0.2cm] at (current bounding box.north east) {
    \node [scale = 2, preaction={color D0}, shape=rectangle, pattern=grid,  label={right:$\Ac_0 \sub \Ac_1$}, line width = 2] {}; \\
    \node [scale = 2, preaction={color D0}, shape=rectangle, pattern=vertical lines,  label={right:$\Ac_0 \subx \Ac_1$}, line width = 2] {}; \\
    \node [scale = 2, preaction={color D0}, shape=rectangle, pattern= horizontal lines,  label={right:$\Ac_0 \suby \Ac_1$}, line width = 2] {}; \\
    };

    \coordinate (A) at (-3.2, 3.5);
    \draw[thick,style = ->] (A) -- ++(.6, 0)  node[right] {$Q_X$} ;
    \draw[thick,style = ->] (A) -- ++(0, -.6)  node[below] {$Q_Y$};
\end{tikzpicture}

}
  }
  \hspace{1em}
  \subfloat[{$\Ac_0 \sub_k \Ac_1$}]{
    \resizebox{!}{5cm}{\begin{tikzpicture}[line width = .07em] 
  \coordinate (C0) at (-.5, .5);
  \coordinate (C1) at (.5,  -0.5);
  \tikzstyle{D shape}=[minimum width=3.5cm, minimum height=1.3cm]
  \tikzstyle{B shape}=[minimum width=2.65cm, minimum height=2.65cm]
  
  \node [draw, rotate=45, ellipse, color D0, D shape, align=center, name = D0, label={[scale = 1, xshift = -0.6cm, yshift = -1.2cm]$\Ac_0$}] at (C0)   {}; 
  \node [draw, ellipse, rotate=45, color D1, D shape, align=center,  name = D1,  label={[scale = 1, xshift = 1cm, yshift = -1.9cm]$\Ac_1$}] at (C1)   {}; 

  \node [draw = none, rectangle, B shape, align=center, name = B0] at   (C0)   {}; 
  \node [draw = none, rectangle, B shape, align=center, name = B1] at   (C1)   {}; 




  \coordinate (P0) at (B1.north west);
  \coordinate (P1) at ($(P0) + (1.43, -1.43)$);
  \coordinate (P2) at ($(P0) + (0.41, -0.41)$);
  \coordinate (P3) at ($(P0) + (1.15, -1.15)$);

  \def\myop{.3}
  \draw [opacity = \myop] (P0) -- (B1.north east);
  \draw [opacity = \myop] (P0) -- (B1.south west);
  \draw (P0) [opacity = \myop] rectangle (P1);
  \draw (P1) [opacity = \myop] rectangle (P2);
  \draw (P2) [opacity = \myop] rectangle (P3);

\newcommand{\drawEllipse}[2]{    \node [draw, ellipse, rotate=45, #2, D shape, align=center] at (#1)   {};}

\begin{scope}
  \clip (P0) rectangle ++ (3, -3);
  \drawEllipse{C0}{pattern=vertical lines} 
\end{scope}
  
  \begin{scope}
    \clip (P1) rectangle ++ (-1, 1); 
    \drawEllipse{C1}{pattern=horizontal lines} 
\end{scope}

  \begin{scope}
    \clip (P2) rectangle ++ (1, -1);
    \drawEllipse{C0}{pattern=horizontal lines}
\end{scope}

  \begin{scope}
    \clip (P3) rectangle ++ (-1, 1); 
    \drawEllipse{C1}{pattern=vertical lines} 
\end{scope}

  \matrix [xshift = 1.1cm, yshift = -1.2cm] at (current bounding box.north east) {
    \node [scale = 1.4, preaction={color D0}, shape=rectangle, pattern=vertical lines, label={[scale = .8]right:$ \Ac_0 \sub_2 \Ac_1$},  line width = 2] {}; \\
    \node [scale = 1.4, preaction={color D1},  shape=rectangle, pattern=horizontal lines, label={[scale = .8]right:$ \Ac_0 \sub_3 \Ac_1$}, line width = 2] {}; \\
    \node [scale = 1.4, preaction={color D0}, shape=rectangle, pattern=grid, label={[scale = .8]right:$ \Ac_0 \sub_4 \Ac_1$}, line width = 2] {}; \\
    \node [
    scale = 1.4, preaction={color D1}, shape=rectangle, pattern=grid, label={[scale = .8]right:$ \Ac_0 \sub_5 \Ac_1$}, line width = 2] {}; \\
    };

  \coordinate (A) at (-2, 1.8);

  \draw[style = ->, scale = .8] (A) -- ++(.6, 0)  node[scale = .8, right] {$Q_X$} ;
  \draw[style = ->, scale = .8] (A) -- ++(0, -.6) node[scale = .8, below] {$Q_Y$};
  
  
\end{tikzpicture}

}
    }    
    \caption{{The relations of binary operators $\sub$ ($\sub_2$), $\subx, \suby$, and $\sub_k, k \geq 2$, where $\Ac_0$ and $\Ac_1$ are subsets of $\pdist = \cP^\X \times \cP^\Y$.}}
    \label{fig:sub}
  \end{figure}
  
{\figref{fig:sub} demonstrates relations of operators $\sub, \subx, \suby$ and $\sub_k$, where the horizontal axis and vertical axis represent the marginal distributions of $X$ and $Y$, respectively, and where each point corresponds to a pair of marginal distributions $(Q_X, Q_Y) \in \pdist$.}

Finally, for sequences $\{a_n\}_{n \geq 1}$ and $\{b_n\}_{n \geq 1}$, we use $a_n = o(b_n)$ to indicate that $\lim_{n \to \infty }\frac{a_n}{b_n} = 0$. We also define $\msgset{M}\defeq \{0, \dots, M - 1\}$ for $M \geq 1$, and $\ibar \defeq 1 - i$ for $i \in \{0, 1\}$.

\subsection{Encoders and Decoders}
We then provide characterizations on encoders and decoders in constant-bit compression regime $(0_{M_X}, 0_{M_Y})$, where $M_X, M_Y \geq 1$. Without loss of generality, we assume  that the corresponding message sets are  $\Mc_X^{(n)} \equiv \msg{M_X}$ and $\Mc_Y^{(n)} \equiv \msg{M_Y}$, respectively.

\subsubsection{Type-based Encoders}
An encoder is called \emph{type-based} if its output depends only on the type of th e input. Specifically, $f_n$ is type-based, when there exist a mapping $\enctx \colon \cP^\X \to \msgset{M_X}$ such that $f_n(x^n) = \enctx(\Ph_{x^n})$, for all $x^n \in \X^n$. Similarly, $g_n$ is type-based if $g_n(y^n) = \encty(\Ph_{y^n})$ for some $\encty \colon \cP^\Y \to \msgset{M_Y}$. Then, the type-based encoders $f_n$, $g_n$ are fully characterized by the mappings $\enctx$ and $\encty$, which we refer to as \emph{type-encoding functions}.

\subsubsection{Decoder Representation and Special Decoders}
Each decoder $\dec$ is a Boolean-valued function on $\msg{M_X} \times \msg{M_Y}$, formalized as follows.
\begin{definition}
  Given $M_X, M_Y \geq 1$, an $M_X \times M_Y$ decoder is a function $\dec: \msg{M_X} \times \msg{M_Y} \to \{0, 1\}$. The \emph{decision matrix} associated with $\dec$ is defined as an $M_Y \times M_X$ Boolean matrix $\Ab$ with entries $A(m_Y, m_X) \defeq \dec(m_X, m_Y)$ for all $(m_X, m_Y) \in \msg{M_X} \times \msg{M_Y}$, and we use  $\dec \lfa \Ab$ to denote this one-to-one correspondence. %
\end{definition}

In addition, we call $\dec$ \emph{trivial} if $\dec \equiv 0$ or $\dec \equiv 1$. For a given decoder $\dec$, we define its \emph{complement} $\bar{\dec}$ as
 $ \bar{\dec}(m_X, m_Y) \defeq 1 - \dec(m_X, m_Y),$  for all $(m_X, m_Y) \in \msg{M_X} \times \msg{M_Y}$.

 Moreover, the threshold decoders will be useful in our analyses, defined as follows.
\begin{definition}%
  \label{def:threshold}
  For given $M_X, M_Y \geq 1$, the $M_X \times M_Y$ \emph{threshold decoders}  are the $M_X \times M_Y$ decoder $\decr_{M_{X}, M_{Y}}$ and  its complement $\bar{\decr}_{M_{X}, M_{Y}}$, where %
$  \decr_{M_{X}, M_{Y}}(m_X, m_Y) \defeq \kron_{\{m_X + m_Y \geq \min\{M_X, M_Y\}\}}$, %
  for all $(m_X, m_Y) \in \msg{M_X} \times \msg{M_Y}$.
\end{definition}

We will sometimes find it convenient to express a decision matrix as filled grids of the same dimensions, with occupied grids and empty grids indicating ``1'' and ``0'', respectively. For example, when $M_X = M_Y = 2$, 
the threshold decoders $\decr_{2, 2}$ and $\bar{\decr}_{2, 2}$ as defined in \defref{def:threshold}
can be represented as 
``\putgrid{\drawgrid{
        {0,0},
        {0,1},}}'' and
``\putgrid{\drawgrid{
        {1,1},
        {1,0},}}'', respectively.

{
The decoder representations allow us to formalize the following fact on error exponent regions.
\begin{fact}
  \label{fact:equiv}
  Suppose $\dec \lfa \Ab$ and $\dec' \lfa \Ab'$. Then, we have $\Ec[\dec'] \subset \Ec[\dec]$ if $\Ab'$ is a submatrix of $\Ab$.  In addition, $\Ec[\dec] = \Ec[\dec']$ if $\Ab'$ can be obtained from $\Ab$ by deleting duplicated rows/columns, or permuting rows/columns. Specifically, for all $M_X > M_Y \geq 1$, we have $\Ec[\decr_{M_X, M_Y}] = \Ec[\bar{\decr}_{M_X, M_Y}]$.
\end{fact}
}

{As an example of \factref{fact:equiv}, the following result is useful for our later further derivations.  }

{
  \begin{example}
      \label{ex:4:2}
  We have $\Ec[\decr_{4, 2}] = \Ec[\decr_{3, 2}] = \Ec[\bar{\decr}_{3, 2}] = \Ec[\bar{\decr}_{4, 2}]$, i.e.,
  $ \Ec[\putgrid{\drawgrid{
      {1,1,0,0},
      {1,0,0,0},}}] =
  \Ec[\putgrid{\drawgrid{
      {1,1,0},
      {1,0,0},}}] %
  =   \Ec[\putgrid{\drawgrid{
      {0,0,1},
      {0,1,1},}}] %
  =   \Ec[\putgrid{\drawgrid{
      {0,0,1,1},
      {0,1,1,1},}}].$
\end{example}
}

  Furthermore, we use $\Dec_{M_X, M_Y}$ to denote the collection of all $M_X \times M_Y$ decoders, and we define %
   $ \Dec \defeq \bigcup_{M_X\geq 1, M_Y\geq 1} \Dec_{M_X, M_Y}$
   as the collection of all decoders. Then, for each
   collection of decoders $\Hc \subset \Dec$,
   we use $\Ec[\Hc]$ to denote its associated error exponent region, defined as $  \Ec[\Hc] \defeq \bigcup_{\dec \in \Hc} \Ec[\dec]$. Specifically, we have the following fact, of which a proof is provide in \appref{app:fact:dec}.
   \begin{fact}
  \label{fact:dec}
  For all $P^{(0)}_{XY}, P^{(1)}_{XY} \in \Pc^{\X \times \Y}$ and $M_X, M_Y \geq 1$, we have
  $  \Ec(0_{M_X}, 0_{M_Y}) = \Ec[\Dec_{M_X, M_Y}].$
\end{fact}

\section{Optimality of Type-based Encoders}
\label{sec:optim-type-based}

%
%
%
%
%

%
This section 
demonstrates the asymptotic optimality of type-based encoders for %
DHT problems satisfying %
zero-rate communication constraints. To formalize this optimality, we first introduce the following result. A proof is provided in \appref{app:lem:type}, via exploiting the celebrated blowing up lemma \cite{ahlswede1976bounds}.\footnote{We adopt the same technique introduced in \cite[Theorem 1]{shalaby1992multiterminal}, which was used to establish the optimal type-II error exponent $E_1$ of DHT under zero-rate communication constraints, with type-I error $\pi_0$ constrained by a constant.} 

\begin{lemma}
  \label{lem:type-based-coding}
  Suppose  $\{(X_i, Y_i)\}_{i = 1}^n$ are i.i.d. generated from a joint distribution $P_{XY}$ with $P_{XY}(x, y) > 0$, for all $x \in \X$, $y \in \Y$. Then, for all zero-rate encoders $f_n \colon \X^n \to \Mc_X^{(n)}$ and $g_n \colon \Y^n \to \Mc_Y^{(n)}$, there exist mappings $\enct_X \colon \Pc^\X \to \Mc_X^{(n)}$ and $\enct_Y \colon \Pc^\Y \to \Mc_Y^{(n)}$, such that
\begin{align}
  \prob{f_n(X^n) = \enct_X(Q_X), g_n(Y^n) = \enct_Y(Q_Y)} \geq \prob{(\Ph_{X^n}, \Ph_{Y^n}) = (Q_X, Q_Y)} \cdot \exp(-n \cdot o(1))
  \label{eq:lem:type}
\end{align}
for all $(Q_X, Q_Y) \in \hat{\cP}_n^{\X} \times \hat{\cP}^{\Y}_n$.
\end{lemma}

%

        %
  %
  %
  %
  %
  %
  %
  %

%

%
%
%
%
%
%

%
%
%
By using \lemref{lem:type-based-coding}, we can establish the following result illustrating the asymptotic optimality of type-based encoder in zero-rate DHT. A proof is provided in \appref{app:thm:optm-type-based}.

\begin{theorem}
  \label{thm:optm-type-based}
  For a given $n \geq 1$ and zero-rate encoders $f_n$ and $g_n$ with ranges $\Mc_X^{(n)}$ and $\Mc_Y^{(n)}$, there exist type-based encoders $\ft_n, \gt_n$ with the same ranges as $f_n$, $g_n$, respectively, such that, for each decoder $\dec_n \colon \Mc_X^{(n)} \times \Mc_Y^{(n)} \to \{0, 1\}$   and the corresponding coding schemes  $\Cc_n \defeq (f_n, g_n, \dec_n), $  $\Cct_n \defeq (\ft_n, \gt_n, \dec_n)$, we have
  \begin{align*}
    \pi_i(\tilde{\Cc}_n) \leq \pi_i(\Cc_n)\cdot \exp(n \zeta_n),\quad \text{for }i \in \{0, 1\},
  \end{align*}
  with $\zeta_n = o(1)$.%
\end{theorem}
\begin{remark}  
  The optimality of type-based decision in non-distributed hypothesis testing can be established by a more straightforward argument, see, e.g., \cite[Lemma 3.5.3]{DemboZ98}. Specifically, suppose $n$ i.i.d samples $x^n\in \X^n$ are generated by $P_X^{(\Hs)}$, and $f_n(x^n)$ is used as our decision for  $\Hs \in \{0, 1\}$, where $f_n \colon \X^n \to \{0, 1\}$. Then, there exists a type-based decision $\ft_n \colon \X^n \to \{0, 1\}$ such that
  \begin{align*}
    \pi_i(\ft_n) \leq 2\cdot \pi_i(f_n), \quad \text{for~}i\in \{0, 1\},
  \end{align*}
  where $\pi_0(\cdot)$ and  $\pi_1(\cdot)$ denote the type-I error and type-II error for corresponding decision functions, respectively. It is also easy to verify that both Neyman--Pearson test \cite{neyman1933problems} and Hoeffding's test \cite{hoeffding1967probabilities} depend only on the types. In particular, Neyman--Pearson test depends only on the empirical mean of log-likelihood ratio $\log \frac{P_X^{(0)}(x)}{P_X^{(1)}(x)}$, see, e.g., \cite[Theorem 11.7.1]{cover06}. And, when only $P^{(0)}_{X}$ is available but $P^{(1)}_{X}$ is unknown, the resulting Hoeffding's test depends only on the KL divergence $D\bigl(\Ph_{x^n}\big\| P^{(0)}_{X}\bigr)$, which is also a function of the type $\Ph_{x^n}$.
\end{remark}

{\begin{remark}
    The type-based encoders have also appeared frequently in previous literature on zero-rate or one-bit DHT problems, e.g., \cite[Theorem 5, Theorem 8]{han1987hypothesis}, \cite[Theorem 6]{han1989exponential}, \cite[Therem 5.5]{han1998statistical}. However, its optimality has not been formalized or discussed in these studies. Specifically, the type-based encoder was merely used for constructing achievability results (i.e., the \emph{direct part} of a proof), while the converse parts ware established by separate arguments. In contrast,  \thmref{thm:optm-type-based} demonstrates the fundamental role of type-based encoders in zero-rate DHT problems, which allows us to focus on the characterization on distribution spaces, instead of the original sequence spaces.
\end{remark}}

\section{A Geometric Characterization of Error Exponent Region}%
\label{sec:geom-char-dht}

With the optimality of type-based encoders, we further illustrate that the error exponent region  $\Ec(0_{M_X}, 0_{M_Y})$ can be characterized as a geometric problem of separating two sets in $\pdist$. For convenience, in the following discussions we will assume that $M_X \geq M_Y$, and the result for $M_X < M_Y$ can be obtained by symmetry arguments.

First, we introduce the notion of separability on $\pdist$.

\begin{definition}
  \label{def:separation}
  Given $M_X, M_Y \geq 1$, a decoder $\dec \in \Dec_{M_X, M_Y}$, and a pair of disjoint subsets $(\Ac_0, \Ac_1)$ of $\pdist$, we say that $\dec$ \emph{separates} $(\Ac_0, \Ac_1)$ [or, $(\Ac_0, \Ac_1)$ \emph{is separable by} $\dec$], denoted by $\dec \mid (\Ac_0, \Ac_1)$, if there exist mappings $\enct_X\colon \Pc^\X \to \msg{M_X}$ and $\enct_Y\colon \Pc^\Y \to \msg{M_Y}$, such that for both $i \in \{0, 1\}$,
  \begin{align}
    \dec(\enct_X(Q_X), \enct_Y(Q_Y)) = i, \quad\text{for all } (Q_X, Q_Y) \in \Ac_i.
    \label{eq:def:dec:enct}
  \end{align}
\end{definition}

Then, our main result is summarized as follows. A proof is provided in \appref{app:thm:separation}.
\begin{theorem}
  \label{thm:separation}
  For each $\dec \in \Dec$, we have
  $\Ec[\dec] = \{(E_0, E_1) \colon \dec \mid (\Dc_0(E_0), \Dc_1(E_1))\}$,
  where $\Dc_0(\cdot)$ and $\Dc_1(\cdot)$ are as defined in \eqref{eq:def:Dc}. In addition, each exponent pair $(E_0, E_1)  \in \interior(\Ec[\dec])$ can be achieved by the coding schemes $\{(f_n, g_n, \dec)\}_{n \geq 1}$ with
type-based encoders $f_n(x^n) \defeq \enct_X(\Ph_{x^n}), g_n(y^n) \defeq \enct_Y(\Ph_{y^n})$, where
 $\interior(\cdot)$ denotes the interior, and where the type-encoding functions $\enct_X$ and $\enct_Y$ correspond to the mappings such that \eqref{eq:def:dec:enct} holds for $\Ac_i = \Dc_i(E_i), i \in \{0, 1\}$.
\end{theorem}

\begin{remark}
  By using a similar argument, we can show that under zero-rate communication constraints $(R_X, R_Y) = (0, 0)$, the error exponent region is
  \begin{align}
    \Ec(0, 0) = \{(E_0, E_1)\colon \Dc_0(E_0) \cap \Dc_1(E_1) = \varnothing \},
    \label{eq:Ec:zero}
  \end{align}
  which coincides with the classical results demonstrated in, e.g., \cite[Theorem 6]{han1989exponential}, \cite[Theorem 5.5]{han1998statistical}. %
  Furthermore, note that \eqref{eq:Ec:zero} also corresponds to a limiting case of \thmref{thm:separation}, and we have %
  $  \Ec[\Dec_{M_X, M_Y}] \to \{(E_0, E_1)\colon \Dc_0(E_0) \cap \Dc_1(E_1) = \varnothing\}$
  as $M_X \to \infty, M_Y \to \infty$. %
\end{remark}

{\thmref{thm:separation} provides a single-letter characterization of the error exponent region, which allows us to focus on studying the separability on the distribution space $\pdist$, rather than the original sequence space $\X^n \times \Y^n$. Later on we will show that existing results on one-bit communication constraints can be recovered immediately by using such geometric characterizations.}%

\subsection{Threshold Decoder Inner Bound}%
\label{sec:inner:bound}

From the geometric characterization in \thmref{thm:separation}, we can establish the error exponent regions under threshold decoders $\decr_{M_X, M_Y}$ and $\bar{\decr}_{M_X, M_Y}$, which also provide an inner bound of $\Ec(0_{M_X}, 0_{M_Y})$.

{
Specifically, our characterization uses the following recursive property of the separability of threshold decoders, a proof is provided in \appref{app:prop:recursive:decr}.  
    \begin{proposition}
     \label{prop:recursive:decr}
     Suppose $\Ac$ and $\Ac'$ are two disjoint subsets of $\pdist$. Given $M \geq 2$, we have
     \begin{align}
       \label{eq:rec:equiv}
       \decr_{M, M} \mid (\Ac, \Ac') \iff \decr_{M-1, M-1} \mid (\Ac', \Ac \sub \Ac')
       \iff \Ac \sub_M \Ac' = \varnothing.
     \end{align}
     In addition, for given $M_X > M_Y \geq 1$, $\decr_{M_X, M_Y}\mid(\Ac, \Ac')$  if and only if $\decr_{M_Y, M_Y} \mid (\Ac, \Ac' \subx \Ac)$. 
   \end{proposition}
 }
 
   By using \propref{prop:recursive:decr}, the error exponent region under threshold decoders can be established as follows. A proof is provided in \appref{app:exp:th}.

   \begin{theorem}
  \label{thm:exp:th}
  Given $M_X \geq M_Y \geq 1 $, the error exponent regions under $M_X \times M_Y$ threshold decoders are%
  \begin{align}
    \Ec[\decr_{M_X, M_Y}] &= \{(E_0, E_1)\colon \Dc_0(E_0) \sub_{M_Y} \Dc_1(E_1) = \varnothing\},
    \label{eq:decr:eq}\\
    \Ec[\bar{\decr}_{M_X, M_Y}] &= \{(E_0, E_1)\colon \Dc_1(E_1) \sub_{M_Y} \Dc_0(E_0) = \varnothing\}
                                  \label{eq:decr:eq:bar}
  \end{align}
  if $M_X = M_Y$, and
  \begin{align}
    &\Ec[\decr_{M_X, M_Y}] =  \Ec[\bar{\decr}_{M_X, M_Y}]%
      = \{(E_0, E_1)\colon%
    \Dc_0(E_0) \sub_{M_Y} \left(\Dc_1(E_1) \subx \Dc_0(E_0)\right) = \varnothing\}
    \label{eq:decr:ieq}
  \end{align}
   if $M_X > M_Y$,  where the operators
   ``$\,\sub_k$'' and ``$\,\subx\,$'' are as defined in \defref{def:sub}.
 \end{theorem}

From \factref{fact:dec} and \thmref{thm:exp:th}, we can readily obtain an inner bound of $\Ec(0_{M_X}, 0_{M_Y})$ as
  \begin{align}
    \Ec(0_{M_X}, 0_{M_Y})
    &= \Ec[\Dec_{M_X, M_Y}] \notag\\
    &\supset \left(\Ec[\decr_{M_X, M_Y}] \cup \Ec[\bar{\decr}_{M_X, M_Y}]\right),
      \label{eq:inner}
  \end{align}
  which we refer to as the \emph{threshold decoder inner bound}. 
  Later on we will discuss several cases where this bound is tight.

  \begin{figure}[!t]
  \centering
  \subfloat[$\decr_{2, 2}\lfa \putgrid{\drawgrid{
        {0,0},
        {0,1},}}$ and $\decr_{3, 3}  \lfa \putgrid{\drawgrid{
        {0,0,0},
        {0,0,1},
        {0,1,1},}}$]{
    \resizebox{!}{6.3cm}{\tikzstyle{D shape}=[minimum width=3.5cm, minimum height=1.7cm]
\tikzstyle{DL shape}=[minimum width=5.4cm, minimum height=2.7cm]

\begin{tikzpicture}[line width = .1em]
  \coordinate (C0) at (-1, .9);
  \coordinate (C1) at (1, -1);
  \node [draw, color D0, ellipse, rotate=30, align=center, name = D02, D shape, label={[xshift = .6cm, yshift = -1.3cm]$\Dc_0(E_0)$}] at (C0)   {}; 
  
  \node [draw, thin, color D0L, ellipse, rotate=30, DL shape, align=center, name = D03, label={[xshift = -0.8cm, yshift = -.8cm, text = black!50!gray,]$\Dc_0(E_0')$}] at (C0)   {}; 

  \node [draw, color D1, ellipse, rotate=30, D shape, align=center,  name = D1,   label={[xshift = 2.1cm, yshift = -2.2cm]$\Dc_1(E_1)$}] at (C1)   {}; 
  
  \node [draw = none, rectangle, minimum width=3.17cm, minimum height=2.3cm, align=center, name = B1] at   (C1)   {}; 
  \node [draw = none, anchor=north west, rectangle, minimum width=.89cm, minimum height=0.57cm, align=center, name = B] at   (B1.north west)   {};

  \draw [dashed] ($(B.north west) + (-3.2, 0)$) -- ($(B.north east) + (2.7, 0)$);
  \draw [dashed] ($(B.north west) + (0, 2.8)$) -- ($(B.south west) + (0, -2.0)$);
  
  \draw [dashed, thin] ($(B.south west) + (-3, 0)$) -- ($(B.south east) + (2.5, 0)$);
  \draw [dashed, thin] ($(B.north east) + (0, 2.6)$) -- ($(B.south east) + (0, -1.8)$);

  \begin{scope}
    \clip (B.north west) rectangle ($(B.north west) + (.89, -0.57)$); 
    \node [draw, thin, 
    my pattern, ellipse, rotate=30, DL shape, align=center] at (C0)   {}; 
  \end{scope}
  
  \matrix [xshift = -1.9cm, yshift = .5cm] at (current bounding box.north east) {
    \node [scale = 2, preaction={color D0L}, shape=rectangle, my pattern,  label={right:$\Dc_0(E_0') \sub \Dc_1(E_1)$}, line width = 2] {}; \\
    };

    \coordinate (A) at (-3.2, 3.5);
    \draw[thick,style = ->] (A) -- ++(.6, 0)  node[right] {$Q_X$} ;
    \draw[thick,style = ->] (A) -- ++(0, -.6)  node[below] {$Q_Y$};
\end{tikzpicture}
}
    \label{fig:sep:a}
  }%
  \hspace{2em}
  \subfloat[$\decr_{3, 2} \lfa \putgrid{\drawgrid{
        {0,0,1},
        {0,1,1},}}$]{
  \resizebox{!}{6.3cm}{\begin{tikzpicture}[line width = .07em] 
  \coordinate (C0) at (-.82, .6);
  \coordinate (C1) at (.82,  -0.6);
  \tikzstyle{D shape}=[minimum width=3.5cm, minimum height=1.3cm]
  \tikzstyle{B shape}=[minimum width=2.65cm, minimum height=2.65cm]
  
  \node [draw, rotate=45, ellipse, color D0, D shape, align=center, name = D0, label={[scale = .9, xshift = -0.0cm, yshift = -0.2cm]$\Dc_0(E_0)$}] at (C0)   {}; 
  \node [draw, ellipse, rotate=45, color D1, D shape, align=center,  name = D1,  label={[scale = .9, xshift = 2.0cm, yshift = -2cm]$\Dc_1(E_1)$}] at (C1)   {}; 

  \node [draw = none, rectangle, B shape, align=center, name = B0] at   (C0)   {}; 
  \node [draw = none, rectangle, B shape, align=center, name = B1] at   (C1)   {}; 




  \draw [dashed] (-2.5, 0) -- (2.5, 0);
  \draw [dashed] ($(B0.north east)+(0, .3)$) -- ($(B0.south east)+(0, -1.5)$);
  \draw [dashed] ($(B1.north west)+(0, 1.5)$) -- ($(B1.south west)+(0, -.3)$);

  \begin{scope}
    \clip ($(-1, 0)!(B0.east)!(1, 0)$) rectangle ++ (-1, -2); 
    \node [draw, ellipse, rotate=45, my pattern, D shape, align=center] at (C1)   {}; 
\end{scope}

  \matrix [xshift = 1.4cm, yshift = .7cm] at (current bounding box.north) {
    \node [scale = 1.8, preaction={color D1}, shape=rectangle, my pattern, label={right:\small$ \Dc_1(E_1) \subx \Dc_0(E_0)$}] {}; \\
    };

  \coordinate (A) at (-2.4, 3);

  \draw[style = ->] (A) -- ++(.6, 0)  node[scale =1, right] {$Q_X$} ;
  \draw[style = ->] (A) -- ++(0, -.6) node[scale = 1, below] {$Q_Y$};
  
  
\end{tikzpicture}

}
    \label{fig:sep:b}
  }    %
  \caption{  %
    Geometric interpretation for achievable error exponent pairs
    under different threshold decoders, with each point representing a pair of marginal distributions $(Q_X, Q_Y) \in \pdist$.
  }  %
  \label{fig:sep}
\end{figure}
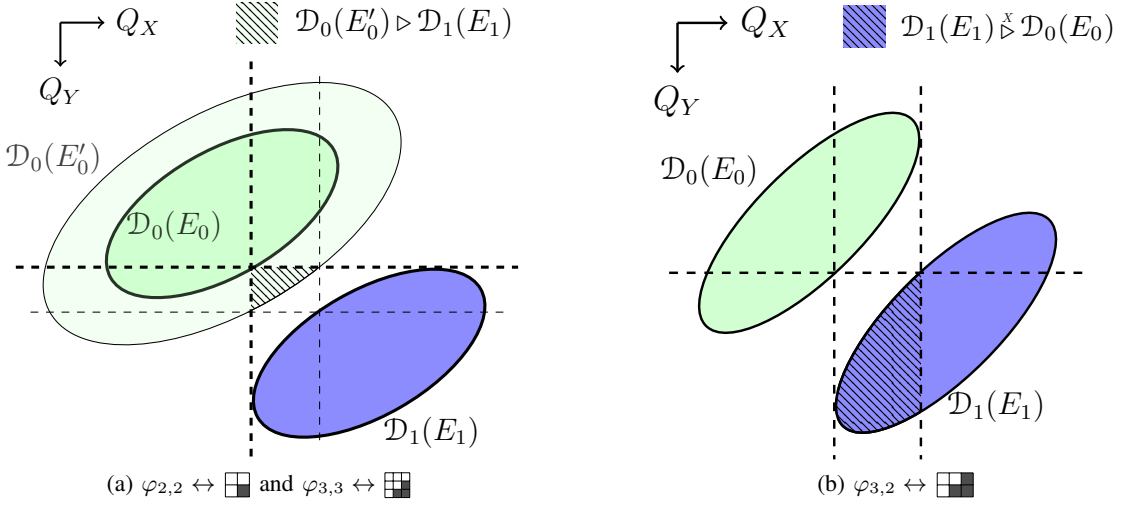

Furthermore, we provide a geometric interpretation of above characterizations in \figref{fig:sep}. To begin, let us first consider the one-bit constraint $(0_2, 0_2)$, with  $\decr_{2,2} \lfa \putgrid{\drawgrid{
        {0,0},
        {0,1},}}$ used as the decoder. 
    \figref{fig:sep:a} demonstrates the case where  $\decr_{2,2}$ separates $(\Dc_0(E_0), \Dc_1(E_1))$, and it follows from \thmref{thm:separation} that $(E_0, E_1) \in \Ec[\decr_{2, 2}]$.  Moreover, with the type-II error exponent $E_1$ fixed, $E_0$ is the optimal type-I error exponent under $\decr_{2, 2}$, since  $(\Dc_0(E_0 + \eps), \Dc_1(E_1))$ is not separable by $\decr_{2, 2}$  for all $\eps > 0$. Now, suppose both nodes are allowed to transmit one-trit messages with $\decr_{3, 3} \lfa \putgrid{\drawgrid{
        {0,0,0},
        {0,0,1},
        {0,1,1},}}$ used as the decoder. Then, the optimal type-I error exponent can be improved to $E_0' > E_0$, as illustrated in the figure. Compared with the one-bit setting, it can be noted that the two additional symbols are used to encode the hatched area $\Dc_0(E_0') \sub \Dc_1(E_1)$, such that $(\Dc_0(E_0'), \Dc_1(E_1))$ is still separable.%

    Similarly, \figref{fig:sep:b} illustrates the separability under decoder  $\decr_{3, 2} \lfa \putgrid{\drawgrid{
        {0,0,1},
        {0,1,1},}}$. It can be noted that $\decr_{3, 2}$ separates $(\Dc_0(E_0), \Dc_1(E_1))$, if and only if $\decr_{2, 2}$ separates $\Dc_0(E_0)$ and $\Dc_1(E_1) \subx \Dc_0(E_0)$ (shown in hatched).

    \subsection{Decoder Comparison and Selection}
    \label{sec:decod-comp-select}
  { From \factref{fact:dec} and the geometric characterization in \thmref{thm:separation}, each point $(E_0, E_1)$    from the error exponent region $\Ec(0_{M_X}, 0_{M_Y}) = \cup_{\dec \in \Dec_{M_X, M_Y}} \Ec[\dec]$ is contributed by some decoder $\dec \in \Dec_{M_X, M_Y}$ that separates $(\Dc_0(E_0), \Dc_1(E_1))$. To select the decoder for a given DHT problem, it will be useful to understand the contribution of each decoder in $\Dec_{M_X, M_Y}$. To this end, we then characterize and compare the performance of different decoders, in terms of their separability.}

  First, note that a simple example of decoder comparison is \factref{fact:equiv}, which can be directly verified by definition. A non-trivial characterization will make use of the following operation.
  \begin{definition}[Decoder Decomposition]
    \label{def:decomp}
    Given $M_X, M_Y \geq 1$, an $M_X \times M_Y$ decoder $\dec$ is called \emph{decomposable} if there exist non-trivial decoders $\dec_0, \dec_1 \in \Dec_{M_X, M_Y}$ and $i \in \{0, 1\}$, such that for all $(m_X, m_Y) \in \msgset{M_X} \times \msgset{M_Y}$,
    \begin{gather}
      \dec(m_X, m_Y) = \dec_0(m_X, m_Y) \oplus \dec_1(m_X, m_Y) \oplus \ibar,
      \label{eq:decomp}\\
      \Icx^{(i)}(\dec_0) \cap \Icx^{(i)}(\dec_1) = \Icy^{(i)}(\dec_0) \cap \Icy^{(i)}(\dec_1) = \varnothing,  \label{eq:icx:icy:empty}      
  \end{gather}
    where ``$\oplus$'' represents the ``exclusive or'' operation, and where, %
   for  $i \in \{0, 1\}$, we have defined %
    \begin{align}
      \Icx^{(i)}(\dec) \defeq \{m_X \in \msg{M_X}&\colon \exists\,m_Y' \in \msg{M_Y}, \dec(m_X, m_Y') = i\},\notag\\
      \Icy^{(i)}(\dec) \defeq \{m_Y \in \msg{M_Y}&\colon  \exists\, m_X' \in \msg{M_X}, \dec(m_X', m_Y) = i\}.
                                                   \label{eq:def:icx:icy}
    \end{align}
  \end{definition}
   We will refer to \eqref{eq:decomp} as a decomposition of $\dec$.

{ \begin{example}
    $\putgrid{\drawgrid{
        {1,0},
        {0,1},}}$ is decomposable, which can be decomposed as %
    $\putgrid{\drawgrid{
        {0,1},
        {1,0},}} = \putgrid{\drawgrid{
        {0,1},
        {0,0},}} \oplus \putgrid{\drawgrid{
        {0,0},
        {1,0},}}$
    or
    $\putgrid{\drawgrid{
        {0,1},
        {1,0},}} = \putgrid{\drawgrid{
        {1,0},
        {0,0},}} \oplus \putgrid{\drawgrid{
        {0,0},
        {0,1},}} \oplus 1$.
    The decoder $\putgrid{\drawgrid{
        {1,1, 0},
        {1,0, 0},
        {0,0, 1},}}$
    and its complement $\putgrid{\drawgrid{
        {0,0, 1},
        {0,1, 1},
        {1,1, 0},}}$
    are decomposable, with decompositions %
    $ \putgrid{\drawgrid{
        {1,1, 0},
        {1,0, 0},
        {0,0, 1},}} =
       \putgrid{\drawgrid{
        {1,1, 0},
        {1,0, 0},
        {0,0, 0},}}
    \oplus
    \putgrid{\drawgrid{
        {0,0, 0},
        {0,0, 0},
        {0,0, 1},}}$,
    and
    $ \putgrid{\drawgrid{
        {0,0, 1},
        {0,1, 1},
        {1,1, 0},}} =
       \putgrid{\drawgrid{
        {0,0, 1},
        {0,1, 1},
        {1,1, 1},}}
    \oplus
    \putgrid{\drawgrid{
        {1,1, 1},
        {1,1, 1},
        {1,1, 0},}}
    \oplus 1$,
    respectively.   
  \end{example}
}

{  The following result demonstrates the weak separability of  decomposable decoders. A proof is provided in \appref{app:lem:decomp}.
  \begin{lemma}
  \label{lem:decomp}
  Suppose $\Ac_0$ and $\Ac_1$ are open convex subsets of $\pdist$, and $\dec$ is a decomposable decoder with the decomposition  [cf. \eqref{eq:decomp}]
    $\dec = \dec_0 \oplus \dec_1 \oplus \ibar$ 
  for some $i \in \{0, 1\}$. If $\dec \mid (\Ac_0, \Ac_1)$, then we have $\dec_j \mid (\Ac_0, \Ac_1)$ for some $j \in \{0, 1\}$.
\end{lemma}
}

In addition, we can formalize the recursive properties of threshold decoders discussed in \secref{sec:inner:bound}
as the \emph{reducibility} of decoders. Specifically, given a decision matrix $\Ab$ and $i\in \{0, 1\}$, its \emph{$i$-dominated rows} (or columns) are defined as the rows (or columns) being all $i$'s. Then, a decoder $\dec$ is called \emph{reducible} if $\Ab \lfa \dec$ has dominated rows or columns. Given a reducible decoder $\dec \lfa \Ab$, we say $\dec$ can be reduced to $\dec'$, if $\dec' = \dec$, or $\Ab' \lfa \dec'$ can be obtained from $\Ab$ by successively deleting dominated rows/columns. 

{
\begin{example}
    The decoder $\decr_{3, 2} \lfa  \putgrid{\drawgrid{
      {0,0,1},
      {0,1,1},}}$ is reducible, which can be reduced to  $\decr_{2, 2} \lfa  \putgrid{\drawgrid{
      {0,0},
      {0,1},}}$ (via deleting the $1$-dominated column), or reduced to $\bar{\decr}_{2, 2} \lfa  \putgrid{\drawgrid{
      {0,1},
      {1,1},}}$ (via deleting the $0$-dominated column).
\end{example}
}

Moreover, we call a decoder $\dec$ \emph{completely reducible} if it can be reduced to trivial decoders. Then, for given $M_X, M_Y \geq 1$, we denote the collections of $M_X\times M_Y$ non-completely-reducible decoders and completely reducible decoders by $\Deccn_{M_X, M_Y}$ and $\Decc_{M_X, M_Y}$, respectively.

It can be verified that all threshold decoders are completely reducible. Furthermore, we have the following result, a proof of which is provided in \appref{app:th:suff}.

\begin{lemma}
  \label{lem:th:suff}
  Given $M_X \geq M_Y \geq 1$, we have $\Ec[\Decc_{M_X, M_Y}] = \Ec[\{\decr_{M_X, M_Y}, \bar{\decr}_{M_X, M_Y}\}]$.
\end{lemma}
\begin{remark}
If $M_X > M_Y$, we can apply \factref{fact:equiv}  to refine the result as $\Ec[\Decc_{M_X, M_Y}] = \Ec[\decr_{M_X, M_Y}]$.
\end{remark}

From \lemref{lem:th:suff}, the threshold decoders $\decr_{M_X, M_Y}, \bar{\decr}_{M_X, M_Y}$ have the same separability as the collection of completely reducible decoders $\Decc_{M_X, M_Y}$. In addition, we have the following useful characterization for decoders in $\Deccn$, a proof of which is provided in \appref{app:fact:reduced}.

\begin{fact}
  \label{fact:reduced}
  Given $M_X, M_Y \geq 1$ and a decoder $\dec \in \Deccn_{M_X, M_Y}$, there exists a unique irreducible decoder that can be reduced from $\dec$, denoted by $\redus(\dec)$, which we refer to as the \emph{reduced form} of $\dec$.
\end{fact}

{\begin{example}
    Let $\dec_1 \lfa \putgrid{\drawgrid{
      {1,0, 0},
      {0,1, 0},
      {0,0, 0},}}, \dec_2 \lfa \putgrid{\drawgrid{
      {1,0, 1},
      {0,1, 1},
      {1,1, 1},}}$. Then we have $\dec_1, \dec_2 \in \Deccn_{3, 3}$, with the same reduced form
  $\redus(\dec_1) = \redus(\dec_2) \lfa \putgrid{\drawgrid{
      {1,0},
      {0,1},}}$.
\end{example}
}

Then, we can further partition $\Deccn_{M_X, M_Y}$ as $\Deccn_{M_X, M_Y} = \Decs_{M_X, M_Y} \union \Decd_{M_X, M_Y}$, where
\begin{align}
  \Decs_{M_X, M_Y} &\defeq \{\dec \in \Deccn_{M_X, M_Y}\colon%
  \redus(\dec)\text{ is indecomposable}\},  \notag\\
  \Decd_{M_X, M_Y} &\defeq \{\dec \in \Deccn_{M_X, M_Y}\colon \redus(\dec)\text{ is decomposable}\}.
                     \label{eq:def:decs:decd}
\end{align}

Then, the following theorem demonstrates that the error exponent region can be obtained by using only threshold decoders and the decoders in $  \Decs_{M_X, M_Y}$. A proof of which is presented in \appref{app:thm:sufficient}.
\begin{theorem}
  \label{thm:sufficient}
  Given $M_X \geq M_Y \geq 1$, we have
  $\Ec[\Decd_{M_X, M_Y}] \subset \Ec[\{\decr_{M_X, M_Y}, \bar{\decr}_{M_X, M_Y}\}] \union \Ec[\Decs_{M_X, M_Y}]$
  and
  \begin{align}
  \Ec(0_{M_X}, 0_{M_Y}) = \Ec[\{\decr_{M_X, M_Y}, \bar{\decr}_{M_X, M_Y}\}] \union \Ec[\Decs_{M_X, M_Y}].\label{eq:region:general}    
  \end{align}
\end{theorem}

\section{Optimal Error Exponents and Coding Schemes}
\label{sec:error-exp}

\subsection{Exact Characterization of Error Exponent Regions}     %
\label{sec:error-expon-regi}
{We then provide exact characterization of error exponent regions under 
one-bit/one-trit communication constraints, or with conditionally independent observations. Specifically, it can be shown that in theses cases the threshold decoder inner bound \eqref{eq:inner} is tight. }

\subsubsection{One-bit/One-trit Communication Constraints}
We first introduce the following result, a proof of which is provided in \appref{app:thm:opt:dec}. %
{ \begin{theorem}
    \label{thm:opt:dec}
    Suppose $ M_X \geq M_Y \geq 1$ and $(M_X - 2)(M_Y - 2) < 2$. Then, there exists no $M_X \times M_Y$ decoder that is both indecomposable and irreducible, 
    and we have
    \begin{align}
      \Ec(0_{M_X}, 0_{M_Y}) = \Ec[\decr_{M_X, M_Y}] \cup \Ec[\bar{\decr}_{M_X, M_Y}].
      \label{eq:mx:my:rate}
    \end{align}
  \end{theorem}
}

  From \thmref{thm:opt:dec}, it suffices to consider threshold decoders in the one-bit compression settings with $M_X \geq M_Y = 2$ or the two-sided one-trit compression ($M_X = M_Y = 3$). In the following, we discuss the error exponent regions under \emph{two-sided one-bit compression} constraint ($\Ec[0_2, 0_2]$), \emph{two-sided one-trit} constraint ($\Ec[0_3, 0_3]$), and the \emph{one-sided one-bit} constraint ($\Ec[0_M, 0_2]$ for $M \geq 3$, or $\Ec[R, 0_2]$ for $R \geq 0$), respectively.

  \paragraph{Two-sided One-bit compression}
   The exponent region $\Ec(0_2, 0_2)$ under \emph{two-sided one-bit compression} regime \cite{han1989exponential, han1998statistical}, can be obtained as a straightforward corollary of \thmref{thm:opt:dec}.

\begin{corollary}[\!\!{\cite[Theorem 5]{han1989exponential}, \cite[Theorem 5.6]{han1998statistical}}]
We have
   $ \Ec(0_2, 0_2) = \Ec[\decr_{2, 2}] \union \Ec[\bar{\decr}_{2, 2}]$, 
  where $\Ec[\decr_{2, 2}]$ and $\Ec[\bar{\decr}_{2, 2}]$ are as given by \thmref{thm:exp:th}, and can be represented as
  \begin{align*}
    \Ec[\decr_{2, 2}] &= \{(E_0, E_1) \colon \Dc_0(E_0) \cap  \Bc_1(E_1) = \varnothing\},\\
    \Ec[\bar{\decr}_{2, 2}] &= \{(E_0, E_1) \colon \Bc_0(E_0) \cap  \Dc_1(E_1) = \varnothing\},
  \end{align*}
 where for $i \in \{0, 1\}$ and $t \geq 0$, we have defined
$  \Bc_i(t) \defeq \{(Q_X, Q_Y) \colon D(Q_X\|P_X^{(i)}) < t,  D(Q_Y\|P_Y^{(i)}) < t \}.$
\end{corollary}
\begin{remark}
  It has been shown in \cite{han1989exponential} that the same result can be established when we relax the strict positive assumption \eqref{eq:pos} to $D(P^{(0)}_{XY} \| P^{(1)}_{XY}) < \infty$.
\end{remark}

\paragraph{Two-sided One-trit Compression} %
The error exponent region can be again obtained as an immediate corollary of \thmref{thm:opt:dec}.
\begin{corollary}
  The exponent region of $M_X = M_Y = 3$ is
    $\Ec(0_3, 0_3) = \Ec[\decr_{3, 3}] \union \Ec[\bar{\decr}_{3, 3}]$, 
  where $\Ec[\decr_{3, 3}]$ and  $\Ec[\bar{\decr}_{3, 3}]$ are as given by \thmref{thm:exp:th}.  
\end{corollary}

\paragraph{One-sided One-bit Compression}
We first introduce the following result, which demonstrates the connection between one-sided and two-sided constant-bit constraints. A proof is provided in \appref{app:prop:single}.

\begin{proposition}
  \label{prop:single}
  Given $M_Y \geq 1$, $M_X > 2^{M_Y}$, and $R_X \in [0, \infty)$, we have
   $ \Ec(R_X, 0_{M_Y}) = \Ec(0_{M_X}, 0_{M_Y})= \Ec( 0_{2^{M_{\!Y}}}, 0_{M_Y}).$
 \end{proposition} 
 Therefore, without loss of generality we may assume that $M_Y \leq M_X \leq 2^{M_Y}$.
 
{
 In addition, by combining \exref{ex:4:2} and \thmref{thm:opt:dec}, we have $\Ec(0_{4}, 0_{2}) = \Ec[\decr_{4, 2}] = \Ec[\decr_{3, 2}] = \Ec(0_{3}, 0_{2})$.
Hence, from \propref{prop:single}, the error exponent region for \emph{one-sided one-bit compression} can be summarized as follows. }
\begin{corollary}  
    For all $M \geq 3$ and $R \in [0, \infty)$, we have
     $ \Ec(R, 0_{2}) = \Ec(0_{M}, 0_{2})= \Ec(0_{3}, 0_{2}) =      \Ec[\decr_{3, 2}].$
\end{corollary}
\begin{remark}
  It is worth noting that in general we have $ \Ec(0_2, 0_2) \subsetneq \Ec(0_3, 0_2) = \Ec(R, 0_{2})$. Therefore, when one distributed node is allowed to transmit only a one-bit message,
to obtain the optimal performance, %
the other node is required to transmit at least a one-trit message. This situation differs from the one appeared in the discussion of the optimal type-II error exponent $E_1$ with type-I error $\pi_0$ constrained by a constant (cf. \cite[Corollary 7]{han1987hypothesis}), where it requires only a one-bit message sent from the other node to achieve the optimal performance.
\end{remark}

\subsubsection{Conditional Independent Observations}
In addition to the one-bit/one-trit cases, when the observations at both nodes are conditionally independent given $\Hs = 0$ or $\Hs = 1$, the inner bound \eqref{eq:inner} is tight for all $M_X \geq M_Y \geq 1$, illustrated as follows. A proof is provide in \appref{app:cond:indep}.

\begin{theorem}
\label{thm:cond:indep}  
  Suppose $P^{(i)}_{XY} = P^{(i)}_{X}P^{(i)}_{Y}$ for some $i \in \{0, 1\}$, then we have
   $ \Ec(0_{M_X}, 0_{M_Y}) = \Ec[\decr_{M_X,M_Y}] \cup \Ec[\bar{\decr}_{M_X,M_Y}]$,
  for all $M_X \geq M_Y \geq 1$, where $\Ec[\decr_{M_X,M_Y}]$ and  $\Ec[\bar{\decr}_{M_X,M_Y}]$ are as given by \thmref{thm:exp:th}.
\end{theorem}

\subsection{Optimal Coding Schemes}
From \thmref{thm:separation}, for each $M_X \times M_Y$ decoder $\dec$, each error exponent pair $(E_0, E_1)$ in the interior of $\Ec[\dec]$ can be achieved by the coding schemes  $\{(f_n, g_n, \dec)\}_{n \geq 1}$, where $f_n$ and $g_n$ are type-based encoders characterized by corresponding type-encoding functions.
Specifically, for the error exponent regions established in \secref{sec:error-expon-regi}, it suffices to consider the coding schemes with threshold decoders, i.e., $\decr_{M, M}, \bar{\decr}_{M, M}$ for $M \geq 1$, and $\decr_{M_X, M_Y}$ for $M_X > M_Y \geq 1$.
For ease of exposition, for each $k \geq 0$, let us define 
\begin{align}
  \chi_k
  \defeq
  \begin{cases}
    1&\text{if $k$ is odd,}\\
    0&\text{if $k$ is even,}
  \end{cases}
       \label{eq:def:chi}
\end{align}
and $\chibar_k \defeq 1 - \chi_k$. Then, for all $M \geq 1$, we define the mapping $r_M \colon \msgset{M} \to \msgset{M}$, such that
  \begin{align}
    r_M(k) \defeq \frac{k}{2} + \left(M - k - \frac12\right) \chi_k, \quad \text{for all $k \in \msgset{M}$}.
    \label{eq:def:rM}
  \end{align}
  For convenience, given subsets $\Qc_X \subset \cP^\X, \Qc_Y \subset \cP^\Y$ and $i \in \{0, 1\} $, we adopt the notation
  \begin{align}
    \divp_i(\Qc_X, \Qc_Y) &\defeq \inf_{\substack{ Q_X \in \Qc_X\\ Q_Y \in \Qc_Y}} \divp_i(Q_X, Q_Y),
    \label{eq:def:divp:inf}
  \end{align}
  and denote $\divp_i(Q_X, \Qc_Y) \defeq  \divp_i(\{Q_X\}, \Qc_Y)$ and $ \divp_i(\Qc_X, Q_Y) \defeq \divp_i(\Qc_X, \{Q_Y\})$  for distributions $Q_X \in \cP^\X$ and $Q_Y \in \cP^\Y$.

  The following result summarizes the error exponent region and the corresponding type-encoding functions,  %
with a proof presented in \appref{app:prop:decr:coding:all}.

{
\begin{proposition}
  \label{prop:decr:coding:all}
  Given $M_X > M_Y = M \geq 1$, for $\dec \in \{\decr_{M, M}, \bar{\decr}_{M, M}, \decr_{M_X, M_Y}\}$ and an error exponent pair $(E_0, E_1)$, let us define sequences of sets $\{\Qc_X^{(k)}\}_{k \geq 0}$  and $\{\Qc_Y^{(k)}\}_{k \geq 0}$ such that
  \begin{align*}
    \Qc_X^{(0)} \defeq    
    \begin{cases}
      \left\{Q_X \in \cP^\X \colon D(Q_X\|P_X^{(0)}) < E_0\right\}, &\text{if }\dec = \decr_{M_X, M_Y},\\
       \cP^\X,&\text{otherwise,}
    \end{cases}
  \end{align*}
  and
  $\Qc_Y^{(0)} \defeq \cP^\Y$, and, for each $k \geq 1$,
\begin{subequations}
\begin{align}
  \Qc_X^{(k)} &\defeq \left\{Q_X \in \Qc_X^{(k - 1)}\colon
 \divp_{\hat{\chi}_k}(Q_X,   \Qc_Y^{(k - 1)}) < E_{\hat{\chi}_k}\right\},\\
  \Qc_Y^{(k)} &\defeq \left\{Q_Y \in \Qc_Y^{(k - 1)}\colon 
  \divp_{\hat{\chi}_k}(  \Qc_X^{(k - 1)}, Q_Y) < E_{\hat{\chi}_k}\right\},
\end{align}\label{eq:def:Qc}
\end{subequations}
where for all $k \geq 1$, we have defined
\begin{align*}
  \hat{\chi}_k \defeq 
    \begin{cases}
      \chibar_k &\text{if }\dec = \bar{\decr}_{M, M},\\
      \chi_k&\text{otherwise.}
    \end{cases}
\end{align*}

Then, for each $M \geq 1$,  $(E_0, E_1) \in \Ec[\dec]$ if and only if
 \begin{align}
   \divp_{\hat{\chi}_M}\bigl(\Qc_X^{(M - 1)},  \Qc_Y^{(M - 1)}\bigr) \geq E_{\hat{\chi}_M}.
   \label{eq:decr:err}
 \end{align}   
Moreover, each error exponent pair $(E_0, E_1) \in \interior(\Ec[\dec])$
can be achieved by the type-encoding functions
  \begin{align}
    \enctx(Q_X) \defeq
    \begin{cases}
      r_{M}(\enctix(Q_X))&\text{if }Q_X \in \Qc_X^{(0)},\\
      M&\text{otherwise,}
    \end{cases}\quad \encty \defeq r_{M} \circ \enctiy,
    \label{eq:def:enct:th}
  \end{align}
  where $r_M$ is as defined in \eqref{eq:def:rM}, and where ``$\circ$'' denotes the composition of functions. In addition, we have defined
  \begin{subequations}
      \begin{align}
    \enctix(Q_X) &\defeq \max\{k \in \msgset{M}\colon Q_X \in \Qc_X^{(k)}\},\quad\text{for all  }Q_X \in \Qc_X^{(0)}\\      \enctiy(Q_Y) &\defeq \max\{k \in \msgset{M}\colon Q_Y \in \Qc_Y^{(k)}\}, \quad\text{for all  }Q_Y \in \cP^\Y.
  \end{align}\label{eq:def:encti:xy}
\end{subequations}
\end{proposition}
}
From \eqref{prop:decr:coding:all}, the decision regions in the distribution space are characterized by the sets $\Qc_X^{(k)}, \Qc_Y^{(k)}$ as defined in \eqref{eq:def:Qc}, which can be regarded as generalizations of the divergence ball used in Hoeffding's test \cite{hoeffding1967probabilities}. For example, when $\decr_{M, M}$ is used as the decoder, from \eqref{eq:def:Qc}, the decision regions for $k = 1$ are the divergence balls
\begin{subequations}
\begin{align}
  \Qc_X^{(1)} &= \bigl\{Q_X \in \cP^\X\colon D(Q_X\|P_X^{(1)}) < E_1\bigr\},\\
  \Qc_Y^{(1)} &= \bigl\{Q_Y \in \cP^\Y\colon D(Q_Y\|P_Y^{(1)}) < E_1\bigr\},
\end{align}\label{eq:qc:xy:1}
\end{subequations}
in $\cP^\X$ and $\cP^\Y$, %
respectively. As a result, from \eqref{eq:def:encti:xy} we have $\enct(Q_X) > 0$ if and only if $D(Q_X\|P_X^{(1)}) < E_1$, and $\enct(Q_Y) > 0$ if and only if $D(Q_Y\|P_Y^{(1)}) < E_1$, which share similar forms as Hoeffding's test \cite{hoeffding1967probabilities}.

For $k > 1$, the decision regions $\Qc_X^{(k)}$ and $\Qc_Y^{(k)}$ %
do not have analytical solutions in general. The error exponent region and the optimal type-encoding functions $\enctx, \encty$ can still be computed via solving related multi-level optimization problems \cite{vicente1994bilevel} obtained from \eqref{eq:def:Qc}--\eqref{eq:decr:err}. A detailed discussion of the computation is provided in \appref{app:comp}.

Specifically, when the observations at nodes $\nx$ and $\ny$ are conditionally independent under both hypotheses, the decision regions $\Qc_X^{(k)}$ and $\Qc_Y^{(k)}$ can be simply represented by KL divergences of some marginal distributions, and the corresponding type-encoding functions %
become
quantization functions of the divergences. %
For simplicity of exposition, we again focus on the decoder $\decr_{M, M}$, and define functions $\lambda_X^{(i)}(\cdot), \lambda_Y^{(i)}(\cdot)$ for $i = 0, 1$, with %
 \begin{subequations}
\begin{align}
  \lambda_X^{(i)}(t) &\defeq \inf_{Q_X\colon D(Q_X\|P_X^{(\ibar)}) < t} D(Q_X\|P_X^{(i)}),\\
  \lambda_Y^{(i)}(t) &\defeq \inf_{Q_Y\colon D(Q_Y\|P_Y^{(\ibar)}) < t} D(Q_Y\|P_Y^{(i)}).
\end{align}
\label{eq:def:lamb} 
\end{subequations}
These functions can be interpreted as the optimal error exponents of local decision at each distributed node. For example, consider the setting where $\nx$ is required to make a local decision based on the observed $x^n$, then $\lambda_X^{(0)}(t)$ is the optimal type-I error exponent when we require type-II error exponent not exceed $t$; similarly, $\lambda_X^{(1)}(\cdot)$ represents the optimal type-II error exponent when type-I error exponent does not exceed $t$. %

Then, we have the following result, a proof of which is provided in \appref{app:prop:optim:cond}.

\begin{proposition}
  \label{prop:optim:cond}
  Suppose we have, for both $i \in \{0, 1\}$, 
  \begin{align}
    P^{(i)}_{XY}(x, y) = P_{X}^{(i)}(x)P_{Y}^{(i)}(y), \quad \text{for all }(x, y) \in \X \times \Y.
    \label{eq:def:cond:ind}
  \end{align}
  Then, for given $E_0, E_1$ and $k \geq 1$, 
  with $\decr_{M, M}$ used as the decoder, the sets $\Qc_X^{(k)}$ and $\Qc_Y^{(k)}$ as defined in \eqref{eq:def:Qc} are
\begin{subequations}
\begin{align}
  \Qc_X^{(k)} &= \left\{Q_X \in \cP^\X\colon D(Q_{X}\|P_{X}^{(0)}) < \gamma_{X}^{(k - \chi_k)}, D(Q_{X}\|P_{X}^{(1)}) < \gamma_{X}^{(k - \bar{\chi}_k)}\right\},\label{eq:Q:k:x}\\
  \Qc_Y^{(k)} &= \left\{Q_Y \in \cP^\Y\colon D(Q_{Y}\|P_{Y}^{(0)}) < \gamma_{Y}^{(k - \chi_k)}, D(Q_{Y}\|P_{Y}^{(1)}) < \gamma_{Y}^{(k - \bar{\chi}_k)}\right\},\label{eq:Q:k:y}
\end{align}
\label{eq:Q:k:xy}
\end{subequations}
where %
$\chi_k$ and $\chibar_k$ are as defined in \eqref{eq:def:chi}, and where we have defined the sequences $\bigl\{\gamma_X^{(k)}\bigr\}_{k \geq 0}$ and $\bigl\{\gamma_Y^{(k)}\bigr\}_{k \geq 0}$ such that $\gamma_{X}^{(0)} \defeq \infty,  \gamma_{Y}^{(0)} \defeq \infty$, and, for all $k \geq 1$,
\begin{subequations}
\begin{align}
  \gamma_{X}^{(k)} &\defeq   E_{\chi_{k}} - \lambda_Y^{(\chi_{k})}(\gamma_{Y}^{(k-1)}),\\
  \gamma_{Y}^{(k)} &\defeq   E_{\chi_{k}} - \lambda_X^{(\chi_{k})}(\gamma_{X}^{(k-1)}),
\end{align}\label{eq:def:gamma}
\end{subequations}
where $\lambda_X^{(i)}$ and $\lambda_Y^{(i)}$ are as defined in \eqref{eq:def:lamb}.

In addition, for each $M \geq 1$, $(E_0, E_1) \in \Ec[\decr_{M, M}]$ if and only if \begin{align}
  \gamma_X^{(M)} + \gamma_Y^{(M)} - E_{\chi_{M}} \leq 0.
  \label{eq:err:gam}
  \end{align}  
\end{proposition}

\section{Numerical Examples}
\label{sec:numerical-examples}
We then provide
the error exponent region and type-encoding functions for two concrete examples.

First, we consider the DHT problem with binary alphabets $\X = \Y = \{0, 1\}$, and the joint distributions
\begin{align}
  P^{(i)}_{XY}(x, y)=
  \begin{cases}
    \frac12 &\text{if~$x = y = 1 - i$},\\
    \frac16 &\text{otherwise,}
  \end{cases}
              \label{eq:dist:ex1}
\end{align}
for $i \in \{0, 1\}$. Then, the corresponding marginal distributions are
   $ (P^{(0)}_{X}(0), P^{(0)}_{X}(1))
  =  ( P^{(0)}_{Y}(0), P^{(0)}_{Y}(1))
  =  \left(\frac13,
    \frac23\right)$ and
   $(P^{(1)}_{X}(0), P^{(1)}_{X}(1)) =   (P^{(1)}_{Y}(0), P^{(1)}_{Y}(1)) = \left(\frac23, \frac13\right).$

\begin{figure}[!t]
  \centering
  \includegraphics[width = .4\textwidth]{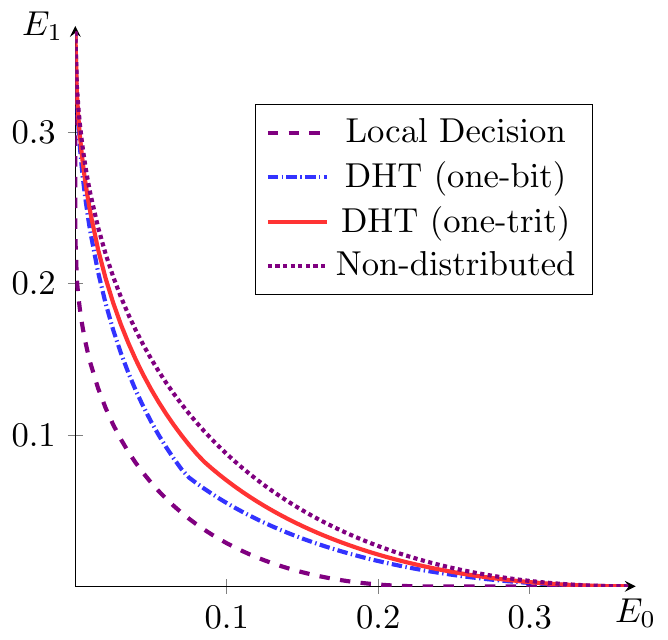}
  \caption{Optimal achievable error exponent pairs $(E_0, E_1)$ for the distribution \eqref{eq:dist:ex1}, under local decision (decision by node $\nx$/$\ny$ only), DHT with one-bit and one-trit communication constraints, and non-distributed decision based on complete observations of both $x^n$ and $y^n$.}
  \label{fig:exp}
\end{figure}

The optimal error exponents under different communication constraints are shown in \figref{fig:exp}. Specifically, the four curves demonstrate the boundaries of error exponent regions in the settings with
\begin{itemize}
\item \emph{Local Decision}: the error exponent pairs obtained by local decision at node $\nx$ based on observed $x^n$, which can also be represented as the region $\Ec(0_2, 0_1)$. Due to the symmetric form of \eqref{eq:dist:ex1}, the error exponent pairs obtained by local decision at node $\ny$ are the same, i.e., $\Ec(0_2, 0_1) = \Ec(0_1, 0_2)$.
\item \emph{DHT (one-bit)}: the error exponent pairs obtained by DHT with two-sided one-bit communication constraints, $\Ec(0_2, 0_2)$.
\item \emph{DHT (one-trit)}: the error exponent pairs obtained by DHT with two-sided one-trit communication constraints, $\Ec(0_3, 0_3)$.
\item \emph{Non-distributed}: the error exponent pairs obtained by complete observations of $x^n$ and $y^n$ sequences, which can also be represented as\footnote{Note that under both hypotheses $\Hs = 0, 1$, we have $H(X) \leq \log |\X| = \log 2$, and similarly, $H(Y) \leq \log 2$, where $H(\cdot)$ denotes the entropy. Therefore, the full sequences $x^n$ and $y^n$ can be transmitted to the center under rate constraints $(\log 2, \log 2)$.} $\Ec(\log 2, \log 2)$.
\end{itemize}

In addition, since the log-likelihood function 
  \begin{align*}
    \log\frac{P^{(0)}_{XY}(x, y)}{P^{(1)}_{XY}(x, y)} = (x + y - 1) \cdot \log 3
  \end{align*}
  can be represented as the superposition of functions of $x$ and $y$, it can be verified that (see, e.g., \cite[Remark 3]{watanabe2017neyman})
  \begin{align*}
    \Ec(0, 0) = \Ec(R_X, R_Y), \quad \text{for all }R_X \geq 0, R_Y \geq 0.
  \end{align*}
  Therefore, the performance of the non-distributed case also coincides with the DHT with zero-rate communication constraints.

\begin{figure}[!t]
  \centering
  \includegraphics[width = .7\textwidth]{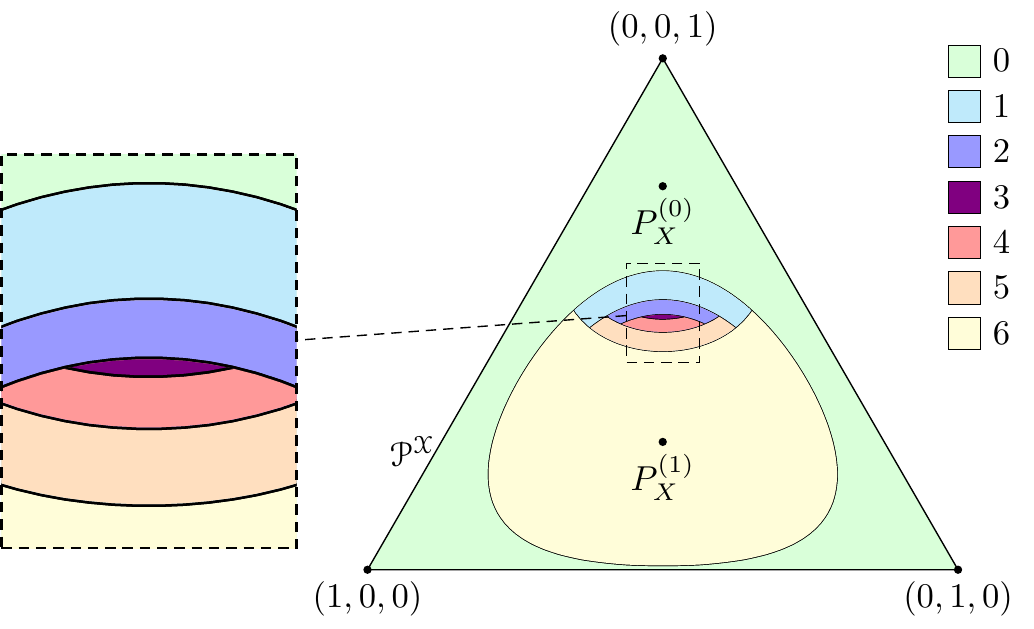}
  \caption{The optimal type-encoding function $\theta_X$ as defined in \eqref{eq:def:enct:th} for DHT with joint distributions \eqref{eq:dist:ex2} and communication constraints $(0_7, 0_7)$.}
  \label{fig:encoding}
\end{figure}

Our second example demonstrates the optimal coding scheme and type-encoding functions. In particular, we consider the DHT problem with alphabets  $\X = \Y = \{0, 1, 2\}$ and assume that $X$ and $Y$ are conditionally independent given both hypotheses, i.e., \eqref{eq:def:cond:ind} holds for both $i \in \{0, 1\}$. Let the marginal distributions be
\begin{align}
  \begin{bmatrix}
    P^{(0)}_{X}(0)\\
    P^{(0)}_{X}(1)\\
    P^{(0)}_{X}(2)
  \end{bmatrix}
  =
  \begin{bmatrix}
    P^{(0)}_{Y}(0)\\
    P^{(0)}_{Y}(1)\\
    P^{(0)}_{Y}(2)
  \end{bmatrix}
  =
  \begin{bmatrix}
    \frac18\\[0.5ex]
    \frac18\\[0.5ex]
    \frac34
  \end{bmatrix},\quad
    \begin{bmatrix}
    P^{(1)}_{X}(0)\\
    P^{(1)}_{X}(1)\\
    P^{(1)}_{X}(2)
  \end{bmatrix}
  =
  \begin{bmatrix}
    P^{(1)}_{Y}(0)\\
    P^{(1)}_{Y}(1)\\
    P^{(1)}_{Y}(2)
  \end{bmatrix}
  =
  \begin{bmatrix}
    \frac38\\[0.5ex]
    \frac38\\[0.5ex]
    \frac14
  \end{bmatrix}. 
\label{eq:dist:ex2}
\end{align}

Specifically, we consider the DHT problem with communication constraints $(0_{7}, 0_{7})$. By applying \propref{prop:optim:cond}, we can verify that the error exponent pair $(E_0, E_1) = (0.3, 0.25)$ can be obtained by the coding scheme with decoder $\decr_{7, 7}$ and type-encoding functions $\enctx\colon \cP^\X \to \{0, \dots, 6\}$ and  $\encty\colon \cP^\Y \to \{0, \dots, 6\}$, where  $\theta_X$ is depicted  in \figref{fig:encoding}. Note that due to $\cX = \cY$ and the symmetry of underlying distributions \eqref{eq:dist:ex2}, the type-encoding function $\encty$ coincides with $\enctx$, i.e., we have
$\encty(Q) = \enctx(Q)$ for all $Q \in \cP^\X = \cP^\Y$, and thus the plot of $\encty$ can also be demonstrated by \figref{fig:encoding}.

\section{{Discussions}}
\label{sec:diss}
{
  Our analysis provides a geometric approach for  constant-bit DHT problems, which reduce the characterization of error exponent regions to the study of separability on the distribution space (cf. \defref{def:separation}). With this approach, we establish the threshold decoder inner bound of error exponent regions. Moreover, we provide exact characterizations when the observations at nodes are conditionally independent or when the constraints are of one-bit/one-trit type. Specifically, these error exponent regions can be obtained by threshold decoders, and can be effectively computed.}

\newcommand{\hfour}{.8em} %
\newcommand{\figheight}{6.5cm} %
\newcommand{\decfour}{\putgrid[\hfour]{\drawgrid{
        {1,0, 0, 1},
        {0,0, 1, 1},
        {0,1, 1, 0},
        {1,1, 0, 0},}}}
\begin{figure}[!t]
  \centering
  \subfloat[$(\Ac_0, \Ac_1)$ is not separable by $\decr_{4, 4}$ ]{
    \resizebox{!}{\figheight}{\begin{tikzpicture}[line width = .07em, box/.style={rectangle,draw=none,thick, minimum size=1cm},] 
  \coordinate (C0) at (-.5, .5);
  \coordinate (C1) at (.5,  -0.5);
  \tikzstyle{D shape}=[minimum width=8cm, minimum height=.69cm]
  \tikzstyle{B shape}=[minimum width=5.7cm, minimum height=5.7cm]
  
  \node [draw, rotate=45, ellipse, color D0, D shape, align=center, name = D0, label={[scale = 1, xshift = 0.6cm, yshift = 1.6cm]left:$\Ac_0$}] at (C0)   {}; 
  \node [draw, ellipse, rotate=45, color D1, D shape, align=center,  name = D1,  label={[scale = 1, xshift = 1.2cm, yshift = .5cm]left:$\Ac_1$}] at (C1)   {}; 

  \node [draw = none, rectangle, B shape, align=center, name = B0] at   (C0)   {}; 
  \node [draw = none, rectangle, B shape, align=center, name = B1] at   (C1)   {}; 

  \def\xshift{-4.35}
  \def\yshift{-3.85}
  \foreach \x in {1,...,8}{
    \foreach \y in {1,..., 7}
    \node[box] at (\xshift + \x, \yshift + \y){};
  }
    \foreach \x[evaluate=\x as \xi using {int(Mod(\x - 1,4))}] in {1,..., 8}{
    \node at (\xshift + \x, \yshift + 7.75)[opacity = 0, scale = .8]{\xi};
  }




  \coordinate (P0) at (B1.north west);
  \def\dista{4.1}
  \def\distb{1.2}
  \def\distc{3}
  \coordinate (P1) at ($(P0) + (\dista, -\dista)$);
  \coordinate (P2) at ($(P0) + (\distb, -\distb)$);
  \coordinate (P3) at ($(P0) + (\distc, -\distc)$);

  \def\myop{.3}
  \draw [opacity = \myop] (P0) -- (B1.north east);
  \draw [opacity = \myop] (P0) -- (B1.south west);
  \draw (P0) [opacity = \myop] rectangle (P1);
  \draw (P1) [opacity = \myop] rectangle (P2);
  \draw (P2) [opacity = \myop] rectangle (P3);

\newcommand{\drawEllipse}[2]{    \node [draw, ellipse, rotate=45, #2, D shape, align=center] at (#1)   {};}

\begin{scope}
  \clip (P0) rectangle ++ (4, -4);
  \drawEllipse{C0}{pattern=vertical lines} %
\end{scope}
  
  \begin{scope}
    \clip (P1) rectangle ++ (-3, 3); 
    \drawEllipse{C1}{pattern=horizontal lines}
\end{scope}

  \begin{scope}
    \clip (P2) rectangle ++ (2, -2);
    \drawEllipse{C0}{pattern=horizontal lines}
\end{scope}

  \begin{scope}
    \clip (P3) rectangle ++ (-1, 1); 
    \drawEllipse{C1}{pattern=vertical lines} 
\end{scope}

  \matrix [xshift = -.9cm, yshift = 1.4cm] at (current bounding box.south east) {
    \node [scale = 1.4, preaction={color D0}, shape=rectangle, pattern=vertical lines, label={[scale = .8]right:$ \Ac_0 \sub_2 \Ac_1$},  line width = 2] {}; \\
    \node [scale = 1.4, preaction={color D1}, shape=rectangle, pattern=horizontal lines, label={[scale = .8]right:$ \Ac_0 \sub_3 \Ac_1$}, line width = 2] {}; \\
    \node [scale = 1.4, preaction={color D0}, shape=rectangle, pattern=grid, label={[scale = .8]right:$ \Ac_0 \sub_4 \Ac_1$}, line width = 2] {}; \\
    \node [
    scale = 1.4, preaction={color D1}, shape=rectangle, pattern=grid, label={[scale = .8]right:$ \Ac_0 \sub_5 \Ac_1$}, line width = 2] {}; \\
    };

  \coordinate (A) at (-3, 3.5);

  \draw[style = ->, scale = .8] (A) -- ++(.6, 0)  node[scale = .8, right] {$Q_X$} ;
  \draw[style = ->, scale = .8] (A) -- ++(0, -.6) node[scale = .8, below] {$Q_Y$};
  
  
\end{tikzpicture}

}\label{fig:baguette:l}}
  \hspace{1em}
  \subfloat[$(\Ac_0, \Ac_1)$ is separable by the decoder   {\decfour}
  ]{
  \resizebox{!}{\figheight}{\def\myop{.3}

\begin{tikzpicture}[line width = .07em, box/.style={rectangle,draw=black, opacity = \myop, minimum size=1cm}, nbox/.style={box, draw = none, fill = gray, fill opacity = .2}, boxfour/.style={box, draw = purple, minimum size=4cm, opacity = .5},] 
  \coordinate (C0) at (-.5, .5);
  \coordinate (C1) at (.5,  -0.5);
  \tikzstyle{D shape}=[minimum width=8cm, minimum height=.69cm]
  \tikzstyle{B shape}=[minimum width=5.7cm, minimum height=5.7cm]
  
  \node [draw, rotate=45, ellipse, color D0, D shape, align=center, name = D0, label={[scale = 1, xshift = 0.6cm, yshift = 1.5cm]left:$\Ac_0$}] at (C0)   {}; 
  \node [draw, ellipse, rotate=45, color D1, D shape, align=center,  name = D1,  label={[scale = 1, xshift = 1.2cm, yshift = .5cm]left:$\Ac_1$}] at (C1)   {}; 

  \node [draw = none, rectangle, B shape, align=center, name = B0] at   (C0)   {}; 
  \node [draw = none, rectangle, B shape, align=center, name = B1] at   (C1)   {}; 




  \def\xshift{-4.35}
  \def\yshift{-3.85}

  \foreach \x in {1,...,8}{
    \foreach \y in {1,..., 7}
    \node[box] at (\xshift + \x, \yshift + \y){};
  }

  \newcommand{\mydraw}[3]{  \foreach \x in {1,...,#1}{
      \node[nbox] at (\xshift + \x + #2, \yshift + \x + #3){};
    }}
  \mydraw{2}{6}{0}
  \mydraw{3}{5}{0}
  \mydraw{6}{2}{0}
  \mydraw{7}{1}{0}
  \mydraw{5}{0}{2}
  \mydraw{4}{0}{3}
  \mydraw{1}{0}{6}

  \def\colorx{cyan}
  \def\colory{orange}
  \foreach \y[evaluate=\y as \yi using {int(3 - Mod(\y,4))}] in {7,..., 1}{
    \node at (\xshift  + .3, \yshift + \y)[\colory, scale = .8]{\yi};
  }

  \foreach \x[evaluate=\x as \xi using {int(Mod(\x - 1,4))}] in {1,..., 8}{
    \node at (\xshift + \x, \yshift + 7.75)[\colorx, scale = .8]{\xi};
  }

  \node[boxfour, ultra thick] at (\xshift + 6.5, \yshift + 5.5){};
      
  \matrix [draw, fill = white, xshift = -1.4cm, yshift = 1cm] at (current bounding box.south east) {
    \node [scale = .8, preaction={},  label={[scale = .8]right:$\enctx(Q_X)$}] [\colorx]{0\, 1}; \\
    \node [scale = .8, preaction={ },  label={[scale = .8]right:$\encty(Q_Y)$}] [\colory]{0\, 1}; \\
    };

  \coordinate (A) at (-3.7, 3.5);

  \draw[style = ->, scale = .8] (A) -- ++(.6, 0)  node[scale = .8, right] {$Q_X$} ;
  \draw[style = ->, scale = .8] (A) -- ++(0, -.6) node[scale = .8, below] {$Q_Y$};
  
  
\end{tikzpicture}

}\label{fig:baguette:r}}    
    \caption{{Threshold decoders are not necessarily optimal under two-bit constraints $(0_4, 0_4)$. \protect\subref{fig:baguette:l}: $(\Ac_0, \Ac_1)$ is not separable by threshold decoders $\decr_{4, 4} \lfa \putgrid[\hfour]{\drawgrid{
        {0,0, 0, 0},
        {0,0, 0, 1},
        {0,0, 1, 1},
        {0,1, 1, 1},}}$ or $ \bar{\decr}_{4, 4} \lfa
    \putgrid[\hfour]{\drawgrid{
        {1,1, 1, 1},
        {1,1, 1, 0},
        {1,1, 0, 0},
        {1,0, 0, 0},}}$; \protect\subref{fig:baguette:r}: $(\Ac_0, \Ac_1)$ is separable by the $4 \times 4$ decoder {\decfour}, with the corresponding mappings $\enct_X(\cdot)$, $\enct_Y(\cdot)$ [cf. \defref{def:separation}]. }}   
    \label{fig:baguette}
  \end{figure}

  {On the other hand, the notion of separability also suggests the intrinsic  complexity of characterizing error exponent region $\Ec(0_{M_X}, 0_{M_Y})$. In fact, even the two-bit setting $(0_4, 0_4)$ can have significantly more complicated behaviors, compared with the one-trit case $(0_3, 0_3)$. To see this, we can show that under $(0_4, 0_4)$ constraint, threshold decoders are not necessarily optimal, as illustrated in \figref{fig:baguette}. In this figure,  $\Ac_1$ is a mirror image of $\Ac_0$, and it can be noted that $(\Ac_0, \Ac_1)$ is not separable by ${\decr}_{4, 4}$ as $\Ac_0 \sub_4 \Ac_1 \neq \varnothing$ (cf. \propref{prop:recursive:decr}). Moreover, $(\Ac_0, \Ac_1)$ is not separable by $\bar{\decr}_{4, 4}$ from the reflection symmetry.  However,  $(\Ac_0, \Ac_1)$  can be separated by the decoder {\decfour}, as shown in \figref{fig:baguette:r}. It is worth mentioning this demonstrating case can appear in DHT characterizations, when we consider the separation of $\Ac_i \defeq \Dc_i(E), i = 0, 1$ for some $E > 0$ (cf. \thmref{thm:separation}). Specifically, one example is the DHT problem with $P_{XY}^{(0)}$ and  $P_{XY}^{(1)}$ with $\X = \Y = \{0, 1\}$, where
    \begin{align}
      P^{(0)}_{XY}(x, y) \defeq
      \begin{cases}
        \frac{1}{2}(1 - \eps)& \text{if $x = y$},\\
        \alpha \eps& \text{if $(x, y) = (0, 1)$},\\
        (1-\alpha)\eps& \text{if $(x, y) = (1, 0)$},
       \end{cases}
    \end{align}
    and $P^{(1)}_{XY}(x, y) \defeq P^{(0)}_{XY}(y, x)$, for all $(x, y) \in \X \times \Y$, where $\alpha \in (0, \frac12)$, and where $\eps > 0$ is chosen to be a small number.}

  {From \figref{fig:baguette}, a main difference between $\decfour$ and threshold decoders $\decr_{4, 4}$ or $\bar{\decr}_{4, 4}$ is that, $\decfour$ can reuse symbols in $\{0, 1, 2, 3\}$, which produces periodic patterns in $\pdist$ to obtain better separability. In particular, in \figref{fig:baguette:r}, $\cP^\X$ is divided into 8 different regions}\footnote{Formally, we can define such regions as  maximal connected subsets of $\{Q_Y \in \cP^\Y\colon \encty(Q_Y) = m_Y\}$ for $m_Y \in \msgset{M}$.} {(columns), with only $M = 4$ different symbols used. In contrast, for threshold decoders $\decr_{M, M}$ and $\bar{\decr}_{M, M}$, the number of different regions in $\cP^\X$ or $\cP^\Y$ is at most $M$. Generally, we can also generate such periodic patterns by an $M \times M$ decoder $\dec$, if its decoded result  $\dec(m_X, m_Y)$ depends only on the value of $((m_X + m_Y)\bmod M)$.  In addition, note that to obtain effective separability using such periodic patterns, it requires at least $M = 4$ symbols, since at least two distinct symbols are needed to encode (``cover'') each of $\Ac_0$ and $\Ac_1$ (cf. \figref{fig:baguette:r}). Therefore, such periodic patterns would not appear when $M = 3$, which also illustrates a fundamental difference between the setting
    $(0_3, 0_3)$ and $(0_M, 0_M)$ for $M \geq 4$. In general, the exact characterization of separability for an arbitrary decoder, including decoders with such periodic  patterns, can also be more difficult than threshold decoders.
  }
    
                        %

%
%
%
%

  
\newpage
\appendices
\section*{Auxiliary Notations and Definitions}

We first present some useful notations and definitions in our proof.

To begin, for two given decoders $\dec, \dec'$ with decision matrices $\Ab \lfa \dec$ and $\Ab' \lfa \dec'$, we call $\dec'$ a \emph{subdecoder} of $\dec$ if $\Ab'$ is a submatrix of $\Ab$. In addition, $\dec, \dec'$ are called \emph{equivalent}, denoted by $\dec \simeq \dec'$, if $\Ab'$ can be obtained from $\Ab$ by %
some row permutations and column permutations.

  Moreover, we refine the reducibility of decoders introduced in \secref{sec:decod-comp-select} as follows.
\begin{definition}
  \label{def:dec:redu}
  Given a non-trivial reducible decoder $\dec \lfa \Ab$, if $\Ab$ has $i$-dominated columns for $i \in \{0, 1\}$, we define  decoder $\redux^{(i)}(\dec)$ such that $\redux^{(i)}(\dec) \lfa \Ab_X^{(i)}$, where $\Ab_X^{(i)}$ denotes the submatrix of $\Ab$ obtained by deleting its $i$-dominated columns; similarly, if $\Ab$ has $i$-dominated rows, we define $\reduy^{(i)}(\dec)$ such that $\reduy^{(i)}(\dec) \lfa  \Ab_Y^{(i)}$, where $\Ab_Y^{(i)}$ is the submatrix of $\Ab$ obtained by deleting $i$-dominated rows.
\end{definition}
We refer to $\redux^{(0)}, \redux^{(1)}, \reduy^{(0)}, \reduy^{(1)}$ as elementary reduction operators. We then use \emph{reduction operators} to refer to the elementary reduction operators and their compositions. 

In addition, when compare two collection of decoders
$\Hc, \Hc' \subset \Dec$, we use $\Hc \preceq \Hc'$ to indicate that $\Ec[\Hc] \subset \Ec[\Hc']$. Specifically, the following fact would be useful in our proofs.
  \begin{fact}
    \label{fact:preceq}
    The relation ``$\preceq$'' is transitive, i.e., for all decoder collections $\Hc_0, \Hc_1$ and $\Hc_2$, if $\Hc_0 \preceq \Hc_1$ and $\Hc_1 \preceq \Hc_2$, then $\Hc_0\preceq \Hc_2$. In addition, given $\Hc_0, \Hc_1 \subset \Dec$ with $\Hc_0 \preceq \Hc_1$, we have $(\Hc_0 \union \Hc') \preceq (\Hc_1 \union \Hc')$ for all $\Hc' \subset \Dec$.
  \end{fact}

\section{Proof of \factref{fact:dec}}
\label{app:fact:dec}

  To begin, suppose $(E_0, E_1) \in \Ec(0_{M_X}, 0_{M_Y})$, then for each $\eps > 0$, there exists a sequence of coding scheme $\{\Cc_n\}_{n \geq 1}$, such that [cf. \eqref{eq:exp:def}]
  \begin{align}
    - \lim_{n\to \infty} \frac1n\log\pi_i(\Cc_n) = E_i - \eps,\quad i = 0, 1,
  \end{align}
  where each coding scheme $\Cc_n$ is equipped with some decoder in $\Dec_{M_X, M_Y}$.

  Note that since the set $\Dec_{M_X, M_Y}$ is finite, there exists a decoder $\phi \in \Dec_{M_X, M_Y}$ and an infinite subsequence $\{m_k\}_{k \geq 1}$ of positive integers, such that for each $k \geq 1$, the corresponding coding scheme $\Cc_{m_k}$ is equipped with $\phi$.

  Moreover, we define a new sequence of coding scheme $\Cc'_n \defeq \Cc_{m_{\hat{k}}}$ where $\hat{k} = \hat{k}(n) \defeq \max\{k \colon m_k \leq n\}$. It can be verified that
  \begin{align}
    - \lim_{n\to \infty} \frac1n\log\pi_i(\Cc'_{n}) &=     - \lim_{k\to \infty} \frac1n\log\pi_i(\Cc_{n_k})\notag\\
    &=  E_i - \eps,\quad \text{for } i = 0, 1,
  \end{align}
  which implies that $(E_0, E_1) \in \Ec[\dec]$.

  Therefore, we obtain
  \begin{align}
    \Ec(0_{M_X}, 0_{M_Y}) \subset \bigcup_{\dec \in \Dec_{M_X, M_Y}} \Ec[\dec] = \Ec[\Dec_{M_X, M_Y}].
    \label{eq:inclu:f}
  \end{align}

  In addition, note that for each decoder $\dec \in \Dec_{M_X, M_Y}$, we have $\Ec[\dec] \subset \Ec(0_{M_X}, 0_{M_Y})$, which implies the reverse inclusion
  \begin{align}
   \Ec[\Dec_{M_X, M_Y}] \subset \Ec(0_{M_X}, 0_{M_Y}).\label{eq:inclu:r}    
  \end{align}

  From \eqref{eq:inclu:f} and \eqref{eq:inclu:r}, we obtain $\Ec(0_{M_X}, 0_{M_Y})= \Ec[\Dec_{M_X, M_Y}]$ as desired.\hfill\IEEEQED

\section{Proof of \lemref{lem:type-based-coding}}
\label{app:lem:type}
We first introduce several useful definitions for a given alphabet $\Z$.
  The \emph{Hamming $d$-neighborhood} of $\Sc_Z \subset \Z^n$ is %
  $  \nbhdh^{d}(\Sc_Z) \defeq \{z^n\in \Z^n \colon d_{\mathrm{H}}(z^n, \tilde{z}^n) \leq k \text{ for some }\tilde{z}^n \in \Sc_Z\}$,  where $d_{\mathrm{H}}(z^n, \tilde{z}^n)$ denotes the Hamming distance between  $z^n, \tilde{z}^n \in \Z^n$, viz., $
  d_{\mathrm{H}}(z^n, \tilde{z}^n) \defeq \frac1n \sum_{i = 1}^n \kron_{\{z_i \neq \tilde{z}_i \}}$,
and where $\kron_{\{\cdot\}}$ denotes the indicator function.

In addition, given a type $Q_Z \in \Pch^\Z_n$, we use $\Tc^n_{Q_Z}$ (or simply $\Tc_{Q_Z}$) to denote the set of sequences $z^n \in \Z^n$ with the type $Q_Z$, i.e.,$  \Tc^n_{Q_Z} \defeq\{z^n\in \Z^n \colon \Ph_{z^n} = Q_Z\}.$
 Moreover, for a given $\eta > 0$, we use $\Tc^{n}_{Q_Z;\eta}$ to denote the sequences with type close to $Q_Z$, viz., 
$\left \{z^n \in \Z^n\colon \dmax(\Ph_{z^n}, Q_Z)
  \leq \eta\right\}$, 
where the metric $\dmax(\cdot, \cdot)$ on $\cP^\Z$ is defined, such that for $P_Z, Q_Z \in \cP^\Z$, 
\begin{align}
\dmax(P_Z, Q_Z)  %
  \defeq \max_{z \in \Z} |P_Z(z) - Q_Z(z)|.
  \label{eq:def:dmax}
\end{align}

Proceeding to the proof of the lemma, for a given pair of marginal distributions $(Q_X, Q_Y) \in \hat{\cP}_n^{\X} \times \hat{\cP}^{\Y}_n$, we first define
  \begin{align*}
    \Sc_X &\defeq  \{x^n \in \X^n \colon f_n(x^n) = \enct_X(Q_X)\},\\
    \Sc_Y &\defeq  \{y^n  \in \Y^n \colon g_n(y^n) = \enct_Y(Q_Y)\}
  \end{align*}
  and\footnote{With slight abuse of notation, we use  $(x^n, y^n)$ or simply $x^ny^n$ to denote the sequence $\{(x_i, y_i)\}_{i = 1}^n \in (\X \times \Y)^n$, and denote the set
$   \left\{(x^n, y^n)\colon x^n \in \Sc_X, y^n \in \Sc_Y \right\} \subset (\X \times \Y)^n$
by $\Sc_X \times \Sc_Y$, for given $\Sc_X \subset \X^n$ and $\Sc_Y \subset \Y^n$.} $\Sc_{XY} \defeq \Sc_X \times \Sc_Y$,  where for given $f_n$ and $g_n$, we have defined $\enct_X \colon \Pc^\X \to \Mc_X^{(n)}$ and $\enct_Y \colon \Pc^\Y \to \Mc_Y^{(n)}$ such that for all $P_X \in \cP^\X$ and $P_Y \in \cP^\Y$,
  \begin{align}
    \enct_X(P_X) &\defeq \argmax_{m_X \in \Mc_X^{(n)}}\,
                   \prob{f_n(X^n) = m_X \middle| X^n \sim P_X^{\otimes n}},                       \label{eq:def:phi:f}
\\
    \enct_Y(P_Y) &\defeq \argmax_{m_Y \in \Mc_Y^{(n)}}\,
    \prob{g_n(Y^n) = m_Y \middle| Y^n \sim P_Y^{\otimes n}},
                   \label{eq:def:phi:g}
  \end{align}
  where $(P_X)^{\otimes n}$ and $(P_X)^{\otimes n}$ represent the $n$-th product of $P_X$ and $P_Y$, respectively.

  By symmetry, it suffices to establish \eqref{eq:lem:type} for $i = 0$. To this end, let $(X^n, Y^n)$ be i.i.d. generated from $P^{(0)}_{XY}$, and define $Q_{XY}\in \cP^{\X \times \Y}$ such that it satisfies $[Q_{XY}]_X = Q_X, [Q_{XY}]_Y = Q_Y$ and $ \divp_0(Q_X, Q_Y) = D(Q_{XY} \| P^{(0)}_{XY})$. In addition,  by applying Sanov's theorem \cite{cover06}, for each $i \in \{0, 1\}$ and $(Q_X, Q_Y) \in \Pch^\X_n \times \Pch^\Y_n$, we have
  \begin{align}
    &\prob{(\Ph_{X^n}, \Ph_{Y^n}) = (Q_X, Q_Y)\middle| \Hs = i} %
        = \exp(-n (\divp_i(Q_X, Q_Y) + o(1))).
       \label{eq:prob:type}
  \end{align}

  Therefore, we can equivalently express   \eqref{eq:lem:type} as %
  \begin{align}
      \prob{(X^n, Y^n) \in \Sc_{XY}} %
      \geq \exp(-n \cdot (D(Q_{XY} \| P^{(0)}_{XY}) + \eps_n))
    \label{eq:lem:type:eq}
  \end{align}
  with $\eps_n = o(1)$.

  We then illustrate that \eqref{eq:lem:type:eq} holds, %
  if there exists a sequence of positive integers  $\{l_n\}_{n \geq 1}$ with $l_n = o(n)$, such that
for $n$ sufficiently large, we have
\begin{align}
  \max_{\Qt_{XY}\in \Qc_n} \beta_n(\Qt_{XY})
  \geq \frac12,
  \label{eq:gtr:half}
\end{align}
where for each $n \geq 1$ and $\Qt_{XY} \in \Pch^{\X \times \Y}_{n}$, we have defined
\begin{align}
  \beta_n(\Qt_{XY}) \defeq   \frac{\left|\Tc^{n}_{\Qt_{XY}} \cap  \nbhdh^{l_n}(\Sc_{XY})\right|}{\left|\Tc^{n}_{\Qt_{XY}} \right|}
\end{align}
with $\eta_n \defeq n^{-\frac13}$, and %
\begin{align}
  \Qc_n \defeq \left\{ \Qt_{XY} \in  \Pch^{\X \times \Y}_{n} \colon \dmax(\Qt_{XY}, Q_{XY}) \leq \eta_n\right\}.
  \label{eq:def:qcn}
\end{align}
To see this, first note that from \cite[Lemma 5.1]{csiszar2011information}, %
we have
\begin{align}
    \prob{(X^n, Y^n) \in \Sc_{XY}}%
    \geq \prob{(X^n, Y^n) \in \nbhdh^{l_n}(\Sc_{XY})} \cdot \exp(-n \eps_n')
    \label{eq:bd:1}
\end{align}
for some $\eps_n' = o(1)$. 

Moreover, from \eqref{eq:gtr:half}, for sufficiently large $n$, there exists $Q'_{XY} \in \Qc_n$, such that $\beta_n(Q'_{XY}) \geq \frac12$. As a result, we have
 \begin{align}
    \prob{(X^n, Y^n) \in  \nbhdh^{l_n}(\Sc_{XY}) }%
  &%
    \geq  \prob{(X^n, Y^n) \in \Tc^{n}_{Q'_{XY}} \cap   \nbhdh^{l_n}(\Sc_{XY})}\notag\\
  &%
    = \prob{(X^n, Y^n) \in \Tc^{n}_{Q'_{XY}}} \cdot \beta_n(Q'_{XY})\notag\\
  &%
    \geq \frac12 \cdot \prob{(X^n, Y^n) \in \Tc^{n}_{Q'_{XY}}}   
  \label{eq:bd:2}
\end{align}
where the equality follows from the fact that different sequences in a type class are equiprobable. %

In addition, it follows from the definition of $\Qc_n$ [cf.  \eqref{eq:def:qcn}] that $\dmax(Q'_{XY}, Q_{XY}) \leq \eta_n$. Hence, from the uniform continuity of KL divergence, there exists $\eps_n'' = o(1)$ such that
\begin{align*}
  \left|D(Q'_{XY}\|P_{XY}^{(0)}) - D(Q_{XY}\|P_{XY}^{(0)})\right| < \eps_n''.
\end{align*}
This implies that
\begin{align}
    \prob{(X^n, Y^n) \in \Tc_{Q'_{XY}}}%
  &%
    \geq (n + 1)^{-|\X||\Y|}\exp(-n D(Q'_{XY}\| P^{(0)}_{XY}))\notag\\
  &%
    \geq (n + 1)^{-|\X||\Y|}\exp(-n \eps_n'') \cdot \exp(-n D(Q_{XY}\| P^{(0)}_{XY})),%
              \label{eq:bd:3}
\end{align}
where the first inequality follows from a lower bound %
for the probability of a type class, see, e.g., \cite[Theorem 11.1.4]{cover06} or \cite[Lemma 2.6]{csiszar2011information}.

Then, it can then be verified from \eqref{eq:bd:1}, \eqref{eq:bd:2} and \eqref{eq:bd:3} that \eqref{eq:lem:type:eq} holds with
\begin{align*}
  \eps_n = \eps_n' + \eps_n'' + \frac1n \log 2 + \frac{|\X||\Y|}{n} \log (n + 1) = o(1).
\end{align*}

Hence, it remains to establish \eqref{eq:gtr:half}. 
To this end, we turn to consider probabilities %
under the measure $Q_{XY}$, and let $(\Xt^n, \Yt^n)$ be i.i.d. generated from $Q_{XY}$. Then, it follows from \cite[Lemma 2.12]{csiszar2011information}
that
\begin{align}
  \prob{(\Xt^n, \Yt^n) \in \Tc^{n}_{Q_{XY};{\eta_n}}}
  &\geq 1 - \frac{|\X||\Y|}{4n\eta_n^2} = 1 - \frac{|\X||\Y|}{4n^{\frac13}}.
\end{align}

Moreover, from \eqref{eq:def:phi:f} we have
  \begin{align}
    \prob{\Xt^n \in \Sc_X}
    &= \prob{f_n(\Xt^n)= \enct_X(Q_X)}\notag\\
    &\geq \frac{1}{\|f_n\|} = \exp\left(-n \cdot \frac{\log \|f_n\|}{n}\right),
  \end{align}
  and, similarly, from \eqref{eq:def:phi:g} we have
   $ \prob{\Yt^n \in \Sc_Y}  \geq \exp\left(-n \cdot \frac{\log \|g_n\|}{n}\right).$

  Then, since $f_n$ and $g_n$ are with zero-rates, both $\frac1n\log \|f_n\|$ and $\frac1n\log \|g_n\|$ vanish as $n$ tends to infinity. Therefore, it follows from the blowing up lemma  (cf. \cite{ahlswede1976bounds}, {\cite[Lemma 5.4]{csiszar2011information}})
that there exist $d_n = o(n)$ and $\nu_n = o(1)$, such that
$\prob{\Xt^n \in \nbhdh^{d_n}(\Sc_X)} \geq 1 - \nu_n$
and $\prob{\Yt^n \in \nbhdh^{d_n}(\Sc_Y)} \geq 1 - \nu_n$.

  Let $l_n \defeq 2 d_n = o(n)$, and it follows from the fact $\nbhdh^{d_n}(\Sc_X)\times \nbhdh^{d_n}(\Sc_Y) \subset \nbhdh^{2d_n}(\Sc_X \times \Sc_Y) =  \nbhdh^{l_n}(\Sc_{XY})$ that
  \begin{align}
      \prob{(\Xt^n, \Yt^n) \in \nbhdh^{l_n}(\Sc_{XY})}%
    & \geq \prob{(\Xt^n, \Yt^n) \in \nbhdh^{d_n}(\Sc_X)\times \nbhdh^{d_n}(\Sc_Y)} \notag\\
    &\geq \prob{\Xt^n \in \nbhdh^{d_n}(\Sc_X) } + \prob{\Yt^n \in \nbhdh^{d_n}(\Sc_Y)} - 1 \notag\\
    &\geq 1 - 2\nu_n\notag\\
    &= 1 - o(1),
  \end{align}
  where the second inequality follows from the elementary fact that, for two events $E_1$ and $E_2$,
  \begin{align}
    \prob{E_1\cap E_2} &= \prob{E_1} + \prob{E_2} - \prob{E_1\cup E_2}\notag\\
    &\geq \prob{E_1} + \prob{E_2} - 1.
    \label{eq:simple:prob}  
  \end{align}

As a result, for sufficiently large $n$, we can obtain 
\begin{align}
   &
    \prob{(\Xt^n, \Yt^n) \in    \Tc^{n}_{Q_{XY};{\eta_n}} \cap \nbhdh^{l_n}(\Sc_{XY})} \notag\\
  & \geq \prob{(\Xt^n, \Yt^n) \in \Tc^{n}_{Q_{XY};{\eta_n}}}%
    +   \prob{(\Xt^n, \Yt^n) \in \nbhdh^{l_n}(\Sc_{XY})}
    - 1 \notag\\
  &\geq
    \frac12.
     \label{eq:half} 
\end{align}
Therefore, with $\Qc_n$ as defined in \eqref{eq:def:qcn}, we obtain
\begin{align}
    \max_{\Qt_{XY}\in \Qc_n} \beta_n(\Qt_{XY})%
  &%
    \geq \sum_{\Qt_{XY}\in \Qc_n} \beta_n(\Qt_{XY}) \cdot \prob{\Ph_{\Xt^n\Yt^n }= \Qt_{XY}}\notag\\
  &%
    = \sum_{\Qt_{XY}\in \Qc_n} \beta_n(\Qt_{XY}) \cdot \prob{(\Xt^n, \Yt^n) \in \Tc^{n}_{\Qt_{XY}}}\notag\\
  &%
    =\sum_{\Qt_{XY}\in \Qc_n}\prob{(\Xt^n, \Yt^n) \in \Tc^{n}_{\Qt_{XY}} \cap   \nbhdh^{l_n}(\Sc_{XY})}\notag\\
  &%
    =\prob{(\Xt^n, \Yt^n) \in  \Tc^{n}_{Q_{XY};{\eta_n}} \cap  \nbhdh^{l_n}(\Sc_{XY}) } \notag\\
  &%
    \geq \frac12,
\end{align}
 where to obtain the second equality we have used the fact that
\begin{align*}
  \beta_n(\Qt_{XY}) &= \prob{(\Xt^n, \Yt^n) \in   \nbhdh^{l_n}(\Sc_{XY}) \middle|(\Xt^n, \Yt^n) \in \Tc^{n}_{\Qt_{XY}} }, 
\end{align*}
and where the last equality follows from that
$  \Tc^{n}_{Q_{XY};{\eta_n}} = \bigcup_{\Qt_{XY}\in \Qc_n}\Tc^{n}_{\Qt_{XY}}.$
\hfill\IEEEQED

\section{Proof of \thmref{thm:optm-type-based}}
\label{app:thm:optm-type-based}
  To begin, note that from \eqref{eq:prob:type}, there exists some $\eps_n = o(1)$, such that for each $i \in \{0, 1\}$, we have
  \begin{align}
    &\prob{X^n \in \Tc^n_{Q_X}, Y^n \in \Tc^n_{Q_Y}|\Hs = i}%
      \leq \exp(-n (\divp_i(Q_{X}, Q_{Y}) - \eps_n)).
    \label{eq:ieq:prob:type}
  \end{align}

  In addition, we construct the type-based encoders $\ft_n$, $\gt_n$ such that 
    \begin{align}
    \ft_n(x^n) \defeq \enct_X(\Ph_{x^n}), \quad     \gt_n(y^n) \defeq \enct_Y(\Ph_{y^n})
    \label{eq:ft:gt}
    \end{align}
    for all $x^n \in \X^n$ and $y^n \in \Y^n$, where $\enct_X(\cdot)$ and $\enct_Y(\cdot)$ are as defined in \lemref{lem:type-based-coding}.  We also define
    \begin{align*}
      \Gamma^n_i \defeq \{(Q_X, Q_Y) \in \Pch^\X_n \times \Pch^\Y_n \colon
      \dec_n(\enct_X(Q_{X}), \enct_Y(Q_Y)) \neq i\}
    \end{align*} 
    for $i = 0, 1$ and $n \geq 1$.
    
 Then, it can be verified that for given sequences $x^n \in \X^n$ and $y^n \in \Y^n$, we have $\dec_n(\ft(x^n), \gt(y^n)) \neq i$ if and only if $(\Ph_{x^n}, \Ph_{y^n}) \in \Gamma^n_i$. Therefore, the error of the type-based coding scheme $\Cct_n$ can be written as
  \begin{align}
    \pi_i(\Cct_n)
    &= \prob{\dec_n(\ft_n(X^n), \gt_n(Y^n)) \neq i\middle|\Hs = i}\notag\\
    &= \prob{(\Ph_{X^n}, \Ph_{Y^n}) \in \Gamma_i^n\middle|\Hs = i}\notag\\
    &= \sum_{(Q_X, Q_Y) \in \Gamma_i^n}\prob{(\Ph_{X^n}, \Ph_{Y^n}) = (Q_X, Q_Y) \middle| \Hs = i }.%
      \label{eq:pi:Cch:0}
  \end{align}

  If $\Gamma^n_i$ is empty, then $ \pi_i(\Cct_n) = 0 \leq \pi_i(\Cc_n)$ is trivially true. Otherwise, for each $n \geq 1$, let us define\footnote{For convenience, the dependencies of $Q^{(i)}_X, Q^{(i)}_Y$ on $n$ are omitted from the notations.}
  \begin{align}
    (Q^{(i)}_X, Q^{(i)}_Y) \defeq
\argmax_{(Q_X, Q_Y) \in \Gamma_i^n} \prob{(\Ph_{X^n}, \Ph_{Y^n}) = (Q_X, Q_Y) \middle| \Hs = i },
    \label{eq:Q:opt}
  \end{align}
and from \eqref{eq:pi:Cch:0} we have
  \begin{align}
    \pi_i(\Cct_n)
    &\leq |\Gamma^n_i| \cdot \prob{(\Ph_{X^n}, \Ph_{Y^n}) = (Q_X^{(i)}, Q_Y^{(i)}) \middle| \Hs = i }\notag\\
    &\leq (n + 1)^{|\X|+|\Y|} \cdot  \prob{(\Ph_{X^n}, \Ph_{Y^n}) = (Q_X^{(i)}, Q_Y^{(i)}) \middle| \Hs = i },%
      \label{eq:pi:CCh}
  \end{align}
  where the second inequality follows from the fact that
  \begin{align}
    |\Gamma^n_i| \leq \left|\Pch^{\X}_n \times \Pch^{\Y}_n\right|
    &\leq (n + 1)^{|\X|} \cdot (n + 1)^{|\Y|}\notag\\
    &= (n + 1)^{|\X|+|\Y|}.
      \label{eq:type:up}
  \end{align}

Then, from \lemref{lem:type-based-coding}, for $i = 0, 1$, there exists $\xi^{(i)}_n = o(1)$, such that
  \begin{align}
      \pi_i(\Cc_n) %
      &=  \prob{\dec_n(f_n(X^n), g_n(Y^n)) \neq i \middle| \Hs = i }\notag\\
    &%
      \geq \prob{f_n(X^n) = \enct_X(Q_X^{(i)}), g_n(Y^n) = \enct_Y(Q_Y^{(i)}) \middle| \Hs = i }\notag\\
    &%
      \geq \prob{(\Ph_{X^n}, \Ph_{Y^n}) = (Q_X^{(i)}, Q_Y^{(i)}) \middle| \Hs = i } \cdot \exp(-n \cdot %
      \xi^{(i)}_n),
    \label{eq:pi:Cc}
  \end{align}
  where the first inequality follows from the fact that $ \dec_n(\enct_X(Q_X^{(i)}), \enct_Y(Q_Y^{(i)})) \neq i$ since  $(Q^{(i)}_X, Q^{(i)}_Y) \in \Gamma^n_i$ [cf. \eqref{eq:Q:opt}].

  Therefore, from \eqref{eq:pi:CCh} and \eqref{eq:pi:Cc} we have     $\pi_i(\Cct_n) \leq \pi_i(\Cc_n) \cdot \exp(n \zeta_n)$ for $i = 0, 1$, where
  \begin{align*}
    \zeta_n \defeq \frac{(|\X| + |\Y|)\log(n + 1)}{n} + \max\{\xi^{(0)}_n, \xi^{(1)}_n\} = o(1).
  \end{align*}
  \hfill\IEEEQED

\section{Proof of \thmref{thm:separation}}
\label{app:thm:separation}
  We first demonstrate that $(E_0, E_1) \in \Ec[\dec]$ if $(\Dc_0(E_0), \Dc_1(E_1))$ is separable by $\dec$. To this end, we consider the error exponents associated with the coding schemes $\{\Cc_n\}_{n \geq 1}$ with $\Cc_n \defeq (f_n, g_n, \dec)$, where $f_n(x^n) \defeq \enct_X(\Ph_{x^n}), g_n(y^n) \defeq \enct_Y(\Ph_{y^n})$, and $\enct_X$ and $\enct_Y$ are the corresponding functions as defined in \defref{def:separation} to separate $(\Dc_0(E_0), \Dc_1(E_1))$.

  To begin, first note that from \eqref{eq:prob:type}, there exists some $\eps_n = o(1)$, such that for each $i \in \{0, 1\}$, we have
  \begin{align}
    &\prob{(\Ph_{X^n}, \Ph_{Y^n}) = (Q_X, Q_Y)|\Hs = i}%
      \leq \exp(-n (\divp_i(Q_{X}, Q_{Y}) - \eps_n)).
    \label{eq:ieq:prob:type:thm}
  \end{align}

  In addition, for each $i = 0, 1$ and $n \geq 1$, let us define
  \begin{align*}
    \Gamma^n_i \defeq \{(Q_X, Q_Y) \in \Pch^\X_n \times \Pch^\Y_n \colon
    \dec(\enct_X(Q_{X}), \enct_Y(Q_Y)) \neq i\},
  \end{align*}
  and it can be verified from \defref{def:separation} that
  \begin{align}
    \divp_i(Q_X, Q_Y) \geq E_i\quad\text{for all}\quad(Q_X, Q_Y) \in \Gamma^n_i.
    \label{eq:divp:ieq}    
  \end{align}
  Therefore, the type-I error $\pi_0$ and type-II error $\pi_1$ can be represented as
  \begin{align}
    \pi_i(\Cc_n) &= \prob{\dec(\enct_X(\Ph_{X^n}), \enct_Y(\Ph_{Y^n})) \neq i\middle|\Hs = i}\notag\\
                 &= \sum_{(Q_X, Q_Y) \in \Gamma_i^n}\prob{X^n \in \Tc_{Q_X}^n, Y^n \in \Tc^n_{Q_{Y}} \middle| \Hs = i }\notag\\
                 &\leq  \sum_{(Q_X, Q_Y) \in \Gamma_i^n} \exp(-n \cdot (\divp_i(Q_{X}, Q_{Y}) - \eps_n))\label{eq:err:up:1}\\
                 &\leq  \sum_{(Q_X, Q_Y) \in \Gamma_i^n} \exp(-n (E_i - \eps_n))\label{eq:err:up:2}\\
                 &\leq |\Gamma^n_i| \cdot \exp(-n (E_i - \eps_n))\label{eq:err:up:3}\\
                 &\leq (n+1)^{|\X|+|\Y|}\exp(-n (E_i - \eps_n)).\label{eq:err:up:4}\\
                 &\leq\exp(-n (E_i - \eps_n')),\label{eq:err:up:4}
  \end{align}
  where \eqref{eq:err:up:1} follows from \eqref{eq:ieq:prob:type:thm}, \eqref{eq:err:up:2} follows from \eqref{eq:divp:ieq}, \eqref{eq:err:up:4} follows from \eqref{eq:type:up}, and where $\eps_n' \defeq \eps_n +
   \frac{\log(n + 1)}n\cdot{(|\X| + |\Y|)}$.

  Note that since
  $\eps_n' = o(1)$, we obtain $(E_0, E_1) \in \Ec[\dec]$.

  In addition, we illustrate that for each $(E_0, E_1) \in \Ec[\dec]$, $(\Dc_0(E_0), \Dc_1(E_1))$ is separable by $\dec$. To this end, first note that from \thmref{thm:optm-type-based}, it suffices to consider coding schemes $\tilde{\Cc}_n = (\ft_n, \gt_n, \dec)$  with type-based encoders $\ft_n \colon x^n \mapsto \hat{\enct}_X^{(n)}(\Ph_{x^n})$ and $\gt_n \colon y^n \mapsto \hat{\enct}_Y^{(n)}(\Ph_{y^n})$, where $\hat{\enct}_X^{(n)} \colon \Pch_n^\X \to \msgset{M_X}$ and $\hat{\enct}_Y^{(n)} \colon \Pch_n^\Y \to \msgset{M_Y}$ are the corresponding type-encoding functions.

  Then, it can  be verified that   for $n$ sufficiently large, the $\hat{\enct}_X^{(n)}$ and $\hat{\enct}_Y^{(n)}$ satisfy that, for both $i = 0, 1$, and each  $(Q_X, Q_Y) \in \Dc_i(E_i) \cap (\Pch^\X_n\times \Pch^\Y_n)$,
  \begin{align}
    \dec(\hat{\enct}^{(n)}_X(Q_X), \hat{\enct}^{(n)}_Y(Q_Y)) = i.%
    \label{eq:sep:enct:hat}    
  \end{align}
  
  By symmetry, it suffices to establish  \eqref{eq:sep:enct:hat} for the case $i = 0$, which can be shown by contradiction. Indeed, suppose that there exists some $(Q_X, Q_Y) \in \Dc_0(E_0) \cap (\Pch^\X_n\times \Pch^\Y_n)$ such that $\dec(\hat{\enct}^{(n)}_X(Q_X), \hat{\enct}^{(n)}_Y(Q_Y)) = 1$, then from \eqref{eq:prob:type}, there exists some $\nu_n = o(1)$, such that the type-I error $ \pi_0(\Cct_n)$ satisfies
  \begin{align*}
    \pi_0(\Cct_n)
    &\geq \prob{X^n \in \Tc^n_{Q_X}, Y^n \in \Tc^n_{Q_Y}| \Hs = 0}\\
    &\geq \exp(-n (\divp_0(Q_X, Q_Y) + \nu_n))    .
  \end{align*}
  Therefore, the type-I error exponent is at most $\divp_0(Q_X, Q_Y)$, which is  strictly less than $E_0$, since $(Q_X, Q_Y) \in \Dc(E_0)$. This contradicts the assumption $(E_0, E_1) \in \Ec[\dec]$.

  Furthermore, let us define functions $\tilde{\enct}_X^{(n)} \colon \Pc^\X \to \msgset{M_X}$ and $\tilde{\enct}_Y^{(n)} \colon \Pch_n^\Y \to \msgset{M_Y}$ such that 
  \begin{align*}
    \tilde{\enct}_X^{(n)}(Q_X) \defeq \hat{\enct}_X^{(n)}(\Qh_X^{(n)})\quad\text{and}\quad
    \tilde{\enct}_Y^{(n)}(Q_Y) \defeq \hat{\enct}_X^{(n)}(\Qh_Y^{(n)})
  \end{align*}
  for all $Q_X \in \cP^\X$ and $Q_Y \in \cP^\Y$, where
  \begin{subequations}
  \begin{align}
    \Qh_X^{(n)} &\defeq \argmin_{Q_X' \in \Pch_n^{\X}}\, \dmax(Q_X', Q_X),\\
    \Qh_Y^{(n)} &\defeq \argmin_{Q_Y' \in \Pch_n^{\Y}}\, \dmax(Q_Y', Q_Y),
  \end{align}
  \label{eq:Qh:def}
\end{subequations}
where $\dmax$ is as defined in \eqref{eq:def:dmax}.

  Note that for each $(Q_X, Q_Y) \in \Dc_0(E_0)$, we have $\divp_0(Q_X, Q_Y) < E_0$. In addition,  from \eqref{eq:Qh:def}, we have
 $ \dmax(\Qh_X^{(n)}, Q_X) \leq \frac1n$ and  $  \dmax(\Qh_Y^{(n)}, Q_Y) \leq \frac1n$. Therefore,  it follows from the uniform continuity of $\divp_0$ that for $n$ sufficiently large, we have $ \divp_0(\Qh_X^{(n)}, \Qh_Y^{(n)}) < E_0$, which implies that
 $(\hat{Q}_X^{(n)}, \hat{Q}_X^{(n)}) \in  \Dc_0(E_0) \cap (\Pch^\X_n\times \Pch^\Y_n)$. Hence, from  \eqref{eq:sep:enct:hat} we obtain
  \begin{align}
    \dec(\tilde{\enct}^{(n)}_X(Q_X), \tilde{\enct}^{(n)}_Y(Q_Y)) = 0.%
    \label{eq:sep:enct:tilde:0}    
  \end{align}
  Similarly, we have
  \begin{align}
    \dec(\tilde{\enct}^{(n)}_X(Q_X), \tilde{\enct}^{(n)}_Y(Q_Y)) = 1%
    \label{eq:sep:enct:tilde:1}    
  \end{align}
  for each $(Q_X, Q_Y) \in \Dc_1(E_1)$. From \eqref{eq:sep:enct:tilde:0} and \eqref{eq:sep:enct:tilde:1}, $\Dc_0(E_0)$ and $\Dc_1(E_1)$ is separable by $\dec$, which completes the proof. \qquad\phantom{Xu}  \hfill\IEEEQED

  \section{Proof of \propref{prop:recursive:decr}}
  \label{app:prop:recursive:decr}
  We first introduce a useful result.
 \begin{proposition}
  \label{prop:rec}
  Suppose $\Ac_0, \Ac_1 \subset \pdist$, and $\dec$ is a reducible decoder. For each $i \in \{0, 1\}$, we have
  \begin{itemize}
  \item when $\redux^{(i)}(\dec)$ exists, %
      $\dec \mid (\Ac_0, \Ac_1)$  if and only if $\redux^{(i)}(\dec) \mid (\Ac_0 \subx \Ac_{\ibar}, \Ac_{1} \subx \Ac_{\ibar})$;    
  \item when $\reduy^{(i)}(\dec)$ exists, $\dec \mid (\Ac_0, \Ac_1)$  if and only if $\reduy^{(i)}(\dec) \mid (\Ac_0 \suby \Ac_{\ibar}, \Ac_{1} \suby \Ac_{\ibar})$.
  \end{itemize}
  Here, ``$\subx$'' and  ``$\suby$'' are as defined in \eqref{eq:def:subx} and  \eqref{eq:def:suby}, respectively. 
\end{proposition}
\begin{IEEEproof}[Proof of \propref{prop:rec}]
It suffices to consider the first statement for $i = 0$, and other statements can be similarly established. To begin, let $\phi' \defeq \redux^{(0)}(\dec)$, and we use $\Ab \lfa \dec$  and $\lfa \Ab_X^{(0)} \lfa \phi' $ denote the corresponding decision matrices (cf. \defref{def:dec:redu}). We also define
  \begin{align}
    \Ac_0' \defeq \Ac_0 \subx \Ac_1, \quad\text{and}\quad \Ac_1' \defeq \Ac_1 \subx \Ac_1 = \Ac_1.
    \label{eq:Ac:prime}
  \end{align}
  Without loss of generality, suppose the $0$-dominated columns of $\Ab$ are its last $d$ columns, i.e., we have
  \begin{align}
    \dec(m_X, m_Y) = 0,
    \label{eq:dec:0}
  \end{align}
  for each $m_X = M_X - d, \dots, M_X - 1$ and $m_Y \in \msg{M_Y}$.

  Moreover, it can be verified that $\dec'$ is the restriction of $\dec$ to $\msg{M_X - d} \times \msg{M_Y}$, and we have
  \begin{align}
    \dec'(m_X, m_Y) = \dec(m_X, m_Y)
    \label{eq:redu:dec}
  \end{align}
  for each $(m_X, m_Y) \in \msg{M_X - d} \times \msg{M_Y}$.
  
  To prove the ``only if'' part of the claim, suppose $(\Ac_0, \Ac_1)$ is separable by $\phi$. Then, from \defref{def:separation}, there exist mappings $\enct_X\colon \Pc^\X \to \msg{M_X}$ and $\enct_Y\colon \Pc^\Y \to \msg{M_Y}$, such that for both $i \in \{0, 1\}$, we have
  \begin{align}
    \dec(\enct_X(Q_X), \enct_Y(Q_Y)) = i, \quad\text{for all } (Q_X, Q_Y) \in \Ac_i.
    \label{eq:sep:def}
  \end{align}
  For each $Q_X \in \projx{\Ac_1}$, it can be verified that  $\enctx(Q_X) \in \msg{M_X - d}$. Otherwise, there exists $Q_Y' \in \cP^\Y$ with $(Q_X, Q_Y') \in \Ac_1$, and it follows from \eqref{eq:dec:0} that $\dec(\enct_X(Q_X), \enct_Y(Q_Y')) = 0$, which contradicts the claim \eqref{eq:sep:def}.

  Then, we define  $\enct' \colon \Pc^\X \to \msg{M_X -d}$ such that
  \begin{align}
    \enct'(Q_X) =
    \begin{cases}
      \enct'(Q_X)& \text{if}~Q_X \in \projx{\Ac_1},\\
      0&\text{otherwise,}
    \end{cases}
  \end{align}
  and
  it follows from \eqref{eq:redu:dec} that, for each $Q_X \in \projx{\Ac_1}$ and $Q_Y \in \cP^\Y$, we have
  \begin{align*}
    \dec(\enct_X(Q_X), \enct_Y(Q_Y)) \equiv \dec'(\enct'_X(Q_X), \enct_Y(Q_Y)).
  \end{align*}
  Moreover, from \eqref{eq:sep:def} we have, for both $i \in \{0, 1\}$,
  \begin{align}
    \dec'(\enct'_X(Q_X), \enct_Y(Q_Y)) = i, \quad\text{for all } (Q_X, Q_Y) \in \Ac_i',
    \label{eq:sep:def:prime}
  \end{align}
  which implies that $(\Ac_0', \Ac_1')$ is separable by $\phi'$.

  For the ``if'' part of the claim, suppose $\dec' \mid (\Ac_0', \Ac_1')$, then  there exist functions $\hat{\enct}_X \colon \cP^\X \to \msgset{M_X - d}$ and $\hat{\enct}_Y \colon \cP^\Y \to \msgset{M_X - d}$, such that for both $i \in \{0, 1\}$, we have
  \begin{align}
    \dec'(\hat{\enct}_X(Q_X), \hat{\enct}_Y(Q_Y)) = i, \quad\text{for all } (Q_X, Q_Y) \in \Ac_i'.
    \label{eq:sep:def:prime}
  \end{align}
  Then, let us define $\hat{\enct}' \colon \Pc^\X \to \msg{M_X}$ such that
  \begin{align}
    \hat{\enct}'(Q_X) =
    \begin{cases}
      \hat{\enct}(Q_X)& \text{if}~Q_X \in \projx{\Ac_1},\\
      M_X - d&\text{otherwise.}
    \end{cases}
  \end{align}
  From \eqref{eq:dec:0}, for both $i \in \{0, 1\}$, we have
    $\dec(\hat{\enct}_X'(Q_X), \hat{\enct}_Y(Q_Y)) = i$ {for all } $(Q_X, Q_Y) \in \Ac_i$,  which implies that $\dec \mid (\Ac_0, \Ac_1)$.
\end{IEEEproof}

     Proceeding to our proof of \propref{prop:recursive:decr}, first note that we have
       $\bar{\decr}_{M-1, M-1} %
       = \reduy^{(0)}\left(\redux^{(0)}(\decr_{M, M})\right)$.
     Therefore, from \propref{prop:rec}, we have
     \begin{align*}
       \decr_{M, M} \mid (\Ac, \Ac')
       &\iff \redux^{(0)}(\decr_{M, M}) \mid (\Ac \subx \Ac', \Ac')\\
       &\iff \bar{\decr}_{M-1, M-1} \mid ((\Ac \subx \Ac') \suby \Ac', \Ac')\\
       &\iff \bar{\decr}_{M-1, M-1} \mid (\Ac \sub \Ac', \Ac')\\
       &\iff \decr_{M-1, M-1} \mid (\Ac',\Ac \sub \Ac'),
     \end{align*}
     where the third ``\!\!$\iff$\!\!'' follows from $(\Ac \subx \Ac') \suby \Ac' = \Ac \sub \Ac'$. To obtain the last  ``\!\!$\iff$\!\!'', we have used the third property of \factref{fact:equiv:empty:cmpmt}.
     
     In addition, 
     by repeatedly applying the first ``$\iff$'' of \eqref{eq:rec:equiv}
     $(M-1)$ times, we know that  first two statements of \eqref{eq:rec:equiv} are  equivalent to
     \begin{align*}
       \decr_{1, 1} \mid \left(\Ac \sub_{M-1} \Ac', \Ac \sub_{M} \Ac'\right)
       \iff \Ac \sub_{M} \Ac' = \varnothing,
     \end{align*}
     where we have used the second property of  \factref{fact:equiv:empty:cmpmt}.

     Similarly, we can establish the second statement of the claim, by noting that
       $\decr_{M_Y, M_Y} = \redux^{(1)}(\decr_{M_X, M_Y})$ for all $M_X > M_Y \geq 1$.
     \hfill\IEEEQED

  \section{Proof of \thmref{thm:exp:th}}
  \label{app:exp:th}
 We first introduce the following fact on the separability, which can be readily verified from \defref{def:separation}.
 \begin{fact}
   \label{fact:equiv:empty:cmpmt}
   Given $\Ac, \Ac'\subset \pdist$, we have
   \begin{itemize}
   \item $\dec \mid (\Ac, \Ac') \iff \dec' \mid (\Ac, \Ac') $, for all $\dec \simeq \dec'$;
   \item $\decr_{1, 1} \mid (\Ac, \Ac') \iff \Ac' = \varnothing$;
   \item $\dec \mid (\Ac, \Ac') \iff \bar{\dec} \mid(\Ac', \Ac)$.
   \end{itemize}
 \end{fact}

 To establish \thmref{thm:exp:th}, we first consider the case $M_X = M_Y$. From \thmref{thm:separation} we have
   \begin{align*}
     (E_0, E_1) \in \Ec[\decr_{M_Y, M_Y}]
     &\iff \text{$\decr_{M_Y, M_Y} \mid (\Dc_0(E_0), \Dc_1(E_1))$}\\
     &\iff \Dc_0(E_0) \sub_M \Dc_1(E_1) = \varnothing,
   \end{align*}
   where the last ``\!\!$\iff$\!\!'' follows from \propref{prop:recursive:decr}. Then, it follows from the third property of \factref{fact:equiv:empty:cmpmt} that
   \begin{align*}
      (E_0, E_1) \in \Ec[\decr_{M_Y, M_Y}]  \iff \Dc_1(E_1) \sub_M \Dc_0(E_0) = \varnothing.   
   \end{align*}

   For the case $M_X > M_Y$, it can be verified that
   \begin{align*}
     \Ec[\decr_{M_X, M_Y}] &= \Ec[\decr_{M_Y + 1, M_Y}]\\
                           &= \Ec[\bar{\decr}_{M_Y + 1, M_Y}]=  \Ec[\bar{\decr}_{M_X, M_Y}],
   \end{align*}
   where the second equality follows from the first property of \factref{fact:equiv:empty:cmpmt} and that $\decr_{M_Y + 1, M_Y} \simeq \bar{\decr}_{M_Y + 1, M_Y}$. To obtain the first equality, note that the decision matrix associated with $\decr_{M_X, M_Y}$ and that associated with  $\decr_{M_Y, M_Y}$ differ only in duplicated columns. The last equality follows from symmetry considerations.

   Then, from \thmref{thm:separation} and \propref{prop:recursive:decr} we can obtain
   \begin{align*}
       (E_0, E_1) \in \Ec[\decr_{M_X, M_Y}]%
     &\iff \decr_{M_X, M_Y} \mid (\Dc_0(E_0), \Dc_1(E_1))\\
     &\iff \decr_{M_Y, M_Y} \mid (\Dc_0(E_0), \Dc_1(E_1) \subx \Dc_0(E_0)) %
     \\
     &\iff \Dc_0(E_0) \sub_{M_Y} (\Dc_1(E_1) \subx \Dc_0(E_0)) = \varnothing,
   \end{align*}
   which completes the proof. \hfill\IEEEQED

\section{Proof of \lemref{lem:decomp}}
\label{app:lem:decomp}
  By symmetry, it suffices to consider the case where \eqref{eq:decomp} holds for $i = 1$, i.e.,
  \begin{align}
    \dec = \dec_0 \oplus \dec_1.
    \label{eq:decomp:oplus}
  \end{align}
  Since $(\Ac_0, \Ac_1)$ is separable by $\dec$, from \defref{def:separation}, there exists $\enct_X\colon \Pc^\X \to \msg{M_X}$ and $\enct_Y\colon \Pc^\Y \to \msg{M_Y}$, such that, we have
  \begin{align}
    \dec(\enct_X(Q_X), \enct_Y(Q_Y)) = 0, \text{ for all }(Q_X, Q_Y) \in \Ac_0,
    \label{eq:sep:0}
  \end{align}  
  and
  \begin{align}
    \dec(\enct_X(Q_X), \enct_Y(Q_Y)) = 1, \text{ for all }(Q_X, Q_Y) \in \Ac_1.
    \label{eq:sep:1}
  \end{align}  

  From  \eqref{eq:icx:icy:empty} and  \eqref{eq:decomp:oplus}, we have, for all $(m_X, m_Y) \in \msgset{M_X} \times \msgset{M_Y}$,
  \begin{gather}
    \dec(m_X, m_Y) = \max\{\dec_0(m_X, m_Y), \dec_1(m_X, m_Y)\},\\
    \dec_0(m_X, m_Y) \cdot \dec_1(m_X, m_Y) = 0.\label{eq:prod:0}
  \end{gather}

  Therefore, we obtain, for each $(Q_X, Q_Y) \in \Ac_0$,
  \begin{align}
    \dec_0(\enct_X(Q_X), \enct_Y(Q_Y)) =  \dec_1(\enct_X(Q_X), \enct_Y(Q_Y)) = 0,
    \label{eq:sep:i:0}
  \end{align}
  and, for each $(Q_X, Q_Y) \in \Ac_1$, 
  \begin{align*}
    \dec_0(\enct_X(Q_X), \enct_Y(Q_Y)) + \dec_1(\enct_X(Q_X), \enct_Y(Q_Y)) = 1.
  \end{align*}

  Furthermore, we can demonstrate that, for either $i = 0$ or $i = 1$,
  \begin{align}
    \dec_i(\enctx(Q_X), \encty(Q_Y)) \equiv 1, \quad\text{for all }(Q_X, Q_Y) \in \Ac_1.
    \label{eq:sep:i:1:new}
  \end{align}
  To see this, we define, for $i \in \{0, 1\}$,
  \begin{align}
    \Ac_1^{(i)} &\defeq \{(Q_X, Q_Y) \in \Ac_1\colon%
      \dec_i(\enctx(Q_X), \encty(Q_Y)) = 1\},
    \label{eq:def:ac:i}
  \end{align}
  from which we obtain the partition $\Ac_1 = \Ac_1^{(0)} \union \Ac_1^{(1)}$ with $\Ac_1^{(0)} \cap \Ac_1^{(1)} = \varnothing$. Then, it suffices to show that $\Ac_1^{(i)} = \varnothing$ for $i = 0$ or $i = 1$, which we will establish by contradiction.

  To begin, suppose we have $(Q_X, Q_Y) \in \Ac_1^{(0)}$ and $(\Qt_X, \Qt_Y) \in \Ac_1^{(1)}$. Then, let us define sequences $\bigl\{(Q_X^{(n)}, Q_Y^{(n)})\bigr\}_{n \geq 0}$ and  $\bigl\{(\Qt_X^{(n)}, \Qt_Y^{(n)})\bigr\}_{n \geq 0}$ such that
   $ (Q_X^{(0)}, Q_Y^{(0)}) = (Q_X, Q_Y)$ and $(\Qt_X^{(0)}, \Qt_Y^{(0)}) = (\Qt_X, \Qt_Y)$.
   Moreover, for each $n \geq 0$, we define
  \begin{align*}
    (Q_X^{(n+1)}, Q_Y^{(n+1)}) \defeq
    \begin{cases}
      (\Qh_X^{(n)}, \Qh_Y^{(n)})&  \text{if $(\Qh_X^{(n)}, \Qh_Y^{(n)}) \in \Ac_1^{(0)}$},\\
      (Q_X^{(n)}, Q_Y^{(n)})&  \text{otherwise,}
    \end{cases}
  \end{align*}
  and
  \begin{align*}
    (\Qt_X^{(n+1)}, \Qt_Y^{(n+1)}) \defeq
    \begin{cases}
      (\Qt_X^{(n)}, \Qt_Y^{(n)})&  \text{if $(\Qh_X^{(n)}, \Qh_Y^{(n)}) \in \Ac_1^{(0)}$}\\
      (\Qh_X^{(n)}, \Qh_Y^{(n)})&  \text{otherwise,}
    \end{cases}
  \end{align*}
  where we have defined
  \begin{align*}
    \Qh_X^{(n)} \defeq \frac12 (Q_X^{(n)} + \Qt_X^{(n)}), \quad
    \Qh_Y^{(n)} \defeq \frac12 (Q_Y^{(n)} + \Qt_Y^{(n)}),
  \end{align*}
  and we have $(\Qh_X^{(n)}, \Qh_Y^{(n)}) \in \Ac_1$  due to the convexity of $\Ac_1$.
  
  Then, for each $n \geq 0$, it can be verified that
  \begin{align}
    (Q_X^{(n)}, Q_Y^{(n)}) \in \Ac_1^{(0)}, \quad (\Qt_X^{(n)}, \Qt_Y^{(n)}) \in \Ac_1^{(1)},
    \label{eq:q:qt}
  \end{align}
  and
  \begin{align*}
      \dmaxp\left(Q_X^{(n)}Q_Y^{(n)}, \Qt_X^{(n)}\Qt_Y^{(n)}\right)%
    &
      = \frac{1}{2} \cdot  \dmaxp\left(Q_X^{(n-1)}Q_Y^{(n-1)}, \Qt_X^{(n-1)}\Qt_Y^{(n-1)}\right)\\
    &%
      = \frac{1}{2^n}  \cdot \dmaxp\left(Q_X^{(0)}Q_Y^{(0)}, \Qt_X^{(0)}\Qt_Y^{(0)}\right)\\
    &%
      = \frac{1}{2^n} \cdot  \dmaxp\left(Q_XQ_Y, \Qt_X\Qt_Y\right).
  \end{align*}

  As a result, we obtain
  \begin{align*}
      \dmaxp\bigl(Q_X^{(n)}\Qt_Y^{(n)}, Q_X^{(n)}Q_Y^{(n)}\bigr)%
    &%
      = \dmax(\Qt_Y^{(n)}, Q_Y^{(n)})\\
    &%
      \leq \dmaxp\left(Q_X^{(n)}Q_Y^{(n)}, \Qt_X^{(n)}\Qt_Y^{(n)}\right)\\
    &%
      \leq \frac{1}{2^n} \cdot  \dmaxp\left(Q_XQ_Y, \Qt_X\Qt_Y\right) = o(1).
  \end{align*}

  Since $\Ac_1$ is open, for sufficiently large $n$ we have $ Q_X^{(n)}\Qt_Y^{(n)} \in \Ac_1$. Thus, it follows from \eqref{eq:sep:1} that
  \begin{align}
    \dec(\enct_X(Q_X^{(n)}), \enct_Y(\Qt_Y^{(n)})) = 1.
    \label{eq:qx:qty:enc}
  \end{align}

  In addition, from \eqref{eq:def:ac:i} and \eqref{eq:q:qt}, we have
  \begin{gather}
    \enctx(Q^{(n)}_X) \in \Icx^{(1)}(\dec_0) %
    \quad\text{and}\quad \encty(\Qt^{(n)}_Y) \in \Icy^{(1)}(\dec_1),\label{eq:enct:q:qt}    
  \end{gather}
  where $\Icx^{(1)}(\cdot)$ and $\Icy^{(1)}(\cdot)$ are as defined in  \eqref{eq:def:icx:icy}. This implies (cf. \defref{def:decomp})
   $ \dec(\enctx(Q^{(n)}_X), \encty(\Qt^{(n)}_Y)) = 0$,
  which contradicts \eqref{eq:qx:qty:enc}.  
  
  Hence, we obtain \eqref{eq:sep:i:1:new} as desired. Finally, it follows from \eqref{eq:sep:i:0} that $(\Ac_0, \Ac_1)$ is separable by $\dec_j$ for  some $j \in \{0, 1\}$.  
\hfill\IEEEQED

\section{Proof of \lemref{lem:th:suff}}
\label{app:th:suff}
\newcommand{\Decm}{\Dec^{\mathrm{m}}}
We first introduce a useful characterization of completely reducible decoders.

\begin{proposition}  
  \label{prop:monotonic}  
  \renewcommand{\theenumi}{S\arabic{enumi}}
  Let $\dec$ denote an $M_X \times M_Y$ decoder with $M_X, M_Y \geq 2$. Then, the following statements are equivalent:
  \begin{enumerate}
  \item\label{mono:S1} $\dec$ is completely reducible;
  \item\label{mono:S2} each $2\times 2$ subdecoder of $\dec$ is reducible;
  \item\label{mono:S3} there exists a monotonic decoder $\dec'$ such that $\dec \simeq \dec'$.
  \end{enumerate}
  Specifically, a decoder $\dec$ is called \emph{monotonic}, if for all $m_X \leq m_X'$ and $m_Y \leq m_Y'$, we have
$\dec(m_X, m_Y) \leq \dec(m_X', m_Y')$.
\end{proposition}
\begin{IEEEproof}[Proof of \propref{prop:monotonic}]
  We will show the equivalences by demonstrating ``\ref{mono:S1} $\implies$ \ref{mono:S2}'', ``\ref{mono:S2} $\implies$ \ref{mono:S3}'', and ``\ref{mono:S3} $\implies$ \ref{mono:S1}''.
  
  First, for the claim ``\ref{mono:S1} $\implies$ \ref{mono:S2}'', note that there are two irreducible $2\times 2$ decoders, which we can denote by
  \begin{align}
\dec_0 \lfa \Ab_0 = \putgrid{\drawgrid{
        {0,1},
        {1,0},}}\quad
    \text{and}\quad\dec_1 \lfa \Ab_1 = \putgrid{\drawgrid{
        {1,0},
        {0,1},}}.\label{eq:irreducible:2x2}
  \end{align}
    We then prove the claim by contradiction. Specifically, we assume that $\dec \lfa \Ab$ has an irreducible subdecoder $\dec_0$.
    Without loss of generality, suppose $\Ab_0$ is the submatrix of $\Ab$ composed of first two rows and first two columns of $\Ab$.  Then, it suffices to show that $\dec$ is not completely reducible, which is trivially true if $\dec$ is irreducible.

    We now consider the case where %
    $\dec$ is reducible. Then, there exists an elementary reduction operator $\redu$, such that  $\redu(\dec)$ exists. Since the first two rows and first two columns of $\Ab$ cannot be dominated, $\Ab_0$ is also a submatrix of $\Ab' \lfa \redu(\dec)$, and thus $\dec_0$ is also a subdecoder of $\redu(\dec)$. As a consequence, for all $\dec'$ that can be reduced from $\dec$, $\dec_0$ is a subdecoder of $\dec'$, which implies that  $\dec$ is not completely reducible. Similarly, $\dec$ is not completely reducible if $\dec_1$ is a subdecoder of $\dec$.

    Then, to prove ``\ref{mono:S2} $\implies$ \ref{mono:S3}'', note that for each decoder $\dec$, we can construct its equivalent decoder  $\dec' \simeq \dec$ such that the functions $\sigma^{(\dec)}_X(\cdot)$ and $\sigma^{(\dec)}_Y(\cdot)$ are both non-decreasing, where for each $\dec \in \Dec_{M_X, M_Y}$, we have defined
    \begin{subequations}
    \begin{align}
      \sigma^{(\dec)}_X(m_X)&\defeq \sum_{m_Y \in \msgset{M_Y}} \dec(m_X, m_Y),\, \forall\,m_X \in \msgset{M_X},\\
      \sigma^{(\dec)}_Y(m_Y)&\defeq  \sum_{m_X \in \msgset{M_X}} \dec(m_X, m_Y), \,\forall\,m_Y \in \msgset{M_Y}.
    \end{align}
    \label{eq:def:sigma}
  \end{subequations}
 We then establish that $\dec'$ is monotonic  if $\dec$ satisfies the statement \ref{mono:S2}. To see this, first note that for all $0\leq m_X < m_X' < M_X$, we have $\sigma_X(m_X) \leq \sigma_X(m_X')$, which implies
    \begin{align}
      \sum_{m_Y \in \msgset{M_Y}} [\dec'(m_X, m_Y) - \dec'(m_X', m_Y)] \leq 0.
      \label{eq:sum:leq:0}
    \end{align}
    Now, suppose $\dec'(m_X, m_Y) - \dec'(m_X', m_Y) > 0$ for some $m_Y \in \msgset{M_Y}$. Since the summation \eqref{eq:sum:leq:0} is non-negative, there exists $m_Y' \in \msgset{M_Y}$ with  $\dec'(m_X, m_Y') - \dec'(m_X', m_Y') < 0$. Therefore,
    \begin{gather*}
      \dec'(m_X, m_Y) = 1, \quad\dec'(m_X', m_Y) = 0,\\
      \dec'(m_X, m_Y') = 0, \quad\dec'(m_X', m_Y') = 1,
    \end{gather*}  
    which implies that $\dec'$ has an irreducible $2\times 2$ subdecoder. Thus, $\dec$ also has an irreducible $2\times 2$ subdecoder, which contradicts the statement \ref{mono:S2}.
    
    As a consequence, we obtain
     $ \dec'(m_X, m_Y) - \dec'(m_X', m_Y) \leq 0$
    for all $m_Y \in \msgset{M_Y}$ and $0\leq m_X < m_X' < M_X$, and, similarly, 
      $ \dec'(m_X, m_Y) - \dec'(m_X, m_Y') \leq 0$      
    for all $m_X \in \msgset{M_X}$ and  $0\leq m_Y < m_Y' < M_Y$. This demonstrates  the statement \ref{mono:S3}. %

      Finally, to establish  ``\ref{mono:S3} $\implies$ \ref{mono:S1}'', note that for equivalent decoders $\dec \simeq \dec'$, $\dec$ is completely reducible if and only if $\dec'$ is completely reducible. Therefore, it suffices to show that monotonic decoders are completely reducible. To this end, we first show that the monotonic decoders are reducible. Indeed, for a given monotonic decoder $\dec \in \Dec_{M_X, M_Y}$, it can be verified from the definition that
  \begin{itemize}
  \item if $\dec(M_X - 1, 0) = 0$, then $\dec(m_X, 0) = 0$ for all $m_X \in \msgset{M_X}$;
  \item if  $\dec(M_X - 1, 0) = 1$, then $\dec(M_X - 1, m_Y) \equiv 1$ for all $m_Y \in \msgset{M_Y}$.
\end{itemize}
Therefore, $\dec$ is reducible.

Moreover, if $\dec$ is non-trivial, there exists an elementary reduction operator $\redu$, such that $\redu(\dec)$ exists. Then, it can be verified that $\redu(\dec)$ is also monotonic, and we can similarly apply reduction operations on $\redu(\dec)$ until obtaining trivial decoders. This establishes the statement \ref{mono:S1}.
\end{IEEEproof}

In addition,  the following simple fact is also useful. %
\begin{fact}
  \label{fact:equiv-sub}If $\dec \simeq \dec'$, then $\Ec[\dec] = \Ec[\dec']$. If $\dec'$ is a subdecoder of $\dec$, then $\Ec[\dec'] \subset \Ec[\dec]$, i.e., $\{\dec'\} \preceq \{\dec\}$.
\end{fact}

Proceeding to the proof of the lemma,  for all given $M_X$ and $M_Y$, we define
  \begin{align}
    \Decm_{M_X, M_Y} \defeq \{\dec \in \Dec_{M_X, M_Y}\colon \dec\text{ is monotonic}\}.
    \label{eq:dec:mono}
  \end{align}
  Then, from \factref{fact:equiv-sub} and the equivalence of statements \ref{mono:S1} and \ref{mono:S3} in \propref{prop:monotonic}, we obtain
  $  \Ec[\Decc_{M_X, M_Y}] = \Ec[\Decm_{M_X, M_Y}]$.
  
  Therefore, it suffices to establish $\{\dec\} \preceq \{\decr_{M_X, M_Y}, \bar{\decr}_{M_X, M_Y}\}$ for each $\dec \in \Decm_{M_X, M_Y}$. To this end, we first establish a useful expression of monotonic decoders via using the functions $\sigma^{(\dec)}_X(\cdot)$ and $\sigma^{(\dec)}_Y(\cdot)$ as defined in \eqref{eq:def:sigma}. In particular, for each $\dec \in \Decm_{M_X, M_Y}$, from the definition of monotonicity we have, for all $(m_X, m_Y) \in \msg{M_X} \times \msgset{M_Y}$,
  \begin{align}
    \dec(m_X, m_Y) 
    &= \kron_{\{m_X + \sigma_Y^{(\dec)}(m_Y) \geq M_X\}}%
      \label{eq:dec:mono:1}\\
    &= \kron_{\{\sigma_X^{(\dec)}(m_X) + m_Y \geq M_Y\}}. \label{eq:dec:mono:2}
  \end{align}

  If $M_X > M_Y$, for each $m_X \in \msgset{M_X}$, we have $\sigma_X^{(\dec)}(m_X) \in \msg{M_X}$. Then, it follows from \eqref{eq:dec:mono:2} that, for all $m_Y \in \msgset{M_Y}$, 
  \begin{align}
    \dec(m_X, m_Y) 
   & = \kron_{\{\sigma_X^{(\dec)}(m_X) + m_Y \geq M_Y\}}\notag\\
   & = \decr_{M_X, M_Y}(\sigma_X^{(\dec)}(m_X), m_Y),
    \label{eq:dec:eq}
  \end{align}  
  which implies that $\dec$ is a subdecoder of $\decr_{M_X, M_Y}$. Therefore, 
  from \factref{fact:equiv-sub} we obtain
  \begin{align}
   \{\dec\} \preceq \{\decr_{M_X, M_Y}\}  \preceq \{\decr_{M_X, M_Y}, \bar{\decr}_{M_X, M_Y}\}. 
    \label{eq:mx:my}
  \end{align}

  For the case $M_X = M_Y$, let $M \defeq M_X$, then %
  $\sigma^{(\dec)}_Y(\cdot)$ is a non-decreasing function on $\msgset{M}$. If $\sigma^{(\dec)}_Y(\cdot)$ is not strictly increasing, then there exists $m_Y' \in \msgset{M-1}$, such that $\sigma^{(\dec)}_Y(m_Y') = \sigma^{(\dec)}_Y(m'_Y + 1)$, and from \eqref{eq:dec:mono:1} we obtain $ \dec(m_X, m'_Y) = \dec(m_X, m'_Y + 1)$, for all $m_X \in \msgset{M}$.  This implies that the $m_Y'$-th and $(m_Y'+1)$-th rows of the associated decision matrix $\Ab \lfa \dec$ are the same. Let $\Ab'$ denote the submatrix of $\Ab$ obtained by deleting its $(m_Y'+1)$-th row. Then, it can be verified that, the decoder $\dec' \lfa \Ab'$ is an $M \times (M-1)$ monotonic decoder with $\Ec[\dec] = \Ec[\dec']$.

  Therefore, we obtain
  \begin{align*}
    \{\dec\} \preceq \Decm_{M, M-1}
             \preceq \{\decr_{M, M-1}\}
             \preceq \{\decr_{M, M}\}
             \preceq \{\decr_{M, M}, \bar{\decr}_{M, M}\},
  \end{align*}
  where the second ``$\preceq$'' follows from \eqref{eq:mx:my}, and where the third ``$\preceq$'' follows from \factref{fact:equiv-sub} and that $\decr_{M, M-1}$ is a subdecoder of $\decr_{M, M}$.

  It remains to establish the claim for the case where $M_X = M_Y = M$ and $\sigma^{(\dec)}_Y(\cdot)$ is strictly increasing on $\msgset{M}$. To this end, first note that if $\sigma^{(\dec)}_Y(0) = 0$, for each $m_X \in \msgset{M}$ we have $\dec(m_X, 0) = 0$. Therefore, we have $\sigma_X^{(\dec)}(m_X) \in \msgset{M}$, and it follows from \eqref{eq:dec:eq}--\eqref{eq:mx:my} that $\{\dec\} \preceq \Decm_{M, M}$. Moreover, if $\sigma^{(\dec)}_Y(\cdot)$ is strictly increasing and $\sigma^{(\dec)}_Y(0) \neq 0$, we have 
   $ \sigma^{(\dec)}_Y(m_Y) = m_Y + 1$ for all $m_Y \in \msg{M}$.

   Hence, from \eqref{eq:dec:mono:1} we have, for all $(m_X, m_Y)\in \msgset{M_X} \times \msgset{M_Y}$,
  \begin{align*}
    \dec(m_X, m_Y) &= \kron_{\{m_X + m_Y \geq M - 1\}}\\
    &= \kron_{\{(M- 1 - m_X) + (M- 1 - m_Y) \leq M - 1\}}\\
    &= \kron_{\{(M- 1 - m_X) + (M- 1 - m_Y) < M\}}\\
    &= \bar{\decr}_{M, M}(M- 1 - m_X, M- 1 - m_Y),
  \end{align*}
  which implies that $\dec \simeq \bar{\decr}_{M, M}$. As a result, we obtain
 $ \{\dec\} \preceq \{\bar{\decr}_{M, M}\} \preceq \{\decr_{M, M}, \bar{\decr}_{M, M}\}$,
  which completes the proof.  \hfill\IEEEQED

\section{Proof of \factref{fact:reduced}}
\label{app:fact:reduced}
  To begin, we consider a decoder $\dec$ that is not completely reducible. If $\dec$ is irreducible, it suffices to let $\dec' = \dec$. Otherwise, since $\dec$ cannot be reduced to trivial decoders, each decoder reduced from $\dec$ is either an irreducible decoder, or a non-trivial reducible decoder. Therefore, we can apply a series of elementary reduction operators on $\dec$, %
 until obtaining some irreducible decoder. 

 It remains only to demonstrate the uniqueness of obtained irreducible decoders. To see this, suppose both $\dec''$ and $\dect''$ are the irreducible decoders obtained from the above procedures.

Note that since $\dec''$ is an irreducible subdecoder of $\dec$, its associated rows and columns in the decision matrix $\Ab \lfa \dec$ cannot be dominated %
during the above reduction procedures. Therefore, it is also a subdecoder of all decoders reduced from $\dec$.

 As a result, $\dec''$ is a subdecoder of $\dect''$, and, similarly, $\dect''$ is a subdecoder of $\dec''$. Hence, we have $\dec'' = \dect''$, corresponding to the unique decoder $\dec'$ reduced from $\dec$.  \hfill\IEEEQED

\section{Proof of \thmref{thm:sufficient}}
\label{app:thm:sufficient}
Our proof makes use of the notion of open sets in $\pdist$, together with discussions on the separability (cf. \defref{def:separation}) under reducible and decomposable decoders.

As a first step, we define the open sets in $\pdist$ as follows. With slight abuse of notation, we use $Q_XQ_Y$ to represent $(Q_X, Q_Y) \in \pdist$. Then, we introduce the metric $\dmaxp$ on $\pdist$, such that for all given $Q_XQ_Y, Q_X'Q_Y' \in \pdist$,
\begin{align*}
    \dmaxp(Q_XQ_Y, Q_X'Q_Y')%
    \defeq \max\left\{\dmax(Q_X, Q_X'), \dmax(Q_Y, Q_Y')\right\}.
\end{align*}
Moreover, $\Ac \subset \pdist$ is open, if for each $Q_XQ_Y \in \Ac$, there exists $\eta > 0$, such that for all $Q_X'Q_Y' \in \pdist$ satisfying $\dmaxp(Q_XQ_Y, Q_X'Q_Y') < \eta$, we have $Q_X'Q_Y' \in \Ac$.

Specifically, with assumption \eqref{eq:pos}, the functions $\divp_0(\cdot)$ and $\divp_1(\cdot)$ as defined in \eqref{eq:def:divp:i} are uniformly continuous, from which we can obtain the following useful fact.

\begin{fact}
  \label{fact:open}
  Suppose the assumption \eqref{eq:pos} holds. Then, for all $t \geq 0$ and $i \in \{0, 1\}$, $\Dc_i(t)$ is open.
\end{fact}

To better illustrate the separability under reducible decoders, we introduce
 notations as follows. %

For all given $\Ac_0, \Ac_1 \subset \pdist$ and a reduction operator $\redu$, we define the sets $\redset_i(\Ac_0, \Ac_1; \redu)$ for $i = 0, 1$, such that for $j \in \{0, 1\}$ and $\jbar \defeq 1 - j$, 
\begin{subequations}
\begin{align}
  \redset_i(\Ac_0, \Ac_1; \redux^{(j)}) &\defeq \Ac_i \subx \Ac_{\jbar},\\
  \redset_i(\Ac_0, \Ac_1; \reduy^{(j)}) &\defeq \Ac_i \suby \Ac_{\jbar},
\end{align}
\label{eq:redset:1}
\end{subequations}
and, for each composite reduction operator $\redu \circ \redu'$, 
\begin{align}
  \redset_i(\Ac_0, \Ac_1; \redu \circ \redu') \defeq
  \redset_i(\Ac_0', \Ac_1'; \redu),
  \label{eq:redset:2}
\end{align}
where $\Ac_j' \defeq \redset_j(\Ac_0, \Ac_1; \redu')$ for $j \in \{0, 1\}$.

Then, we have the following useful fact, which can be verified by definition.
\begin{fact}
  \label{fact:conv:open}
  If $\Ac_0, \Ac_1 \subset \pdist$ are open and convex, then for each reduction operator $\redu$ and $i \in \{0, 1\}$, $\redset_i(\Ac_0, \Ac_1; \redu)$ is open and convex. 
\end{fact}

The following fact, as an immediate consequences of \propref{prop:rec}, is also useful.
\begin{fact}
  \label{fact:reducible}
  Suppose $\Ac_0, \Ac_1 \subset \pdist$, and $\dec$ is a reducible decoder which can be reduced to $\deca \defeq \redu(\dec)$ by some reduction operator $\redu$. Then, we have $ \dec \mid (\Ac_0, \Ac_1)$  if and only if  $\deca \mid (\redset_0(\Ac_0, \Ac_1; \redu), \redset_1(\Ac_0, \Ac_1; \redu))$.
\end{fact}

In addition, our proof will make use of the following result.
\begin{lemma}
  \label{lem:Decd}
  Suppose $\Ac_0, \Ac_1 \subset \pdist$ are open and convex. Then, for each $\dec \in \Decd_{M_X, M_Y}$ that separates $(\Ac_0, \Ac_1)$, there exists $\dec' \in \Dec_{M_X, M_Y}$ with $\kappa(\dec') <  \kappa(\dec)$, such that $\dec' \mid (\Ac_0, \Ac_1)$, where  $\kappa\colon \Dec_{M_X, M_Y} \to \mathbb{N}$ such that for each $\kappa(\dec) = 0$ if  $\dec \in \Decc_{M_X, M_Y}$, and $\kappa(\dec) \defeq \min\{L_X, L_Y\}$ for $\dec \in \Deccn_{M_X, M_Y}$, where we have assumed that $\redus(\dec) \in \Dec_{L_X, L_Y}$ for some $L_X, L_Y \geq 2$, and where $\redus(\dec)$ denotes the reduced form of $\dec$.
\end{lemma}

\begin{IEEEproof}[Proof of \lemref{lem:Decd}]
To begin, suppose $\dec \in \Decd_{M_X, M_Y}$ for some $M_X, M_Y \geq 2$, and let $\deca \defeq \redus(\dec)$ denote the reduced form of $\dec$, as defined in \propref{fact:reduced}. Furthermore, suppose $\deca$ can be reduced from $\dec$ by a reduction operator $\redu$, i.e., $\deca = \redu(\dec)$, and that $\deca \in \Dec_{L_X, L_Y}$ for some $L_X \leq M_X$ and $L_Y \leq M_Y$. Without loss of generality, we assume that, for all $(m_X, m_Y) \in \msgset{L_X} \times \msgset{L_Y}$, 
 $ \deca(m_X, m_Y) = \dec(m_X, m_Y)$.

Then, for $i \in \{0, 1\}$, we define $\Ac_i' \defeq \redset_i(\Ac_0, \Ac_1; \redu)$, with $\redset_i$ as defined in \eqref{eq:redset:1}--\eqref{eq:redset:2}. Then, since $(\Ac_0, \Ac_1)$ is separable by $\dec$, from \factref{fact:reducible} we know that $(\Ac_0', \Ac_1')$ is separable by $\deca$.

In addition, as both $\Ac_0$ and $\Ac_1$ are convex and open, from \factref{fact:conv:open} that, $\Ac_0'$ and $\Ac_1'$ are also convex and open. Then, from the definition of $\Decd_{M_X, M_Y}$ [cf. \eqref{eq:def:decs:decd}], $\deca$ is decomposable with the decomposition
\begin{align}
  \deca = \deca_0 \oplus \deca_1 \oplus \ibar \label{eq:decomp:deca}    
\end{align}
for some $i \in \{0, 1\}$, where $\deca_0, \deca_1 \in \Dec_{L_X, L_Y}$ satisfy \eqref{eq:icx:icy:empty}.

Therefore, it follows from \lemref{lem:decomp} that $(\Ac_0', \Ac_1')$ is separable by $\deca_0$  or $\deca_1$. Furthermore, let us define $\dec_0, \dec_1 \in \Dec_{M_X, M_Y}$ such that, for each $j \in \{0, 1\}$, 
\begin{align*}
  &  \dec_j(m_X, m_Y)%
   \defeq\begin{cases}
     \deca_j(m_X, m_Y)
& \text{if }(m_X, m_Y) \in \msgset{L_X} \times \msgset{L_Y},\\
     \dec(m_X, m_Y)
& \text{otherwise.}
  \end{cases}
\end{align*}
Then, it can be verified that $(\Ac_0, \Ac_1)$ is separable by  $\dec_0$  or $\dec_1$.%

It remains to verify that $\kappa(\dec_j) < \kappa(\dec)$. To see this, note that
from the definition of $\kappa(\cdot)$, for both $j \in \{0, 1\}$, we have
\begin{align}
  \kappa(\dec_j) &= \kappa(\deca_j) \label{eq:kappa:1}\\
                 &\leq \min\bigl\{|\Icx^{(i)}(\deca_j)|, |\Icy^{(i)}(\deca_j)|\bigr\}\label{eq:kappa:2}\\
                 &< \min\{L_X, L_Y\}\label{eq:kappa:3}\\
                 &= \kappa(\deca) = \kappa(\dec), \label{eq:kappa:4}
\end{align}
where $\Icx^{(1)}(\cdot)$ and $\Icy^{(1)}(\cdot)$ are as defined in  \eqref{eq:def:icx:icy}, and where to obtain \eqref{eq:kappa:2}--\eqref{eq:kappa:3} we have used \eqref{eq:icx:icy:empty}. 

\end{IEEEproof}

Our proof of \thmref{thm:sufficient} proceeds as follows. To begin, when $M_Y = 1$, we have $\Decd_{M_X, M_Y} \subset \Deccn_{M_X, M_Y} = \varnothing$, and \thmref{thm:sufficient} is trivially true. Thus, it suffices to consider the case $M_X, M_Y \geq 2$. In particular, note that for all $t > 0$, $\Dc_i(t)$ is convex and open, where the openness follows from \factref{fact:open}. To see the convexity, first note that from the convexity of KL divergence, $\divp_i$ is also convex (see, e.g., \cite[Example 3.17]{boyd2004convex}). Therefore, $\Dc_i(t)$, as a strict sublevel set of $\divp_i$, is also convex.

Therefore, it follows from \thmref{thm:separation} and \lemref{lem:Decd} that, for each $\dec \in \Decd_{M_X, M_Y}$ and error exponent pair $(E_0, E_1) \in \Ec[\dec]$, there exists $\dec' \in \Dec_{M_X, M_Y}$, such that
\begin{align}
 (E_0, E_1) \in \Ec[\dec']\quad \text{and} \quad
  \kappa(\dec') <  \kappa(\dec).
  \label{eq:to:prove}
\end{align}

Therefore, from \eqref{eq:to:prove}, for each $\dec \in \Decd_{M_X, M_Y}$ and $(E_0, E_1) \in \Ec[\dec]$, we can obtain some $\dec'$ satisfying \eqref{eq:to:prove}. Similarly, if $\dec' \in \Decd_{M_X, M_Y}$, we can again apply \eqref{eq:to:prove} to obtain an $M_X \times M_Y$ decoder $\dec''$ with $\kappa(\dec'') < \kappa(\dec')$ and  $(E_0, E_1) \in \Ec[\dec'']$.

In addition, since $\kappa(\cdot)$ is non-negative, for each $\dec \in \Decd_{M_X, M_Y}$ and error exponent pair $(E_0, E_1) \in \Ec[\dec]$, we can repeatedly apply these procedures to obtain
\begin{align*}
\dect \in  \Dec_{M_X, M_Y} \setminus \Decd_{M_X, M_Y} =  \Decc_{M_X, M_Y} \cup \Decs_{M_X, M_Y},
\end{align*}
such that $(E_0, E_1) \in \Ec[\dect]$, which demonstrates $\Ec[\Decd_{M_X, M_Y}] \subset \Ec[\{\decr_{M_X, M_Y}, \bar{\decr}_{M_X, M_Y}\}] \union \Ec[\Decs_{M_X, M_Y}]$.

Therefore, from \lemref{lem:th:suff} we obtain
\begin{align}
   \Ec[\{\decr_{M_X, M_Y}, \bar{\decr}_{M_X, M_Y}\}] \union \Ec[\Decs_{M_X, M_Y}] \subset
    \Ec[\Dec_{M_X, M_Y}] &= \Ec[\Decc_{M_X, M_Y}] \union \Ec[\Decs_{M_X, M_Y}] \union \Ec[\Decd_{M_X, M_Y}]\\
    &\subset  \Ec[\{\decr_{M_X, M_Y}, \bar{\decr}_{M_X, M_Y}\}] \union \Ec[\Decs_{M_X, M_Y}] \label{eq:preceq:1},
\end{align}
which implies $\Ec[\Decs_{M_X, M_Y}] = \Ec[\{\decr_{M_X, M_Y}, \bar{\decr}_{M_X, M_Y}\}]$, i.e., \eqref{eq:region:general}.
\hfill\IEEEQED

  \section{Proof of \thmref{thm:opt:dec}}
\label{app:thm:opt:dec}
  To begin, for each $\dec \in \Dec_{M_X, M_Y}$, we define the bipartite graph $G_{\dec} = (U, V, E_{\dec})$ with the vertex sets
    $U \defeq \{\vtx_{m_X} \colon m_X \in \msg{M_X} \}$ and $ V \defeq \{\vty_{m_Y} \colon m_Y \in \msg{M_Y} \}$,
    and the edge sets
    $  E_{\dec} \defeq \{(\vtx_{m_X}, \vty_{m_Y}) \colon \dec(m_X, m_Y) = 1\}$,
    where $(\vtx_{m_X}, \vty_{m_Y})$ represents the undirected edge connecting $\vtx_{m_X}$ and $\vty_{m_Y}$. This establishes the one-to-one correspondence between decoders and bipartite graphs, and it can be verified that, the decision matrix $\Ab$ associated with $\dec$ corresponds to the biadjacency matrix of $G_{\dec}$. %

    We then illustrate that if $\dec$ is indecomposable and irreducible, then both $G_{\dec}$ and $G_{\bar{\dec}}$ are connected. To this end, first note that since $\dec$ is irreducible, there exists no isolated vertex in $G_{\dec}$.

    Now, suppose $G_{\dec}$ is disconnected and can be divided into bipartite graphs $G^{(0)} = (U_0, V_0, E^{(0)})$ and $G^{(1)} = (U_1, V_1, E^{(1)})$, with non-empty vertex sets $U_0, U_1, V_0, V_1$ satisfying
    \begin{gather*}
      U = U_0 \union U_1,\quad U_0 \cap U_1 = \varnothing,\\
      V = V_0 \union V_1,\quad V_0 \cap V_1 = \varnothing.
    \end{gather*}

    Let $\dec_0$ and $\dec_1$ be the decoders associated with $G^{(0)}$ and $G^{(1)}$, respectively. Then, it can be verified that $\dec$ satisfies \eqref{eq:decomp} with $i = 0$, and thus is decomposable, which contradicts our assumption.  Therefore, $G_{\dec}$ is connected. Via a symmetry argument, we can show that $G_{\bar{\dec}}$ is also connected.

    Therefore, we obtain
    \begin{subequations}
    \begin{align}
      |E_{\dec}| &\geq |U| + |V| - 1\\
      |E_{\bar{\dec}}| &\geq |U| + |V| - 1,
    \end{align}\label{eq:edge}
  \end{subequations}
    where we have used the simple fact that each connected graph with $k$ vertices has at least $k - 1$ edges.
    
    From \eqref{eq:edge}, we obtain
    \begin{align*}
      M_XM_Y &= |E_{\dec}| + |E_{\bar{\dec}}|\\
      &\geq 2(|U| + |V| - 1 )= 2(M_X + M_Y - 1),
    \end{align*}
    which is equivalent to
        \begin{align}
      (M_X - 2)(M_Y - 2) \geq 2.
      \label{eq:mx:my:geq2}      
        \end{align}
     As a result, if $(M_X - 2)(M_Y - 2) < 2$, no $M_X\times M_Y$ decoder is both indecomposable and irreducible.

     It suffices to establish \eqref{eq:mx:my:rate}. To this end, we first demonstrate that $\Decs_{M_X, M_Y} = \varnothing$. Otherwise, for each $\dec \in \Decs_{M_X, M_Y}$, let $\deca \defeq \redus(\dec)$. Then, we have $\deca \in \Dec_{L_X, L_Y}$ for some $L_X \leq M_X, L_Y \leq M_Y$. This implies that $\deca$ is both irreducible and indecomposable, and $(L_X - 2)(L_Y - 2) \geq 2$, which contradicts previous argument.

     Hence, from \thmref{thm:sufficient}, we have
\begin{align*}
  \Ec(0_{M_X}, 0_{M_Y}) 
  &= \Ec[\{\decr_{M_X, M_Y}, \bar{\decr}_{M_X, M_Y}\}] \union \Ec[\Decs_{M_X, M_Y}]\\
  &= \Ec[\{\decr_{M_X, M_Y}, \bar{\decr}_{M_X, M_Y}\}]\\
  &= \Ec[\decr_{M_X, M_Y}] \union \Ec[\bar{\decr}_{M_X, M_Y}].
\end{align*}

\hfill\IEEEQED

 \section{Proof of \propref{prop:single}}
 \label{app:prop:single}
  First, we define $R_{\max} \defeq \max\{H(P_X^{(0)}), H(P_X^{(1)})\}$ with $H(\cdot)$ representing the entropy. Then, due to the inclusion chain
  \begin{align}
    \Ec(0_{2^{M_{\!Y}}}, 0_{M_Y}) &\subset \Ec(0_{M_X}, 0_{M_Y})\notag\\
    &\subset \Ec(R_X, 0_{M_Y})  \subset \Ec(R_{\max}, 0_{M_Y}),
    \label{eq:chain}    
  \end{align}
  it suffices to demonstrate $\Ec(0_{2^{M_{\!Y}}}, 0_{M_Y}) = \Ec(R_{\max}, 0_{M_Y})$.
  
  Specifically, note that under the constraints $(R_{\max}, 0_{M_Y})$, the decoder can obtain the full side information of the $X$ sequence. Then, for each $n \geq 1$, the corresponding coding scheme can be characterized as a encoder $g_n$ that encodes $Y^n$, and a central decoder  $\dec_n\colon \X^n \times \msgset{M_Y} \to \{0, 1\}$. When nodes $\nx$ and $\ny$ observe sequences $X^n = x^n$ and $Y^n = y^n$, respectively, the decision at the center can be represented as $\Hhs = \dec_n(x^n, g_n(y^n))$.  

  Then, we introduce a new encoder $f_n \colon \X^n \to \msgset{2^{M_X}}$ for encoding $X^n$, such that
  \begin{align*}
    f_n(x^n) \defeq \sum_{j \in \msgset{M_Y}} \dec_n(x^n, j) \cdot 2^{j}, \quad\text{for all }x^n \in \X^n.
  \end{align*}
  We also define decoder $\dec' \colon \msgset{2^{M_Y}}\times  \msgset{M_Y}  \to \{0, 1\}$ as
  \begin{align*}
    \dec'(m_X, m_Y) \defeq b_{m_Y}, \quad (m_X, m_Y) \in \msg{2^{M_Y}}\times \msgset{M_Y},
  \end{align*}
  where for each $j \in \msgset{M_Y}$, $b_j \in \{0, 1\}$ denotes the $(j+1)$-th digit of the binary representation of $m_X$, such that
  \begin{align*}
    m_X = \left(b_{M_Y-1} \cdots b_1b_0\right)_2 \defeq \sum_{j \in \msgset{M_Y}} b_{j} \cdot 2^{j}.
  \end{align*}
  It can be verified that for each $x^n \in \X^n$ and $y^n \in \Y^n$, the decision $\Hhs'$ associated with the coding scheme $(f_n, g_n, \phi')$ is
  \begin{align*}
    \Hhs' = \dec'(f_n(x^n), g_n(y^n)) \equiv \dec_n(x^n, g_n(y^n)) = \Hhs.
  \end{align*}
  Therefore, for each coding scheme under the rate constraints $(R_{\max}, 0_{M_Y})$, there exists a coding scheme satisfying constraints $(0_{2^{M_{\!Y}}}, 0_{M_Y})$ which obtains the same decision result. Hence, we have $\Ec(R_{\max}, 0_{M_Y}) \subset \Ec(0_{2^{M_{\!Y}}}, 0_{M_Y})$, and it follows from \eqref{eq:chain} that $\Ec(0_{2^{M_{\!Y}}}, 0_{M_Y}) = \Ec(R_{\max}, 0_{M_Y})$.
  \hfill\IEEEQED

\section{Proof of \thmref{thm:cond:indep}}
\label{app:cond:indep}
For given $M_X, M_Y$, note that if $\Deccn_{M_X, M_Y}$ and $\Decc_{M_X, M_Y}$ satisfy
    \begin{align}
    \Deccn_{M_X, M_Y} \preceq \Decc_{M_X, M_Y},
    \label{eq:deccn:decc}    
  \end{align}
  from \factref{fact:preceq} we have
  \begin{align}
    \Dec_{M_X, M_Y} = \Deccn_{M_X, M_Y} \union \Decc_{M_X, M_Y}
    \preceq \Decc_{M_X, M_Y},
    \label{eq:dec:decc}
  \end{align}
  and thus
  \begin{align}
    \Ec(0_{M_X}, 0_{M_Y}) = \Ec[\Dec_{M_X, M_Y}]
    &= \Ec[\Decc_{M_X, M_Y}]\\
    &= \Ec[\{\decr_{M_X, M_Y}, \bar{\decr}_{M_X, M_Y}\}],
  \end{align}
  where the first equality follows from \factref{fact:dec}, where the second equality follows from \eqref{eq:dec:decc}, and where the last equality follows from \lemref{lem:th:suff}.

  Therefore, it suffices to establish \eqref{eq:deccn:decc}. Note that if $M_Y = 1$, then $\Deccn_{M_X, M_Y} = \varnothing$, and \eqref{eq:dec:decc} is trivially true. We then establish \eqref{eq:deccn:decc} for $M_X \geq M_Y \geq 2$. To this end, we show that for each $\dec \in \Deccn_{M_X, M_Y}$, there exists $\dec' \in \Decc_{M_X, M_Y}$, such that $\Ec[\dec] \subset \Ec[\dec']$.

  To begin, note that from statement \ref{mono:S2} of \propref{prop:monotonic}, $\dec$ has at least one irreducible $2\times 2$ subdecoder [cf. \eqref{eq:irreducible:2x2}]. Without loss of generality, we assume
  \begin{align*}
    \dec(0, 0) = \dec(1, 1) = 0,\\
    \dec(1, 0) = \dec(1, 0) = 1.
  \end{align*}

  By symmetry, it suffices to consider the case
  \begin{align}
    P^{(0)}_{XY} =  P^{(0)}_{X}P^{(0)}_{Y}.\label{eq:cond:ind}    
  \end{align}
  Let $\dec^{(0)} \defeq \dec$, and suppose $f_n \colon \X^n\to \msg{M_X}$ and $g_n \colon \Y^n\to \msg{M_Y}$ are some given encoders. Then, we define $\dec^{(1)}$ as
  \begin{align}
    &\dec^{(1)}(m_X, m_Y)%
      \defeq
    \begin{cases}
      0 & \text{if } (m_X, m_Y) = (j_X, \jbar_X),\\
      \dec^{(0)}(m_X, m_Y) & \text{otherwise},
    \end{cases}
                              \label{eq:def:dec1}
  \end{align}
  where we have defined %
  \begin{align}
    j_X \defeq \argmin_{j \in \{0, 1\}}  \prob{f_n(X^n) = j|\Hs = 0}%
    \label{eq:def:jx}
  \end{align}
  and $\jbar_X \defeq 1 - j_X$.

  For $k = 0, 1$, let $\Cc_n^{(k)} \defeq (f_n, g_n, \dec^{(k)})$ denote the corresponding coding schemes. Then, it can be verified that the type-I and type-II errors for $\Cc_n^{(1)}$ satisfy
  \begin{subequations}
  \begin{align}
    \pi_0(\Cc_n^{(1)}) &\leq 2\cdot \pi_0(\Cc_n^{(0)}),     \label{pi:0:ind}   \\
   \pi_1(\Cc_n^{(1)}) &\leq \pi_1(\Cc_n^{(0)}).\label{pi:1:ind}
  \end{align}
  \label{eq:pi:ind}
\end{subequations}
  To establish \eqref{pi:0:ind}, note that
  \begin{align}
      \pi_0(\Cc_n^{(1)}) - \pi_0(\Cc_n^{(0)})%
    &= \prob{(f_n(X^n), g_n(Y^n)) = (j_X, \jbar_X)|\Hs = 0}\\
    &= \prob{f_n(X^n) = j_X|\Hs = 0}\prob{g_n(Y^n) = \jbar_X|\Hs = 0}\label{eq:bd:pi:1}\\
    &\leq \prob{f_n(X^n) = \jbar_X|\Hs = 0}\prob{g_n(Y^n) = \jbar_X|\Hs = 0}\label{eq:bd:pi:2}\\
    &= \prob{(f_n(X^n), g_n(Y^n)) = (\jbar_X, \jbar_X)|\Hs = 0}\label{eq:bd:pi:3}\\
    &\leq \pi_0(\Cc_n^{(0)}),
      \label{eq:bd:pi:4}
  \end{align}
  where \eqref{eq:bd:pi:1} and \eqref{eq:bd:pi:3} follow from \eqref{eq:cond:ind}, and where \eqref{eq:bd:pi:2} follows from \eqref{eq:def:jx}.
  
  Moreover, \eqref{pi:1:ind} follows from the simple fact that, for all $(m_X, m_Y) \in \msgset{M_X} \times \msgset{M_Y}$,
  \begin{align*}
    \dec^{(1)}(m_X, m_Y) = 1 \quad\text{implies}\quad  \dec^{(0)}(m_X, m_Y) = 1.
  \end{align*}

  Furthermore, if $\dec^{(1)} \notin \Decc_{M_X, M_Y}$, we can define $\dec^{(2)}$ similar to \eqref{eq:def:dec1}. Similarly, for each $k \geq 0$, we define $\dec^{(k+1)}$ if $\dec^{(k)} \notin \Decc_{M_X, M_Y}$. Then we can demonstrate that, there exists $k' \leq M_XM_Y - 1$, such that $\dec^{(k')} \in \Decc_{M_X, M_Y}$. Indeed, note that we have, for all $k \geq 0$, 
  \begin{align*}    
   0 \leq \sigma_{XY}(\dec^{(k)}) &= \sigma_{XY}(\dec^{(0)}) - k\leq M_XM_Y - 1 - k,
  \end{align*}
 where we have defined, for each $\dec \in \Dec_{M_X, M_Y}$,
  \begin{align*}
    \sigma_{XY}(\dec) \defeq \sum_{m_X \in \msgset{M_X}}\sum_{m_Y\in \msgset{M_Y}}\dec(m_X, m_Y).
  \end{align*}

  In addition, similar to \eqref{eq:def:dec1}, for each $k$ we have
    \begin{subequations}
  \begin{align}
    \pi_0(\Cc_n^{(k)}) &\leq (k + 1)\cdot \pi_0(\Cc_n^{(0)}),     \label{pi:0:ind:k}   \\
   \pi_1(\Cc_n^{(k)}) &\leq \pi_1(\Cc_n^{(0)}).\label{pi:1:ind}
  \end{align}
  \label{eq:pi:ind:k}
\end{subequations}
This implies that
\begin{subequations}
\begin{gather}
    \pi_0(\Cc_n^{(k')}) \leq  (k' + 1) \pi_0(\Cc_n^{(0)}) \leq M_XM_Y \cdot \pi_0(\Cc_n^{(0)}),      \\
   \pi_1(\Cc_n^{(k')}) \leq \pi_1(\Cc_n^{(0)}).
 \end{gather}
 \label{eq:sub:pi:01}
\end{subequations}
Finally, let $\dec' \defeq \dec^{(k')} \in \Decc_{M_X, M_Y}$. Then, since the encoders $f_n$ and $g_n$ can be arbitrarily chosen, it follows from \eqref{eq:sub:pi:01} that $\Ec[\dec'] \subset \Ec[\dec]$, which completes the proof.   \hfill\IEEEQED

\section{Proof of \propref{prop:decr:coding:all}} %
\label{app:prop:decr:coding:all} 

Our proof makes use of the following fact.
\begin{fact}
  \label{fact:inclusion}
  For all $\Ac_0, \Ac_1 \subset \pdist$ and $k \geq 2$, we have
  \begin{align}
    \Ac_0 \sub_k \Ac_1 \subset \Ac_0 \sub_{k-2} \Ac_1,        \label{eq:sub:k-2}    
  \end{align}
  \begin{subequations}
    \begin{align}
      \projx{\Ac_0 \sub_{k} \Ac_1} &\subset \projx{\Ac_0 \sub_{k-1} \Ac_1},\\
      \projy{\Ac_0 \sub_{k} \Ac_1} &\subset \projy{\Ac_0 \sub_{k-1} \Ac_1},                                     \label{eq:inclusion:proj}
    \end{align}
  \end{subequations}
    and
    \begin{align}
      \Ac_0 \sub_k \Ac_1 = \Ac_{\chi_k} \sub\, (\Ac_0 \sub_{k-1} \Ac_1).
      \label{eq:Ac:sub}
    \end{align}
  \end{fact}

  \begin{IEEEproof}[Proof of \factref{fact:inclusion}]
    From \defref{def:sub}, we have
    \begin{align}
      \Ac_0 \sub_k \Ac_1
      &=  (\Ac_0 \sub_{k-2} \Ac_1) \sub\, (\Ac_0 \sub_{k-1} \Ac_1)\notag\\
      &=  (\Ac_0 \sub_{k-2} \Ac_1) \cap  \left(\projx{\Ac_0 \sub_{k-1} \Ac_1} \times \projy{\Ac_0 \sub_{k-1} \Ac_1}\right),\label{eq:subk:def}
    \end{align}
    from which we can obtain \eqref{eq:sub:k-2} and
    \begin{gather}
          \Ac_0 \sub_k \Ac_1
          \subset \projx{\Ac_0 \sub_{k-1} \Ac_1} \times \projy{\Ac_0 \sub_{k-1} \Ac_1}.
          \label{eq:sub:k}
    \end{gather}

    From \eqref{eq:sub:k} and \eqref{eq:def:proj}, we can readily obtain \eqref{eq:inclusion:proj}. As a result, we can rewrite \eqref{eq:subk:def} as
    \begin{align}
      \Ac_0 \sub_k \Ac_1
      &= (\Ac_0 \sub_{k-2} \Ac_1) \cap  \left(\projx{\Ac_0 \sub_{k-1} \Ac_1} \times \projy{\Ac_0 \sub_{k-1} \Ac_1}\right)\notag\\
      &= (\Ac_0 \sub_{k-4} \Ac_1) \cap  \left(\projx{\Ac_0 \sub_{k-3} \Ac_1} \times \projy{\Ac_0 \sub_{k-3} \Ac_1}\right)\notag\\
      &\qquad\cap  \left(\projx{\Ac_0 \sub_{k-1} \Ac_1} \times \projy{\Ac_0 \sub_{k-1} \Ac_1}\right) \label{eq:subk:rec:1}\\
      &= (\Ac_0 \sub_{k-4} \Ac_1) \cap  \left(\projx{\Ac_0 \sub_{k-1} \Ac_1} \times \projy{\Ac_0 \sub_{k-1} \Ac_1}\right) \label{eq:subk:rec:2}\\
      &= \dots = (\Ac_0 \sub_{\chi_k} \Ac_1) \cap  \left(\projx{\Ac_0 \sub_{k-1} \Ac_1} \times \projy{\Ac_0 \sub_{k-1} \Ac_1}\right)\\
      &= \Ac_{\chi_k}\cap  \left(\projx{\Ac_0 \sub_{k-1} \Ac_1} \times \projy{\Ac_0 \sub_{k-1} \Ac_1}\right)\label{eq:subk:rec:4}\\
      &= \Ac_{\chi_k} \sub\, (\Ac_0 \sub_{k-1} \Ac_1),
      \label{eq:subk:rec}      
    \end{align}
    where to obtain \eqref{eq:subk:rec:2} we have used \eqref{eq:sub:k-2}, and to obtain \eqref{eq:subk:rec:4} we have used the fact that $\Ac_{\chi_k} = \Ac_0 \sub_{\chi_k} \Ac_1$.
\end{IEEEproof}

The following simple fact is also useful.
\begin{fact}
    \label{fact:rm}
    Given $M \geq 1$, for all $k_0, k_1 \in \msgset{M}$, we have $\decr_{M,M}(r_M(k_0), r_M(k_1)) = \chi_{ k_0 \wedge k_1}$, where $k_0 \wedge k_1 \defeq \min\{k_0, k_1\}$.
  \end{fact}
  \begin{IEEEproof}[Proof of \factref{fact:rm}]
    If $M = 1$, we have $k_0 = k_1 = 0$ and $\decr_{1, 1}(k_0, k_1) = 0$, and the claim is trivially true.

    If $M = 2$, we have $r_M(k) = k$ for $k \in \msgset{M} = \{0, 1\}$. Then, for all $k_0, k_1 \in \{0, 1\}$, we have $k_0 \wedge k_1 \in \{0, 1\}$ and
    \begin{align*}
      \decr_{M, M}(r_M(k_0), r_M(k_1)) = 
      \decr_{M, M}(k_0, k_1)
      &= \kron_{\{k_0 + k_1 \geq 2\}}\\
      &= \kron_{\{k_0 \wedge k_1 = 1\}}\\
      &= \chi_{k_0 \wedge k_1}.                               
    \end{align*}

    For the general case with $M > 2$, we will make use of the following properties of $r_M(\cdot)$, which can be verified by definition.
    \begin{itemize}
    \item For each $k \in \msgset{M-2}$, we have
      \begin{subequations}
        \begin{align}
          r_M(k + 2) > r_M(k) \quad&\text{if $\chi_k = 0$},\\
          r_M(k + 2) < r_M(k) \quad&\text{if $\chi_k = 1$}.
        \end{align}
        \label{eq:mono:rm}
      \end{subequations}
    \item For each $k \in \msgset{M-1}$, we have
      \begin{align}
        r_M(k) + r_M(k+1) = M - 1 + \chi_k.
        \label{eq:k:k+1:sum} 
      \end{align}
    \item For each $k \in \msgset{M}$, we have
      \begin{alignat}{3}
        &r_M(k) \leq \frac12 (M + \chi_M) - 1  &\quad&\text{if $\chi_k = 0$},      \label{eq:rM:chi:0}
        \\
        &r_M(k) \geq \frac12 (M + \chi_M)  &&\text{if $\chi_k = 1$}.      \label{eq:rM:chi:1}
      \end{alignat}
    \end{itemize}

    To establish \factref{fact:rm}, without loss of generality we assume $k_0 \leq k_1$. To begin, we consider the case $\chi_{k_0} = \chi_{k_1}$. If $\chi_{k_0} = \chi_{k_1} = 0$, from \eqref{eq:rM:chi:0} we have
    \begin{align*}
      r_M(k_0) + r_M(k_1) \leq M + \chi_M - 2 < M,
    \end{align*}
    which implies that $\decr_{M, M}(r_M(k_0), r_M(k_1)) = 0 = \chi_{k_0}$. Similarly, when $\chi_{k_0} = \chi_{k_1} = 1$, from  \eqref{eq:rM:chi:1} we have
    \begin{align*}
      r_M(k_0) + r_M(k_1) \geq M + \chi_M \geq M,
    \end{align*}
    and thus $\decr_{M, M}(r_M(k_0), r_M(k_1)) = 1 = \chi_{k_0}$.

    Moreover, if $\chi_{k_0} \neq \chi_{k_1}$, then we have $\chi_{k_1} = \chi_{k_0 + 1}$ and  $k_1 \geq k_0 + 1$. Specifically, if $(\chi_{k_0}, \chi_{k_1}) = (0, 1)$, then 
    \begin{align*}
      r_M(k_0) + r_M(k_1) \leq r_M(k_0) + r_M(k_0 + 1) = M - 1,
    \end{align*}
    where the inequality follows from \eqref{eq:mono:rm}, and where the equality follows from \eqref{eq:k:k+1:sum}. Hence, we obtain $\decr_{M, M}(r_M(k_0), r_M(k_1)) = 0 = \chi_{k_0}$. Similarly, if $(\chi_{k_0}, \chi_{k_1}) = (1, 0)$, then from \eqref{eq:mono:rm} and \eqref{eq:k:k+1:sum} we have
    \begin{align*}
      r_M(k_0) + r_M(k_1) \geq r_M(k_0) + r_M(k_0 + 1) = M,
    \end{align*}
    which implies  $\decr_{M, M}(r_M(k_0), r_M(k_1)) = 1 = \chi_{k_0}$.
  \end{IEEEproof}
In addition, the following proposition is also useful. %
\begin{proposition}
  \label{prop:opt:type:enc}
  Given $\Ac_0, \Ac_1 \subset \pdist$, let us define  
  \begin{align}
    \enctix'(Q_X)
    &\defeq
      \begin{cases}
        \max \{k \geq 0 \colon Q_X \in \projx{\Ac_0 \sub_k \Ac_1}\}& \text{if $Q_X \in \projx{\Ac_0 \cup \Ac_1}$},\\
        0& \text{otherwise},\\
      \end{cases}
\\ %
    \enctiy'(Q_Y)
    &\defeq
      \begin{cases}
        \max \{k \geq 0 \colon Q_Y \in \projy{\Ac_0 \sub_k \Ac_1}\}& \text{if $Q_Y \in \projy{\Ac_0 \cup \Ac_1}$},\\
        0& \text{otherwise},\\
      \end{cases}
  \end{align}
  for all $Q_X \in \cP^\X$ and $Q_Y \in \cP^\Y$,  where $\projx{\cdot}$ and $\projy{\cdot}$ are as defined in \eqref{eq:def:proj}, and ``$\,\sub_k$'' is as defined in \defref{def:sub}.  Then, if $ \decr_{M, M} \mid (\Ac_0, \Ac_1)$ for some $M \geq 1$, we have
  \begin{align}
    \enctix'(Q_X), \enctiy'(Q_Y) \in \msgset{M},\quad\text{for all $(Q_X, Q_Y) \in \pdist$},
    \label{eq:in:M}
  \end{align}
    and [cf. \eqref{eq:def:dec:enct}]
    \begin{align}
      \decr_{M, M}(\enct'_X(Q_X), \enct'_Y(Q_Y)) = i, \quad\text{for all } (Q_X, Q_Y) \in \Ac_i,
      \label{eq:sep:decr}
    \end{align}
    for both $i \in \{0, 1\}$, where we have defined $\enctx' \defeq r_{M} \circ \enctix'$ and $\encty' \defeq r_{M} \circ \enctiy'$.
  \end{proposition}

  \begin{IEEEproof}[Proof of \propref{prop:opt:type:enc}]
    First, from \eqref{eq:sub:k-2},
    we can obtain the sequences of nested sets%
    \begin{align*}
      \Ac_0 &= (\Ac_0 \sub_0 \Ac_1) \supset %
              \dots \supset (\Ac_0 \sub_{2k} \Ac_1) \supset (\Ac_0 \sub_{2k+2} \Ac_1) \supset\cdots
    \end{align*}
    and    
    \begin{align*}
    \Ac_1 &= (\Ac_1 \sub_1 \Ac_1)  \supset %
              \dots \supset (\Ac_0 \sub_{2k+1} \Ac_1) \supset (\Ac_0 \sub_{2k+3} \Ac_1) \supset\cdots.
    \end{align*}
    Suppose $(Q_X, Q_Y) \in \Ac_i$ for some $i \in \{0, 1\}$. Let us define
  \begin{align*}
    k' \defeq \max\{k \geq 0 \colon (Q_X, Q_Y) \in \Ac_0 \sub_{2k + i} \Ac_1 \},
  \end{align*}
  then we have
  \begin{gather}
    (Q_X, Q_Y) \in \Ac_0 \sub_{2k' + i} \Ac_1,
    \label{eq:in}    \\
    (Q_X, Q_Y) \notin \Ac_0 \sub_{2(k'+1) + i} \Ac_1.
    \label{eq:notin}
  \end{gather}

  From \factref{fact:inclusion}, we obtain
  \begin{align*}
    \Ac_0 \sub_{2(k'+1) + i} \Ac_1
    &= \Ac_i \sub\, (\Ac_0 \sub_{2k' + i + 1} \Ac_1)\\
    &= \Ac_i \cap \left(\projx{\Ac_0 \sub_{2k' + i + 1} \Ac_1} \times \projy{\Ac_0 \sub_{2k' + i + 1} \Ac_1}\right),
  \end{align*}
  and it follows from $(Q_X, Q_Y) \in \Ac_i$ and \eqref{eq:notin} that
  \begin{align}
    Q_X \notin \projx{\Ac_0 \sub_{2k' + i + 1} \Ac_1}\quad\text{or}\quad
    Q_Y \notin \projy{\Ac_0 \sub_{2k' + i + 1} \Ac_1}.
    \label{eq:notin:xy}
  \end{align}

   In addition, from \eqref{eq:in}, we have
   \begin{align}
     Q_X \in \projx{\Ac_0 \sub_{2k' + i} \Ac_1} \quad\text{and}\quad Q_Y \in \projy{\Ac_0 \sub_{2k' + i} \Ac_1}.
     \label{eq:in:xy}
   \end{align}

  Combining \eqref{eq:notin:xy}--\eqref{eq:in:xy} and \eqref{eq:inclusion:proj}, we obtain
  \begin{align*}
    \min\bigl\{\enctix'(Q_X), \enctix'(Q_Y)\bigr\} =  2k' + i.
  \end{align*}
  Therefore, 
  \begin{align*}
    \decr_{M, M}(\enctx'(Q_X), \enctx'(Q_Y))
    &= \decr_{M, M}(r_M(\enctix'(Q_X)), r_M(\enctiy'(Q_Y))) \\
    &= \chi_{2k' + i}\\
    &= i,
  \end{align*}
  where to obtain the second equality we have used \factref{fact:rm}.
\end{IEEEproof}

Proceeding to our proof of \propref{prop:decr:coding:all}, we first show that for each given $(E_0, E_1)$ and all $k \geq 1$, the $\Qc_X^{(k)}$ and $\Qc_Y^{(k)}$ as defined in \eqref{eq:def:Qc} satisfy
\begin{subequations}
  \begin{align}
    \Qc_X^{(k)} = \projx{\Ac_0 \sub_k \Ac_1},  \label{eq:proj:Qc:x}
\\
    \Qc_Y^{(k)} = \projy{\Ac_0 \sub_k \Ac_1},\label{eq:proj:Qc:y}
  \end{align}
  \label{eq:proj:Qc}
\end{subequations}
and where we have defined
\begin{align}
  \Ac_0 \defeq
  \begin{cases}
    \Dc_1(E_1)& \text{if $\dec = \bar{\decr}_{M, M}$,}\\
    \Dc_0(E_0)& \text{otherwise,}
  \end{cases}
                \quad\text{and}\quad
                \Ac_1\defeq
  \begin{cases}
    \Dc_1(E_1)&\text{if $\dec = \decr_{M, M}$,}\\
    \Dc_0(E_0)&\text{if $\dec = \bar{\decr}_{M, M}$,}\\    \Dc_1(E_1) \subx \Dc_0(E_0)&\text{if $\dec = \dec_{M_X, M_Y}$}.
  \end{cases}
                            \label{eq:ac:dc}
\end{align}

We then verify \eqref{eq:proj:Qc} for decoder $\dec = \decr_{M, M}$, and the other two cases can be similarly established. First, when $k = 1$, we have
\begin{align}
  \projx{\Ac_0 \sub_1 \Ac_1}
  = \projx{\Ac_1}
  &= \projx{\Dc_1(E_1)}\notag\\
  &= \bigl\{Q_X \in \cP^\X\colon D(Q_X\|P_X^{(1)})< E_1\bigr\}= \Qc_X^{(1)}
    \label{eq:qc:k:1}
\end{align}
and, similarly, $\projy{\Dc_0(E_0) \sub_1 \Dc_1(E_1)} = \Qc_Y^{(1)}$. 

Suppose \eqref{eq:proj:Qc} holds for $k = \ell \geq 1$. For $k = \ell + 1 \geq 2$, it follows from  \factref{fact:inclusion} that
\begin{align}
  \Ac_0 \sub_k \Ac_1
  &= \Ac_{\chi_k} \sub\,(\Ac_0 \sub_{k-1} \Ac_1)\notag\\
  &= \Ac_{\chi_k} \cap (\projx{\Ac_0 \sub_{k-1} \Ac_1} \times \projy{\Ac_0 \sub_{k-1} \Ac_1})\notag\\
  &= \Ac_{\chi_k} \cap \bigl(\Qc_X^{(k - 1)} \times \Qc_Y^{(k - 1)}\bigr).
    \label{eq:ac:sub}
\end{align}
As a result, we have
\begin{align*}
    \projx{\Ac_0 \sub_k \Ac_1}%
    &= \bigl\{Q_X \in \Qc_X^{(k - 1)} \colon (Q_X, Q_Y) \in \Ac_{\chi_k} \text{ for some } Q_Y \in  \Qc_Y^{(k - 1)} 
    \bigr\} \\
    &= \bigl\{Q_X \in \Qc_X^{(k - 1)} \colon \divp_{\chi_k}(Q_X, Q_Y) < E_{\chi_k} \text{ for some } Q_Y \in  \Qc_Y^{(k - 1)} %
    \bigr\}\\ %
    &= \bigl\{Q_X \in \Qc_X^{(k - 1)} \colon \divp_{\chi_k}(Q_X, \Qc_Y^{(k - 1)}) < E_{\chi_k}        \bigr\}\\ %
  &= \Qc_X^{(k)}.
\end{align*}
By a symmetry argument, we can also obtain \eqref{eq:proj:Qc:y}. Hence, \eqref{eq:proj:Qc} holds for all $k \geq 1$.

Then, from \thmref{thm:exp:th}, $(E_0, E_1) \in \Ec[\dec]$ if and only if
\begin{align}
  \Ac_0 \sub_{M} \Ac_1 = \varnothing.
  \label{eq:sub:nothing}
\end{align}
In addition, from \eqref{eq:ac:sub}, it can be verified that
\begin{align}
\Ac_0 \sub_M \Ac_1
  &= \Ac_{\chi_M} \cap \bigl(\Qc_X^{(M - 1)} \times \Qc_Y^{(M - 1)}\bigr),\notag\\
  &= \Dc_{\hat{\chi}_M}(E_{\hat{\chi}_M}) \cap \bigl(\Qc_X^{(M - 1)} \times \Qc_Y^{(M - 1)}\bigr).
    \label{eq:sub:M:Qc}
\end{align}

Hence, \eqref{eq:sub:nothing} is equivalent to \begin{align}
  \divp_{\hat{\chi}_M}(\Qc_X^{(M - 1)}, \Qc_Y^{(M - 1)}) \geq E_{\hat{\chi}_M}, 
\end{align}
which is \eqref{eq:decr:err}.

Finally, with the correspondence \eqref{eq:ac:dc}, it follows from \propref{prop:opt:type:enc} that, for all $(E_0, E_1) \in \Ec[\dec]$ and both $i = 0, 1$, we have
\begin{align}
  \decr_{M, M}(\enctx(Q_X), \encty(Q_Y)) = i, \quad\text{for all } (Q_X, Q_Y) \in \Ac_i, %
  \label{eq:decr:enct}
\end{align}
where $\enctx$ and $\encty$ are as defined in \eqref{eq:def:enct:th}. This implies that [cf. \eqref{eq:def:dec:enct}]
\begin{align}
  \dec(\enctx(Q_X), \encty(Q_Y)) = i, \quad\text{for all } (Q_X, Q_Y) \in \Dc_i(E_i).
  \label{eq:decr:enct}
\end{align}
Therefore, it follows from \thmref{thm:separation} that each exponent pair $(E_0, E_1) \in \interior(\Ec[\dec])$ can be achieved by the type-encoding functions  $\enctx$ and $\encty$. 
\hfill\IEEEQED

\section{Computation of Error Exponent Region and Type-encoding Functions}
\label{app:comp}

For convenience, we focus on the decoder $\decr_{M, M}$, and the computation of $\bar{\decr}_{M, M}$ and ${\decr}_{M_X, M_Y}$ is similar. From \eqref{eq:decr:err}, for all $M \geq 1$, $(E_0, E_1) \in \Ec[\decr_{M,M}]$ if and only if
\begin{align*}
  \divp_{\chi_M}\bigl(\Qc_X^{(M - 1)},  \Qc_Y^{(M - 1)}\bigr) \geq E_{\chi_M}.
\end{align*}
From the definition of $\divp_i$ [cf. \eqref{eq:def:divp:i}], this is equivalent to
\begin{align}
  \left\{Q_{XY}\in \cP^{\X\times\Y}\colon D(Q_{XY}\|P_{XY}^{(\chi_M)}) \geq E_{\chi_M},  [Q_{XY}]_X \in \Qc_X^{(M - 1)}, [Q_{XY}]_Y \in \Qc_Y^{(M - 1)}\right\} \neq \varnothing.
  \label{eq:err:exp:equiv}
\end{align}

Moreover, from \eqref{eq:def:Qc}, for all $k \geq 1$ and $Q_X \in \cP^\X$, $Q_X \in \Qc_X^{(k)}$ if and only if
\begin{align*}
  Q_X \in \Qc_X^{(k-1)}\quad\text{and}\quad \divp_{\chi_k}(Q_X, \Qc_Y^{(k -1)}) < E_{\chi_k},
\end{align*}
which is equivalent to
\begin{align}
  \left\{Q_{XY}\in \cP^{\X\times\Y}\colon  D(Q_{XY}\|P_{XY}^{(\chi_k)}) < E_{\chi_k},
[Q_{XY}]_X = Q_X \in \Qc_X^{(k-1)}, [Q_{XY}]_Y \in \Qc_Y^{(k -1)}\right\} \neq \varnothing.
\label{eq:Qc:x}
\end{align}
Similarly, $Q_Y \in \Qc_Y^{(k)}$ if and only if
\begin{align}
  \left\{Q_{XY}\in \cP^{\X\times\Y}\colon    D(Q_{XY}\|P_{XY}^{(\chi_k)}) < E_{\chi_k},
  [Q_{XY}]_Y = Q_Y \in \Qc_Y^{(k-1)}, [Q_{XY}]_X \in \Qc_X^{(k -1)}\right\} \neq \varnothing.
\label{eq:Qc:y}
\end{align}

We first consider the computation of error exponent regions $\decr_{M, M}$. Specifically, when $M = 2$, combining \eqref{eq:err:exp:equiv} and \eqref{eq:qc:xy:1}, we have $(E_0, E_1) \in \Ec[\decr_{2, 2}]$ if and only if there exists $Q_{XY} \in \cP^{\X \times \Y}$, such that
\begin{gather*}
  D(Q_{XY}\|P_{XY}^{(0)}) \geq E_0,\\
  D([Q_{XY}]_X \| P_{X}^{(1)}) < E_1, \\
  D([Q_{XY}]_Y \| P_{Y}^{(1)}) < E_1.
\end{gather*}

Therefore, for each given $E_0$, the optimal $E_1$ achieved by $\decr_{2, 2}$ is given by the optimal value of the convex programming problem
\begin{subequations}
  \begin{alignat}{2}
    &\minimize_{t, Q_{XY}}& \qquad & t\\
    &\st  &&   Q_{XY} \in \cP^{\X \times \Y},\\
    &  &&   D(Q_{XY} \| P^{(0)}_{XY}) \leq E_0,\\
    &  &&   D([Q_{XY}]_X \| P_{X}^{(1)}) \leq t,\\
    &  &&   D([Q_{XY}]_Y \| P_{Y}^{(1)}) \leq t,
  \end{alignat}
\end{subequations}

Similarly, for $M = 3$, we have
 $(E_0, E_1) \in \Ec[\decr_{3, 3}]$ if and only if there exists $Q_{XY} \in \cP^{\X \times \Y}$, such that
\begin{gather}
  D(Q_{XY}\|P_{XY}^{(1)}) \geq E_1,\label{eq:qc:xy:2}\\
  [Q_{XY}]_X \in \Qc_X^{(2)} \quad\text{and}\quad   [Q_{XY}]_Y \in \Qc_Y^{(2)}.\label{eq:qc:x:y:2}
\end{gather}
In addition, from \eqref{eq:Qc:x}, \eqref{eq:qc:x:y:2} is equivalent to
\begin{align}
  \left\{Q'_{XY}\in \cP^{\X\times\Y}\colon  D(Q'_{XY}\|P_{XY}^{(0)}) < E_0,
[Q'_{XY}]_X = [Q_{XY}]_X \in \Qc_X^{(1)}, [Q'_{XY}]_Y \in \Qc_Y^{(1)}\right\} \neq \varnothing
  \label{eq:Qc:x:2'}
\end{align}
and
\begin{align}
  \left\{Q''_{XY}\in \cP^{\X\times\Y}\colon  D(Q''_{XY}\|P_{XY}^{(0)}) < E_0,
[Q''_{XY}]_Y = [Q_{XY}]_Y \in \Qc_Y^{(1)}, [Q''_{XY}]_X \in \Qc_X^{(1)}\right\} \neq \varnothing,
  \label{eq:Qc:y:2'}
\end{align}
respectively.

As a result, combining \eqref{eq:qc:xy:2},  \eqref{eq:qc:xy:1}, and \eqref{eq:Qc:x:2'}--\eqref{eq:Qc:y:2'}, for each given $E_0$, the optimal $E_1$ achieved by $\decr_{3, 3}$ is given by the optimal value of the convex programming problem
\begin{subequations}
  \begin{alignat}{2}
    &\minimize_{t, Q_{XY}, Q'_{XY}, Q''_{XY}}& \qquad & t\\
    &\quad\st  &&   Q_{XY}, Q'_{XY}, Q''_{XY} \in \cP^{\X \times \Y},\\
    &  &&   D(Q_{XY} \| P^{(1)}_{XY}) \leq t,\\
    &  &&   [Q'_{XY}]_X = [Q_{XY}]_X, \quad D([Q'_{XY}]_Y \| P_{Y}^{(1)}) \leq t, \quad D(Q'_{XY} \| P^{(0)}_{XY}) \leq E_0,\\
    &  &&   [Q''_{XY}]_Y = [Q_{XY}]_Y,\quad  D([Q''_{XY}]_X \| P_{X}^{(1)}) \leq t, \quad D(Q''_{XY} \| P^{(0)}_{XY}) \leq E_0.
  \end{alignat}
\end{subequations}

The computation of error exponent region $\Ec[\decr_{M, M}]$ for general $M$ can be obtained similarly. Moreover, the type-encoding functions $\enctx, \encty$ as defined in \eqref{eq:decr:err} can also be computed in a similar manner. As an illustrative example,
we consider the computation of $\enctx$ with $\decr_{3, 3}$ used as the decoder. It can be verified that
\begin{align*}
  \enctx(Q_X) =
  \begin{cases}
    0&\text{if }Q_X \notin \Qc_X^{(1)},\\
    1&\text{if }Q_X \in \Qc_X^{(2)},\\
    2&\text{if }Q_X \in \Qc_X^{(1)}\setminus\Qc_X^{(2)}.
  \end{cases}
\end{align*}

From \eqref{eq:qc:xy:1}, for each given $Q_X$, it is straightforward to decide whether $Q_X \in \Qc_X^{(1)}$ or not, and it suffices to verify if $Q_X \in \Qc_X^{(2)}$. From \eqref{eq:Qc:x},  $Q_X \in \Qc_X^{(2)}$ if and only if
\begin{align}
  \left\{Q_{XY}\in \cP^{\X\times\Y}\colon  D(Q_{XY}\|P_{XY}^{(0)}) < E_0,
[Q_{XY}]_X = Q_X \in \Qc_X^{(1)}, [Q_{XY}]_Y \in \Qc_Y^{(1)}\right\} \neq \varnothing,
\end{align}
which is equivalent to
\begin{subequations}
  \begin{align}
    D(Q_X \| P_{X}^{(1)}) \leq E_1
    \label{eq:in:qc:1}
  \end{align}
  and
  \begin{align}
    \min_{\substack{Q_{XY} \in \cP^{\X \times \Y}\colon
    [Q_{XY}]_X = Q_X,\\
    D([Q_{XY}]_Y \| P_{Y}^{(1)}) \leq E_1}} D(Q_{XY}\|P_{XY}^{(0)}) \leq E_0,
    \label{eq:in:qc:2}
  \end{align}
\end{subequations}
where \eqref{eq:in:qc:1} is due to $Q_X \in \Qc_X^{(1)}$.

Therefore, for each given $Q_X \in \cP^\X$, we have $\enctx(Q_X) = 0$ if and only if \eqref{eq:in:qc:1} cannot be satisfied; In addition, $\enctx(Q_X) = 1$ if and only if both \eqref{eq:in:qc:1} and \eqref{eq:in:qc:2} hold, and $\enctx(Q_X) = 2$ if and only if $Q_X$ satisfies \eqref{eq:in:qc:1} but does not satisfy \eqref{eq:in:qc:2}.

%
%
%
%
%
%
%

%
%

\section{Proof of \propref{prop:optim:cond}}
\label{app:prop:optim:cond}
Our proof makes use of the following two facts.
\begin{fact}
  \label{fact:cond-ind}
  Suppose $P^{(i)}_{XY} = P^{(i)}_{X}P^{(i)}_{Y}$ for some $i \in \{0, 1\}$. Then, for all $(Q_X, Q_Y) \in \pdist$, we have
  \begin{align*}
    \divp_i(Q_X, Q_Y) = D(Q_X\|P_X^{(i)}) + D(Q_Y\|P_Y^{(i)}).
  \end{align*}
\end{fact}
\begin{IEEEproof}[Proof of \factref{fact:cond-ind}]
  For all $Q_{XY}$ satisfying $[Q_{XY}]_X = Q_X$ and $[Q_{XY}]_Y = Q_Y $, we have
  \begin{align*}
    D(Q_{XY}\|P^{(i)}_{XY})
    &= D(Q_X \| P^{(i)}_{X}) + \sum_{x \in \X} Q_X(x) D(Q_{Y|X=x}\|P_{Y}^{(i)})\\
    &\geq D(Q_X \| P^{(i)}_{X}) +  D([Q_{Y|X}Q_X]_Y\|P_{Y}^{(i)})\\
     &= D(Q_X \| P^{(i)}_{X}) +  D(Q_Y\|P_{Y}^{(i)})\\
     &= D(Q_XQ_Y \| P^{(i)}_{XY}),
  \end{align*}
  where the inequality follows from Jensen's inequality.
  
  As a result, from the definition \eqref{eq:def:divp:i}, we have
  \begin{align*}
    \divp_i(Q_X, Q_Y)
    &=  \min_{\substack{Q_{XY}\colon [Q_{XY}]_X = Q_X \\\qquad ~[Q_{XY}]_Y = Q_Y}}  D(Q_{XY}\|P^{(i)}_{XY})\\
    &= D(Q_XQ_Y \| P^{(i)}_{XY})\\
    &= D(Q_X\|P^{(i)}_{X}) + D(Q_Y\|P^{(i)}_Y).
  \end{align*}
\end{IEEEproof}

\begin{fact}
  \label{fact:gam}
  For all $k \geq 1$, the $\Qc_X^{(k)}$ and $\Qc_Y^{(k)}$ given by \eqref{eq:Q:k:xy} satisfy
  \begin{subequations}
  \begin{gather}
    \inf_{Q_X \in \Qc_X^{(k)}} D(Q_X \| P_X^{(\chi_{k + 1})}) = \lambda_X^{(\chi_{k+1})}(\gamma_X^{(k)}),\label{eq:fact:gam:x}
    \\
    \inf_{Q_Y \in \Qc_Y^{(k)}} D(Q_Y \| P_Y^{(\chi_{k + 1})}) = \lambda_Y^{(\chi_{k+1})}(\gamma_Y^{(k)}).\label{eq:fact:gam:y}    
  \end{gather}
\end{subequations}
\end{fact}
\begin{IEEEproof}[Proof of \factref{fact:gam}]
  To establish \eqref{eq:fact:gam:y}, note that
  \begin{align*}
    \inf_{Q_Y \in \Qc_Y^{(k)}} D(Q_Y \| P_Y^{(\chi_{k + 1})})
    &=
    \inf_{
    \substack{
      Q_Y \colon D(Q_{Y}\|P_{Y}^{(0)}) < \gamma_{Y}^{(k - \chi_k)} \\\qquad D(Q_{Y}\|P_{Y}^{(1)}) < \gamma_{Y}^{(k - \bar{\chi}_k)}}
    }  D(Q_{Y}\|P_{Y}^{(\chi_{k + 1})})\\
    &=\inf_{
    \substack{
      Q_Y \colon D(Q_{Y}\|P_{Y}^{(\chi_k)}) < \gamma_{Y}^{(k)} \\\qquad D(Q_{Y}\|P_{Y}^{(\bar{\chi}_k)}) < \gamma_{Y}^{(k - 1)}}
    }  D(Q_{Y}\|P_{Y}^{(\bar{\chi}_{k})})\\
    &=\inf_{Q_Y \colon D(Q_{Y}\|P_{Y}^{(\chi_k)}) < \gamma_{Y}^{(k)} }  D(Q_{Y}\|P_{Y}^{(\bar{\chi}_{k})})\\
    &= \lambda_Y^{(\bar{\chi}_{k})}(\gamma_Y^{(k)})\\
    &= \lambda_Y^{(\chi_{k+1})}(\gamma_Y^{(k)}).
  \end{align*}
  Similarly, \eqref{eq:fact:gam:x} can be proved via a symmetry argument.
\end{IEEEproof}

Proceeding to our proof of \propref{prop:optim:cond}, first note that from \eqref{eq:def:gamma}, 
we have
$  \gamma_X^{(1)} = \gamma_X^{(1)} = E_1$, 
and it follows from \eqref{eq:qc:xy:1} that \eqref{eq:Q:k:xy} holds for $k = 1$.

Suppose \eqref{eq:Q:k:xy} holds for some $k \geq 1$. Then, it suffices to  establish \eqref{eq:Q:k:xy} for $k + 1$, i.e., to demonstrate that
  \begin{align}
    \Qc_{X}^{(k + 1)} &= \left\{Q_X \in \cP^\X \colon D(Q_{X}\|P_{X}^{(0)}) < \gamma_{X}^{(k + 1 - \chi_{k + 1})},  D(Q_{X}\|P_{X}^{(1)}) < \gamma_{X}^{(k + 1 - \bar{\chi}_{k + 1})}\right\},    \label{eq:Q:k+1:x}\\
    \Qc_{Y}^{(k + 1)} &= \left\{Q_Y \in \cP^\Y \colon D(Q_{Y}\|P_{Y}^{(0)}) < \gamma_{Y}^{(k + 1 - \chi_{k + 1})},  D(Q_{Y}\|P_{Y}^{(1)}) < \gamma_{Y}^{(k + 1 - \bar{\chi}_{k + 1})}\right\}.
    \label{eq:Q:k+1:y}
  \end{align}

  In fact, it follows from \eqref{eq:def:Qc} that for all $Q_X \in \cP^\X$, $Q_X \in \Qc_X^{(k+1)}$ is equivalent to
  \begin{subequations}
  \begin{gather}
    Q_X \in \Qc_X^{(k)},\label{eq:cond:k+1:a}\\
  \divp_{\chi_{k+1}}(Q_X,\Qc_Y^{(k)}) < E_{\chi_{k+1}}.\label{eq:cond:k+1:b}
  \end{gather}
    \label{eq:cond:k+1}    
  \end{subequations}
  From \eqref{eq:def:divp:inf} and \factref{fact:cond-ind}, we obtain
  \begin{align}
  \divp_{\chi_{k+1}}(Q_X,\Qc_Y^{(k)})
    &= \inf_{Q_Y \in \Qc_Y^{(k)}} \divp_{\chi_{k+1}}(Q_X, Q_Y)\\
    &= D(Q_X\|P_X^{(\chi_{k + 1})}) + \inf_{Q_Y \in \Qc_Y^{(k)}} D(Q_Y \| P_Y^{(\chi_{k + 1})})\\
    &=  D(Q_X\|P_X^{(\chi_{k + 1})}) + \lambda_Y^{(\chi_{k + 1})}(\gamma_Y^{(k)})%
  \end{align}
  where to obtain the second equality we have used \factref{fact:gam}.

  Therefore, \eqref{eq:cond:k+1:b} is equivalent to
  \begin{align}
    D(Q_X\|P_X^{(\chi_{k + 1})}) < E_{\chi_{k + 1}} - \lambda_Y^{(\chi_{k + 1})}(\gamma_Y^{(k)}) = \gamma_X^{(k+1)},
    \label{eq:div:x:k+1:new}
  \end{align}
  where the equality follows from the definition \eqref{eq:def:gamma}.
  
  Moreover, from \eqref{eq:Q:k:xy}, \eqref{eq:cond:k+1:a} is equivalent to
  \begin{align}
    D(Q_{X}\|P_{X}^{(\chi_k)}) < \gamma_{X}^{(k)}, \quad D(Q_{X}\|P_{X}^{(\chi_{k+1})}) < \gamma_{X}^{(k - 1)}.
    \label{eq:div:x:new}
  \end{align}

  In addition, from the fact that $\lambda_X^{(i)}$ and $\lambda_Y^{(i)}$ are monotonically decreasing functions for $i \in \{0, 1\}$, it can be verified that we have $\gamma_X^{(k+1)} \leq \gamma_{X}^{(k - 1)}$. Hence, combining \eqref{eq:div:x:new} and \eqref{eq:div:x:k+1:new} yields
   $ D(Q_{X}\|P_{X}^{(\chi_k)}) < \gamma_{X}^{(k)}$ and $D(Q_{X}\|P_{X}^{(\chi_{k+1})}) < \gamma_{X}^{(k + 1)}$,
  which imply  \eqref{eq:Q:k+1:x}. By a symmetry argument, we can establish \eqref{eq:Q:k+1:y}.

  Finally, the equivalence between \eqref{eq:decr:err} and \eqref{eq:err:gam} follows from that
  \begin{align}
    \divp_{\chi_M}\bigl(\Qc_X^{(M - 1)},  \Qc_Y^{(M - 1)}\bigr) - E_{\chi_M}
    &=  \inf_{Q_X \in \Qc_X^{(M - 1)}, Q_Y \in \Qc_Y^{(M - 1)}}\divp_{\chi_M}(Q_X,Q_Y) - E_{\chi_M}      \label{eq:prove:gam:1}
\\
    &=  \inf_{Q_X \in \Qc_X^{(M - 1)}} D(Q_X\|P_X^{(\chi_M)}) + \inf_{Q_Y \in \Qc_Y^{(M - 1)}} D(Q_Y\|P_Y^{(\chi_M)}) - E_{\chi_M}      \label{eq:prove:gam:2}
\\
    &= \lambda_X^{(\chi_{M})}(\gamma_X^{(M-1)}) + \lambda_Y^{(\chi_{M})}(\gamma_Y^{(M-1)}) - E_{\chi_M}      \label{eq:prove:gam:3}
\\
    &= (E_{\chi_M} - \gamma_X^{(M)}) + (E_{\chi_M} - \gamma_Y^{(M)}) - E_{\chi_M}      \label{eq:prove:gam:4}
\\
    &= -(\gamma_X^{(M)} + \gamma_Y^{(M)} - E_{\chi_M}),
      \label{eq:prove:gam:1}
  \end{align}
  where to obtain \eqref{eq:prove:gam:1} we have used \eqref{eq:def:divp:inf},  
  to obtain  \eqref{eq:prove:gam:2} we have used \factref{fact:cond-ind},
  to obtain  \eqref{eq:prove:gam:3} we have used \factref{fact:gam},
  and to obtain  \eqref{eq:prove:gam:4} we have used \eqref{eq:def:gamma}.  
\hfill\IEEEQED

\bibliographystyle{IEEEtran}
\bibliography{ref}

\end{document}